\documentclass[aps,prb,twocolumn,showpacs,superscriptaddress,floatfix,nofootinbib,10pt]{revtex4-1}
\usepackage{graphicx}
\usepackage{amsmath}
\usepackage{epsfig}
\usepackage{helvet}
\usepackage{amssymb}

\newcommand{\bra}[1]{\mbox{$\langle #1 |$}}
\newcommand{\ket}[1]{\mbox{$| #1 \rangle$}}
\newcommand{\vecb}[1]{\mathord{\buildrel{\lower3pt\hbox{$\scriptscriptstyle\rightarrow$}} \over #1} }

\newcommand{\Wilson}{\mbox{\tiny W}}
\newcommand{\homog}{{}}
\newcommand{\defect}{\textrm{dfct}}
\newcommand{\impur}{\textrm{imp}}
\newcommand{\bound}{\textrm{bnd}}
\newcommand{\inter}{\textrm{intf}}
\newcommand{\dimpur}{2\times{ \textrm{imp}}}
\newcommand{\dbound}{{2\times \textrm{bnd}}}
\newcommand{\Yjun}{\textrm{YI}}

\def\tr{ \mbox{tr}}
\def\R{ \mathcal R}
\def\B{ \mathcal B}
\def\L{ \mathcal L}
\def\C{ \mathcal C}
\def\A{ \mathcal A}
\def\S{ \mathcal S}
\def\D{ \mathcal D}

\begin{document}

\title{Algorithms for entanglement renormalization: \\boundaries, impurities and interfaces}

\author{G. Evenbly}
\affiliation{Institute for Quantum Information and Matter, California Institute of Technology, MC 305-16, Pasadena CA 91125, USA}
\email{evenbly@caltech.edu}

\author{G. Vidal}
\affiliation{Perimeter Institute for Theoretical Physics, Waterloo, Ontario N2L 2Y5, Canada}  \email{gvidal@perimeterinstitute.ca}

\begin{abstract}
We propose algorithms, based on the multi-scale entanglement renormalization ansatz, to obtain the ground state of quantum critical systems in the presence of boundaries, impurities, or interfaces. By exploiting the theory of minimal updates [Ref. \onlinecite{Evenbly13}: G. Evenbly and G. Vidal, arXiv:1307.0831], the ground state is completely characterized in terms of a number of variational parameters that is independent of the system size, even though the presence of a boundary, an impurity, or an interface explicitly breaks the translation invariance of the host system. Similarly, computational costs do not scale with the system size, allowing the thermodynamic limit to be studied directly and thus avoiding finite size effects e.g. when extracting the universal properties of the critical system. 
\end{abstract}


\maketitle
\tableofcontents

%%%%%%%%%%%%%%%%%%%%%%%%%%%%%%%%%%%%%%%%%%%%%%%%%%%%%%%%%%%%%%%%%%%%%%%%%%%%%%%%%%%%%%%%
%%%%%%%%%%%%%%%%%%%%%%%%%%%%%%%%%%%%%%%%%%%%%%%%%%%%%%%%%%%%%%%%%%%%%%%%%%%%%%%%%%%%%%%%
\section{Introduction} \label{sect:Intro}

%%%%%% ER
Entanglement renormalization\cite{Vidal07} is a renormalization group (RG) approach to quantum many-body systems on a lattice. As with most RG methods \cite{Fisher98}, it proceeds by coarse-graining the microscopic degrees of freedom of a many-body system, and thus also their Hamiltonian $H$, to produce a sequence of effective systems, with Hamiltonians $\{H,H', H'', \cdots\}$ that define a flow towards larger length scale/lower energies. Entanglement renormalization operates in \textit{real space} (it does not rely on Fourier space analysis) and it is a \textit{non-perturbative} approach (that is, it can handle interactions of any strength). As a result, it has a wide range of applicability, from quantum criticality \cite{Evenbly10, Cincio08, Evenbly10b, Evenbly09, Pfeifer09, Evenbly09b, Montangero09, Evenbly10e,Silvi10, Evenbly10d, Vidal10, Evenbly11b} to emergent topological order \cite{Aguado08,Konig09,Aguado11,Tagliacozzo11,Buerschaper13,Haah13,Kao13}, from frustrated antiferromagnets \cite{Evenbly10c,Harada12,Lou12} to interacting fermions \cite{Corboz10,Pineda10,Corboz09} and even to interacting anyons \cite{Pfeifer10,Konig10}. Entanglement renormalization produces an efficient (approximate) representation of the ground state of the system in terms of a variational tensor network, the multi-scale entanglement renormalization ansatz (MERA) \cite{Vidal08}, from which one can extract expectation values of arbitrary local observables.

Most applications of the MERA have so far focused on systems that are translation invariant. Here we will consider instead systems where translation invariance is explicitly broken by the presence of a defect. For simplicity, we assume that the defect is placed on an infinite quantum critical system that, in the absence of the defect, would be both homogeneous (that is, translation invariant) and a fixed point of the RG (that is, scale invariant). Under that assumption, the MERA offers a shockingly simple description: in the absence of the defect it is completely characterized by a single pair of tensors ${u,w}$ and, in the presence of the defect, by just one additional tensor $v$ if the defect is also itself at a (scale invariant) fixed point of the RG flow; or by a sequence of a few additional tensors $\{v,v',v'', \cdots\}$ that describe its flow towards an RG fixed point.

In this paper we propose and benchmark algorithms for quantum critical systems in the presence of defects that exploit the simple description afforded by the MERA. We start by briefly reviewing the required background material on entanglement renormalization, including a recently proposed theory of minimal updates \cite{Evenbly13} that is at the core of the surprisingly compact MERA description of defects in quantum critical systems.

\subsection{RG with a variational tensor network}

Two distinctive aspects of entanglement renormalization are the \textit{tensor network} structure of the coarse-graining transformation and the \textit{variational} nature of the approach.

\begin{figure}[!tbhp]
\begin{center}
\includegraphics[width=8.5cm]{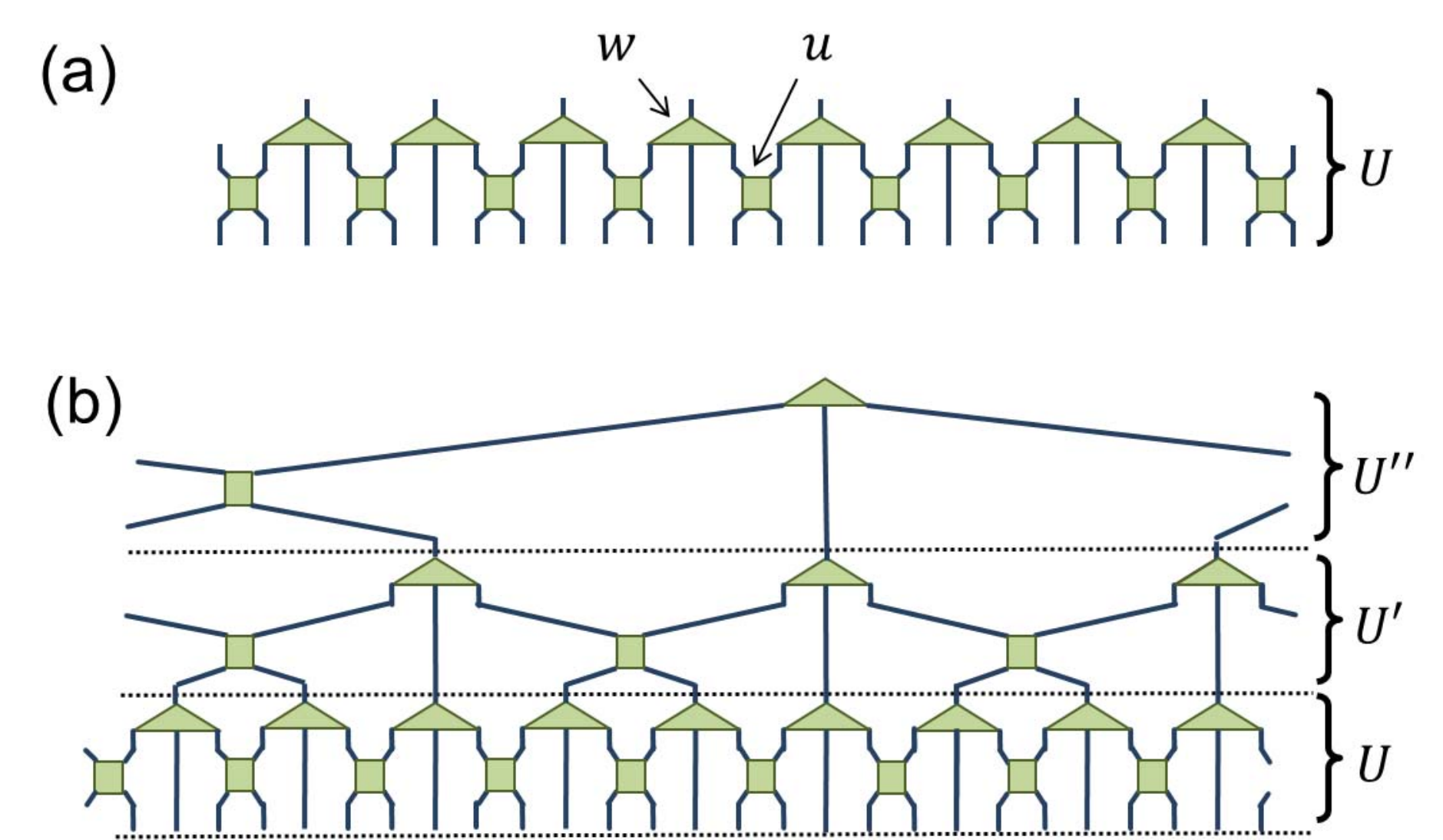}
\caption{(a) Coarse-graining transformation $U$ for a lattice in $D=1$ dimensions that decomposes as a tensor network made of disentanglers $u$, depicted as squares, and isometries $w$, depicted as triangles.
(b) The MERA on a $D=1$ dimensional lattice made of $27$ sites, obtained by collecting together a sequence of coarse-graining transformations $\{U, U', U''\}$.}
\label{fig:ER}
\end{center}
\end{figure}

The coarse-graining transformation is implemented by a linear (isometric) map $U$, relating the Hilbert spaces of the lattice system before and after coarse-graining. As illustrated in Fig. \ref{fig:ER}(a), the linear map $U$ decomposes as a network of tensors, called disentanglers $u$ and isometries $w$. The structure of the network has been designed with the important property that $U$ preserves locality: local operators are mapped into local operators. Thus, if $H$ is a short-ranged Hamiltonian, then the effective Hamiltonians $H'$,$H''$, etc, are also short-ranged.

On the other hand, the approach is variational. The disentanglers $u$ and isometries $w$ are loaded with variational parameters, which are determined through energy minimization. This ensures that the coarse-graining transformation $U$ is properly adapted to the system under consideration. That is, instead of deciding a priori which degrees of freedom should be kept and which should be thrown away, the method proceeds by asking the Hamiltonian $H$ which part of many-body Hilbert space corresponds to low energies and proceeds to safely remove the rest.

%%%%%%%%%%%%%%%%%%%%%%%%%%%%%%%%%%%%%
%%%%%%%%%%%%%%%%%%%%%%%%%%%%%%%%%%%%%
%%%%%%%%%%%%%%%%%%%%%%%%%%%%%%%%%%%%%%%
%%%%%%%%%%%%%%%%%%%%%%%%%%%%%%%%%%%%%%%

However, the most prominent feature of entanglement renormalization, setting it apart from other real space RG approaches, is its handling of short-range entanglement. While isometries $w$ map a block of sites into an effective site, and thus play a rather standard role in a coarse-graining transformation, disentanglers $u$ perform a more singular task: the removal of short-range entanglement from the system. Thanks to this removal, the coarse-graining transformation $U$ constitutes a proper implementation of the RG \cite{Proper}, in that the sequence of effective systems, with Hamiltonians $\{H, H', H'', \cdots\}$, only retain degrees of freedom corresponding to increasing length scales. In particular, at fixed-points of the RG flow, entanglement renormalization explicitly realizes scale invariance: the system before coarse-graining and the system after coarse-graining are seen to be locally identical.

\subsection{MERA and quantum criticality}

The MERA \cite{Vidal08} is the class of tensor network state \cite{{Cirac09a,Evenbly11}} that results from joining the sequence of coarse-graining transformations $\{U, U', U'', \cdots\}$, see Fig. \ref{fig:ER}(b). It is a variational ansatz for ground states (or, more generally, low energy states) of many-body systems on a lattice in $D$ spatial dimensions. By construction, the MERA extends in $D+1$ dimensions, where the additional dimension corresponds to length scale or RG flow. As a result, it is distinctly well suited to study systems where several length scales are relevant, because the information related to each length scale is stored in a different part of the network.

In particular, the MERA offers an extremely compact description of ground states of homogeneous systems at fixed points of the RG flow, that is, in systems with both translation invariance and scale invariance. These encompass both stable (gapped) RG fixed points, which include topologically ordered systems \cite{Aguado08,Konig09,Aguado11,Tagliacozzo11,Buerschaper13,Haah13,Kao13}, and unstable (gapless) RG fixed points, corresponding to quantum critical systems \cite{Evenbly10, Cincio08, Evenbly10b, Evenbly09, Pfeifer09, Evenbly09b, Montangero09, Evenbly10e,Silvi10, Evenbly10d, Vidal10, Evenbly11b}. Indeed, translation invariance leads to a position-independent coarse-graining transformation $U$, made of copies of a single pair of tensors $\{u,w\}$, whereas scale invariance implies that the same $U$ can be used at all scales. As a result, the single pair $(u,w)$ completely characterizes the state of an infinite system.

The study of quantum critical systems is therefore among the natural targets of the MERA. Until now, most applications of the MERA to quantum criticality have focused on systems that are invariant under translations (see, however, Refs. \onlinecite{Evenbly10e, Silvi10}). In translation invariant systems, the MERA provides direct access to the universal information of the quantum phase transition, as often encoded in the conformal data of an underlying conformal field theory\cite{Francesco97,Henkel99} (CFT) (see Appx. \ref{sect:MERA} for a review). In particular, in one spatial dimension one can extract the central charge and identify the set of primary scaling operators $\phi_i$ (both local \cite{Pfeifer09,Montangero09} and non-local \cite{Evenbly10d,Evenbly11b}) together with their scaling dimensions $\Delta_i$ (from which most critical exponents of the theory follow) as well as the corresponding operator product expansion coefficients. This data completely characterizes the underlying CFT.

\subsection{Defects in quantum critical systems}

%%%%%% Defects
The goal of this manuscript is to address quantum critical systems where the translation invariance of a system is explicitly broken by the presence of a \textit{boundary}, an \textit{impurity}, an \textit{interface}, etc. We refer to any such obstruction to translation invariance generically as a \textit{defect}, and to the system in the absence of the defects as the \textit{host} system. Methods for simulating quantum critical systems with such defects are important in order to understand and model their effects in realistic settings.

%%%%%%%%% computational cost
A major difficulty in addressing such systems is that, since the presence of a defect manifestly breaks the translation invariance of the host Hamiltonian, the ground state is no longer homogeneous. Instead, expectation values of local observables differ from the homogeneous case throughout the whole system by an amount that only decays as a power law with the distance to the defect. In this scenario a natural option (which we will not follow here) would be to choose a coarse-graining map $U$ with position-dependent disentanglers and isometries that adjust to the power law profile of ground state expectation values. Notice that the resulting MERA would be made of a large number (proportional to the system size) of inequivalent disentanglers and isometries, and would therefore incur much larger computational costs (again, proportional to the system size) than in a homogeneous system. Importantly, we would not be able to study infinite systems directly, and when extracting the low energy properties of the defect, these would be significantly contaminated by ubiquitous finite size effects, which vanish as a power law with the system size.

\subsection{A theory of minimal updates}

%%%%% Modularity
What one would like, then, is a MERA description of many-body systems with defects that is nearly as compact as in the homogeneous case. Fortunately, a recent theory of \emph{minimal updates} in holography \cite{Evenbly13} provides us with a recipe to obtain such a description. Let $H$ denote a local Hamiltonian for an extended many-body system on a $D$-dimensional lattice, and let $\tilde{H}$
\begin{equation}
\tilde{H} = H + J_{\R}, \label{s2e1}
\end{equation}
denote the Hamiltonian for the same system after we added a new term $J_{\R}$ localized in region $\R$. In addition, let $\ket{\psi}$ and $\ket{\tilde{\psi}}$ denote the ground states of the Hamiltonian $H$ and of Hamiltonian $\tilde{H}$ (the modified Hamiltonian), respectively. Then, the theory of minimal updates in holography \cite{Evenbly13} argues in favor of the following conjecture. \newline

\noindent \textit{\textbf{Conjecture (Minimal update):} A \textsc{MERA} for $\ket{\tilde{\psi}}$ can be obtained from a \textsc{MERA} for $\ket{\psi}$ by modifying the latter only in the causal cone $\C(\R)$ of region $\R$.}
\newline

\begin{figure}[!tbhp]
\begin{center}
\includegraphics[width=8.5cm]{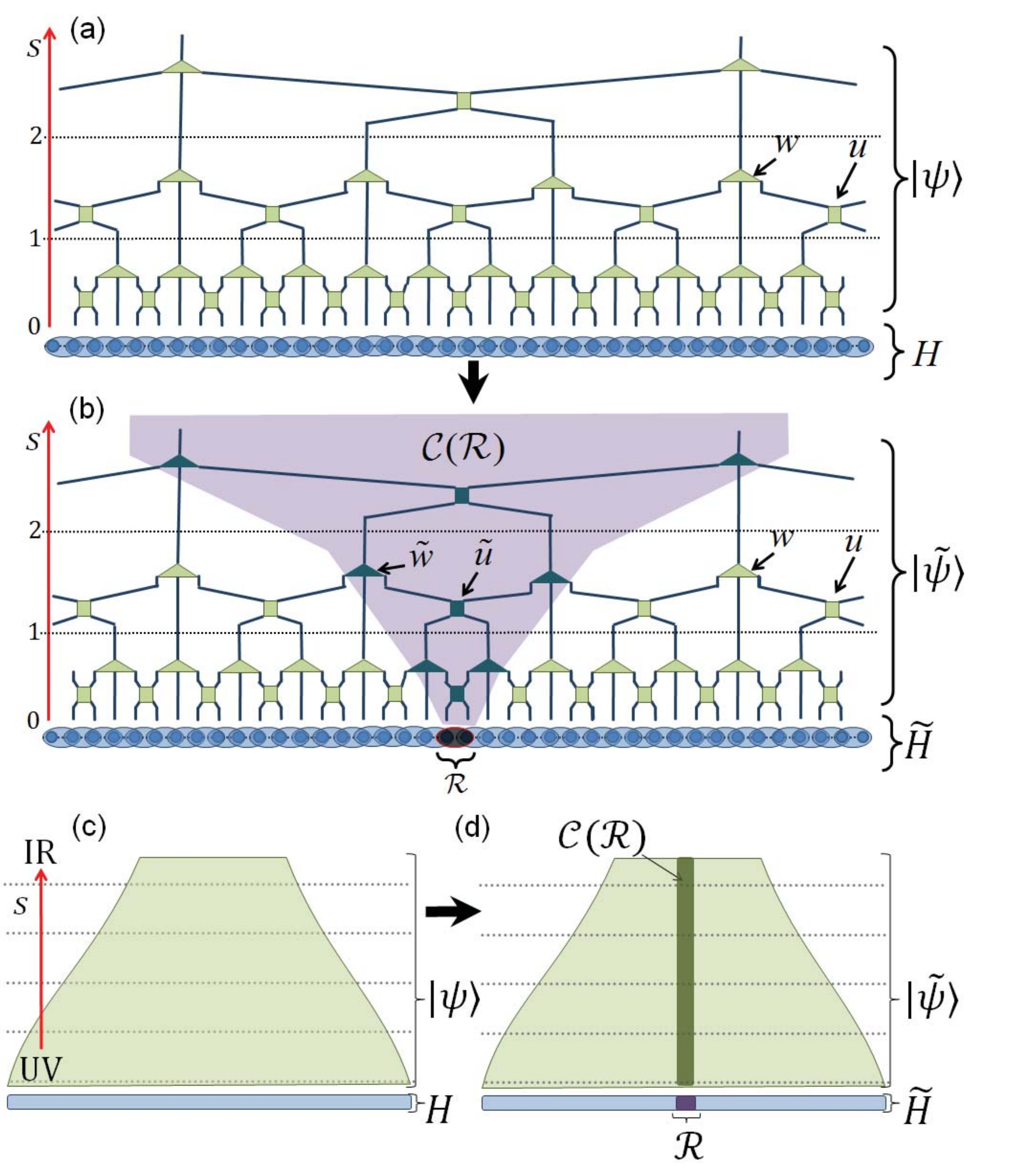}
\caption{(a) MERA tensor network for the ground state $\ket{\psi}$ of a lattice Hamiltonian $H$ in $D=1$ space dimensions. Scale and translation invariance result in a compact description: two tensors $\{ u,w \}$ are repeated throughout the infinite tensor network. (b) The theory of minimal updates dictates that the ground state $\ket{\tilde{\psi}}$ of the Hamiltonian $\tilde{H} = H + J_{\mathcal{R}}$ is represented by a MERA with the same tensors $\{ u,w \}$ outside the causal cone $\mathcal C(\mathcal R)$ (shaded), whereas inside $\mathcal C(\mathcal R)$ two new tensors $\{ \tilde{u},\tilde{w} \}$ are repeated throughout the semi-infinite causal cone. (c-d) The same illustrations, without drawing the tensors of the network.}
\label{fig:DirectedMERA}
\end{center}
\end{figure}

Here, the causal cone $\C(\R)$ of region $\R$ is the part of the MERA that describes the successive coarse-graining of region $\mathcal{R}$. For instance, for a region $\R$ consisting of two contiguous sites, Fig. \ref{fig:DirectedMERA} illustrates the causal cone $\C(\R)$. The figure also shows how a MERA for $\ket{\psi}$ should be modified to obtain a MERA for $\ket{\tilde{\psi}}$.

%%%%%%%%%%%%%%%%%%%%%%%%%%%%%%%%%%%%%
%%%%%%%%%%%%%%%%%%%%%%%%%%%%%%%%%%%%%
%%%%%%%%%%%%%%%%%%%%%%%%%%%%%%%%%%%%%%%
%%%%%%%%%%%%%%%%%%%%%%%%%%%%%%%%%%%%%%%

\subsection{Algorithms for critical systems with defects}

In this paper we propose and benchmark MERA algorithms for quantum critical system with one or several defects. The theoretical foundation of the algorithms is the above conjecture on minimal updates, specialized to a Hamiltonian of the form
\begin{equation}\label{eq:minimal}
    H^{\defect} = H + J_{\R}^{\defect},
\end{equation}
where $H$ is the Hamiltonian for the host system and $J_{\R}^{\defect}$ is the Hamiltonian describing the localized defect. More specifically, we will assume that the host Hamiltonian $H$, which describes an infinite system on a lattice, is a homogeneous, critical, fixed-point Hamiltonian, so that its ground state $\ket{\psi}$ can be succinctly described by a MERA that is characterized in terms of just a single pair of tensors $\{u,w\}$. Region $\R$ will typically consists of one or two sites.

Then, following the above conjecture, a MERA for the ground state $\ket{\psi^{\defect}}$ of the Hamiltonian $H^{\defect}$, which we call \textit{modular} MERA and will be further described in Sect. \ref{sect:Modularity}, is completely characterized in terms of two sets of tensors, see Fig. \ref{fig:DirectedMERA}. First, the pair of tensors $\{u,w\}$ corresponding to the (scale and translation invariant) host system, is repeated throughout the outside of the causal cone of the defect. Second, (for a defect that is scale invariant, that is, a fixed point of the RG flow) another pair of tensors $\{\tilde{u}, \tilde{w}\}$ is repeated throughout the inside of the causal cone of the defect. After some rewiring of the modular MERA, this second pair $\{\tilde{u},\tilde{w}\}$ will be replaced by a single tensor $v$.

[Some settings will require slight modifications of this simple description. For instance, in the case of interfaces involving several types of system, each system will contribute a different pair of tensors for the outside of the causal cone. On the other hand, if the defect is not yet at a fixed-point of the RG flow, then instead of a single tensor $v$, a sequence of scale-dependent tensors $\{v,v',v'', \cdots\}$ will be used to account for the flow of the defect into the RG fixed-point.]

The modular MERA leads to simple numerical algorithms for quantum critical systems in the presence of one of several defects, which complement and generalize those discussed in Ref. \onlinecite{Evenbly09} for homogeneous systems. As in the homogeneous case, the computational cost of the new algorithms is independent of the system size, allowing us to address infinite systems. In this way, we can extract the universal, low energy properties associated to a defect directly in the thermodynamic limit, where they are free of finite-size effects. Although in this paper we restrict our attention to systems in $D=1$ dimensions for simplicity, the key idea of the algorithms can also be applied to systems in $D>1$ dimensions. In the discussion in Sect. \ref{sect:Conclusion} we will also address how to lift the assumption, present throughout this work, that the host system is both translation and scale invariant.

The algorithms proposed in this paper are thus based on assuming the validity of the conjectured theory of minimal updates in holography of Ref. \onlinecite{Evenbly13}. We contribute to that theory in two ways. First, by applying the above conjecture recursively, we will investigate applications that go well beyond the simple scenario described in Ref. \onlinecite{Evenbly13}, namely that of a single impurity. Specifically, the \textit{modular} MERA describes the ground state of a complex system, such as an interface between two systems $A$ and $B$, by combining `modules' obtained by studying simpler systems, such as homogeneous versions of system $A$ and system $B$, separately. Modularity is central to the algorithms proposed in this work and key to their computational efficiency. Second, the benchmark results presented here constitute solid evidence that the conjectured minimal updates are indeed sufficient to accurately represent a large variety of defects. This contributes significantly to establishing the theory of minimal updates, which so far was supported mostly by the theoretical arguments provided in Ref. \onlinecite{Evenbly13}.

\subsection{Structure of the rest of the paper}

%%%%% Outline

In this paper we assume that the reader is already familiar with the scale invariant MERA for translation invariant systems (a detailed introduction to which can be found in Ref. \onlinecite{Evenbly11b}). However, for completeness, we have also included a brief review to the MERA in the presence of scale and translation invariance in Appx. \ref{sect:MERA}.

Sect. \ref{sect:Modularity} introduces the modular MERA and describes how they can be applied to quantum critical systems with an impurity, boundary, interface, and more complex settings, such as several defects or Y-interfaces involving three systems (also called Y-junctions). It also explains how to extract the low energy, universal properties of the defect.

Sect. \ref{sect:OptMod} discusses how to optimize the modular MERA. This is illustrated with the paradigmatic case of a single impurity. The first step involves optimizing a MERA for the homogeneous system (Refs. \onlinecite{Evenbly09,Evenbly11b}) so as to obtain the pair of tensors $\{u,w\}$. Then an effective Hamiltonian for the causal cone of the impurity, or \textit{Wilson chain}, is produced by properly coarse-graining the host Hamiltonian $H$ and adding the impurity term $J_{\R}$. Finally, a simplified tensor network ansatz for the ground state of the Wilson chain is optimized by energy minimization, from which one would be able to extract tensor $v$ (or tensors $\{v,v',v'',\cdots\}$.

Sect. \ref{sect:Bench} benchmarks the modular MERA algorithm for a number of quantum critical systems in $D=1$ spatial dimension. These include systems with one and several impurities, systems with one or two boundaries, interfaces between two systems, and Y-interfaces between three systems. For each type of defect, we outline how the basic algorithm of Sect. \ref{sect:OptMod} needs to be modified. The approach is seen to provide accurate numerical results for ground state properties, both for expectation values of local observables and for low energy, universal properties (e.g. in the form of conformal data describing an underlying CFT, including the critical exponents associated to the defect).

Finally, Sect. \ref{sect:Conclusion} concludes the paper with a discussion and a summary of results. We have also included three appendices. Appx. \ref{sect:MERA} provides a basic introduction to key aspects of ER and MERA used throughout the manuscript, and reviews how to extract universal properties (conformal data) from a translation and scale invariant MERA. Appx. B and C provide technical details on certain aspects of the modular MERA.

%%%%%%%%%%%%%%%%%%%%%%%%%%%%%%%%%%%%%%%%%%%%%%%%%%%%%%%%%%%%%%%%%%%%%%%%%%%%%%%%%%%%%%%%
%%%%%%%%%%%%%%%%%%%%%%%%%%%%%%%%%%%%%%%%%%%%%%%%%%%%%%%%%%%%%%%%%%%%%%%%%%%%%%%%%%%%%%%%
%%%%%%%%%%%%%%%%%%%%%%%%%%%%%%%%%%%%%%%%%%%%%%%%%%%%%%%%%%%%%%%%%%%%%%%%%%%%%%%%%%%%%%%%
%%%%%%%%%%%%%%%%%%%%%%%%%%%%%%%%%%%%%%%%%%%%%%%%%%%%%%%%%%%%%%%%%%%%%%%%%%%%%%%%%%%%%%%%
%%%%%%%%%%%%%%%%%%%%%%%%%%%%%%%%%%%%%%%%%%%%%%%%%%%%%%%%%%%%%%%%%%%%%%%%%%%%%%%%%%%%%%%%
%%%%%%%%%%%%%%%%%%%%%%%%%%%%%%%%%%%%%%%%%%%%%%%%%%%%%%%%%%%%%%%%%%%%%%%%%%%%%%%%%%%%%%%%
%%%%%%%%%%%%%%%%%%%%%%%%%%%%%%%%%%%%%%%%%%%%%%%%%%%%%%%%%%%%%%%%%%%%%%%%%%%%%%%%%%%%%%%%
%%%%%%%%%%%%%%%%%%%%%%%%%%%%%%%%%%%%%%%%%%%%%%%%%%%%%%%%%%%%%%%%%%%%%%%%%%%%%%%%%%%%%%%%
%%%%%%%%%%%%%%%%%%%%%%%%%%%%%%%%%%%%%%%%%%%%%%%%%%%%%%%%%%%%%%%%%%%%%%%%%%%%%%%%%%%%%%%%
%%%%%%%%%%%%%%%%%%%%%%%%%%%%%%%%%%%%%%%%%%%%%%%%%%%%%%%%%%%%%%%%%%%%%%%%%%%%%%%%%%%%%%%%

%\section{Modularity in the holographic description of ground states} \label{sect:Modularity}
\section{Modular MERA} \label{sect:Modularity}

In this section we introduce the modular MERA for homogeneous systems with one or several defects. We also explain how to extract the universal properties of a defect, including its set of scaling dimensions, from which one can derive all critical exponents associated to the defect. For simplicity, we only consider lattice systems in one spatial dimension.

The modular MERA is built upon the conjecture that the presence of a defect can be accurately accounted for by only updating the interior of the causal cone $\C(\R)$ of the region $\R$ on which the defect is supported. Below we will argue that, when applied recursively, this minimal update implies that we can describe e.g. an interface between two semi-infinite quantum critical spin chains by combining `modules' that describe the two systems individually, that is, in the absence of an interface. We refer to this property as \textit{modularity} in the holographic description of quantum states. Next we describe the modular MERA for systems with a single impurity, an open boundary, or an interface of two different quantum systems (notice that the impurity system can be considered as an interface of two identical systems, while the open boundary can be considered as an interface with a trivial system), before discussing more general applications of modularity, such as systems with multiple impurities or Y-interfaces of three quantum chains.

A note on terminology.--- We call \textit{modular} MERA any MERA for a system with one or several defects that, following the theory of minimal updates of Ref. \onlinecite{Evenbly13}, has been obtained from a MERA for the host system (that is, without the defects) by modifying only the tensors in the causal cone of the defects. On the other hand, for specific types of defects, such as an impurity, a boundary, etc, we also occasionally use the more specific terms \textit{impurity} MERA, \textit{boundary} MERA, etc, to denote the corresponding specific type of modular MERAs.

Throughout this section, the quantum critical, homogeneous host system is described by an infinite lattice $\mathcal{L}$ in one dimension, with a fixed-point Hamiltonian
\begin{equation}
   H^{\homog} \equiv \sum\limits_{ r =  -\infty}^{\infty} h^\homog(r,r+1),
\end{equation}
made of constant nearest neighbor couplings $h$, such that its the ground state $\ket{\psi}$ of $H$ can be represented by a (scale invariant and translation invariant) MERA with a single pair of tensors $\{u,w\}$.

\subsection{Impurities} \label{sect:ImpurityMERA}

\begin{figure}[!ptbh]
\begin{center}
\includegraphics[width=8.5cm]{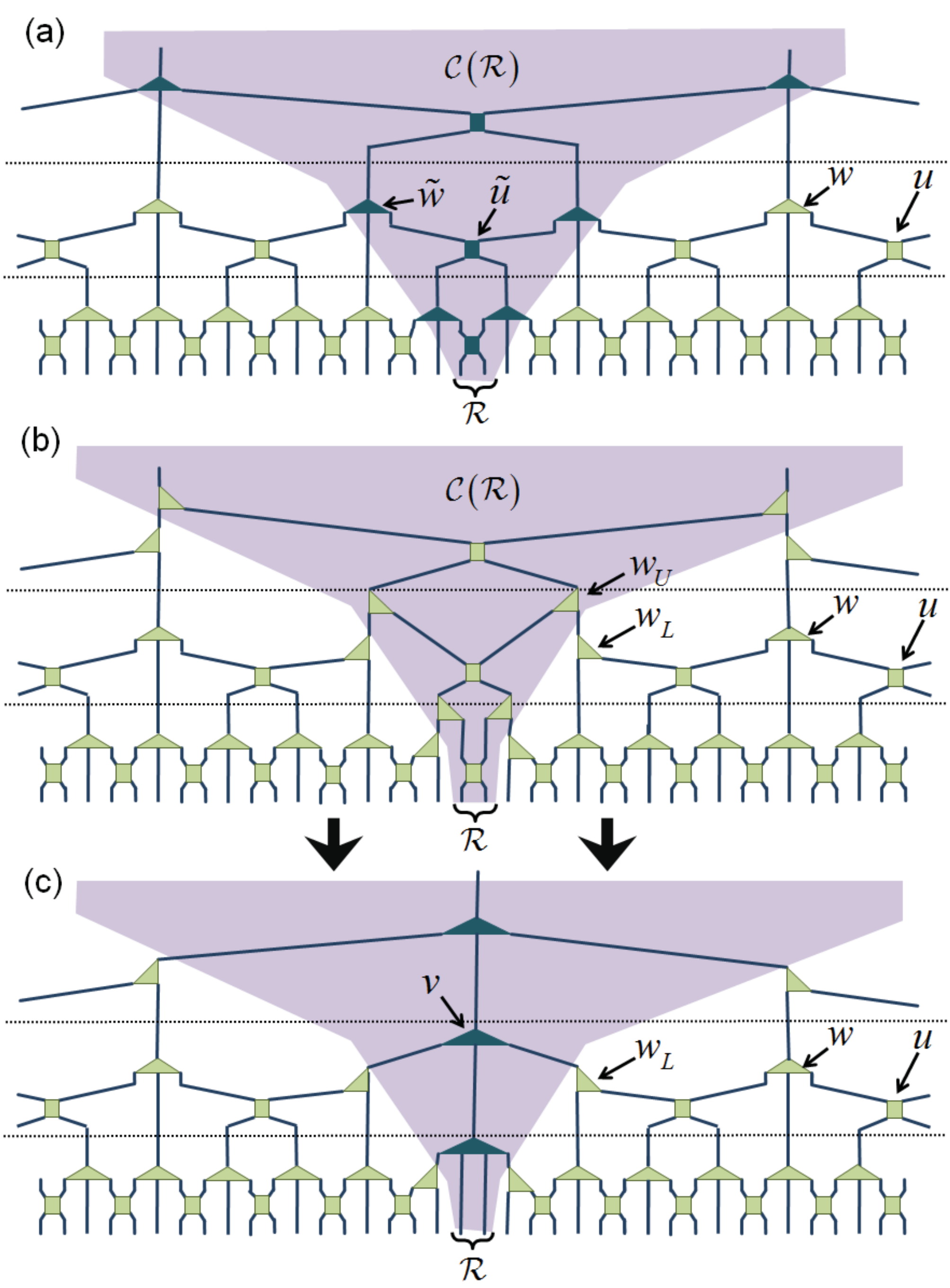}
\caption{Impurity MERA for the ground state $\ket{\psi^{\impur}}$ of Hamiltonian $H^{\impur}$, Eq. \ref{s3e2}. (a) Regular form of an impurity MERA for $\ket{\psi^{\impur}}$, originating in the MERA for a scale-invariant, translation invariant state $\ket{\psi}$ described by a pair of tensors $\{u,w\}$, and that has a different pair of tensors $\{\tilde{u}, \tilde{w}\}$ inside the causal cone $\C(\R)$ (shaded) of the local region $\R$ associated to the impurity. (b) Prior to modifying the homogeneous MERA, we can decompose some of its isometries $w$ into upper $w_U$ and lower $w_L$ isometries, as described in Appendix \ref{sect:IsoDecomp}. (c)  A slightly different impurity MERA for the same ground state $\ket{\psi^{\impur}}$ is obtained by replacing the tensors within the causal cone $\C(\R)$ of the tensor network in (b) with a new set of isometric tensors $v$.}
\label{fig:DefectMERA}
\end{center}
\end{figure}

Let us first consider an impurity problem in one spatial dimension, with Hamiltonian
\begin{equation}
H^{\impur} =  H^\homog + J_{\R}^\impur, \label{s3e2}
\end{equation}
where $H_{\R}^{\impur}$ accounts for an impurity that is supported on a small region $\R$, which in the following is supposed to be made of two contiguous sites. Let $\ket{\psi^\impur}$ denote the ground state of Hamiltonian $H^{\impur}$. Then, the theory of minimal updates in holography \cite{Evenbly13} asserts that a MERA for the ground state $\ket{\psi^\impur}$ can be obtained by modifying the MERA for $\ket{\psi^\homog}$ only in the causal cone $\C(\R)$ of region $\R$, which we assume to also be scale invariant. Accordingly, the impurity MERA is fully described by two pairs of tensors $\{ u, w\}$ and $\{ \tilde u, \tilde w \}$. [If the impurity is not scale invariant, then additional pairs of scale-dependent tensors $\{\tilde{u},\tilde{w}, \tilde{u}',\tilde{w}', \tilde{u}'',\tilde{w}'', \cdots\}$ inside the causal cone will be required in order to describe the non-trivial RG flow of the impurity to a scale invariant, RG fixed point.] Fig. \ref{fig:DefectMERA}(a) depicts the impurity MERA.

In practical computations, we find it more convenient to apply cosmetic changes inside the causal cone of the tensor network, as described in Fig. \ref{fig:DefectMERA}(b-c), and work instead with the impurity MERA depicted in Fig. \ref{fig:DefectMERA}(c). This requires first splitting the isometries $w$ within the causal cone $\C(\R)$ into pairs of binary isometries $w_U$ and $w_L$, as described in Appendix \ref{sect:IsoDecomp}, and then further simplifying the tensor network inside the causal cone replacing the pair of tensors $\{\tilde{u}, \tilde{w}\}$ by a single tensor $v$. [If the impurity is not scale invariant, then additional scale-dependent tensors $\{v,v',v'', \cdots\}$ will be required].

Notice that Figs. \ref{fig:DefectMERA}(a) and \ref{fig:DefectMERA}(c) represent two essentially equivalent forms of the modular MERA. However, the latter form is slightly simpler and, accordingly, we will use it in the theoretical discussion of Sect. \ref{sect:CritMod} and in the benchmark results of Sect. \ref{sect:BenchImpurity}.

%%%%%%%%%%%%%%%%%%%%%%%%%%%%%%%%%%%%%
%%%%%%%%%%%%%%%%%%%%%%%%%%%%%%%%%%%%%
%%%%%%%%%%%%%%%%%%%%%%%%%%%%%%%%%%%%%%%
%%%%%%%%%%%%%%%%%%%%%%%%%%%%%%%%%%%%%%%

\subsection{Boundaries} \label{sect:BoundMERA}

Let us now consider a modular MERA for a semi-infinite chain with a boundary.

Notice that a special case of the impurity Hamiltonian of Eq. \ref{s3e2} corresponds to an impurity that cancels out the interaction between the two sites in region $\mathcal{R}$,
\begin{equation}
J_\R^\impur \equiv -H_\R^\homog, \label{s3e3}
\end{equation}
where $H_\R^\homog$ denotes the part of the homogeneous Hamiltonian $H^{\homog}$ that is supported on $\R$. [More generally, $J_\R^\impur$ could also contain additional single-site terms, such as a single-site magnetic field, etc.]

\begin{figure}[!tbh]
\begin{center}
\includegraphics[width=8.5cm]{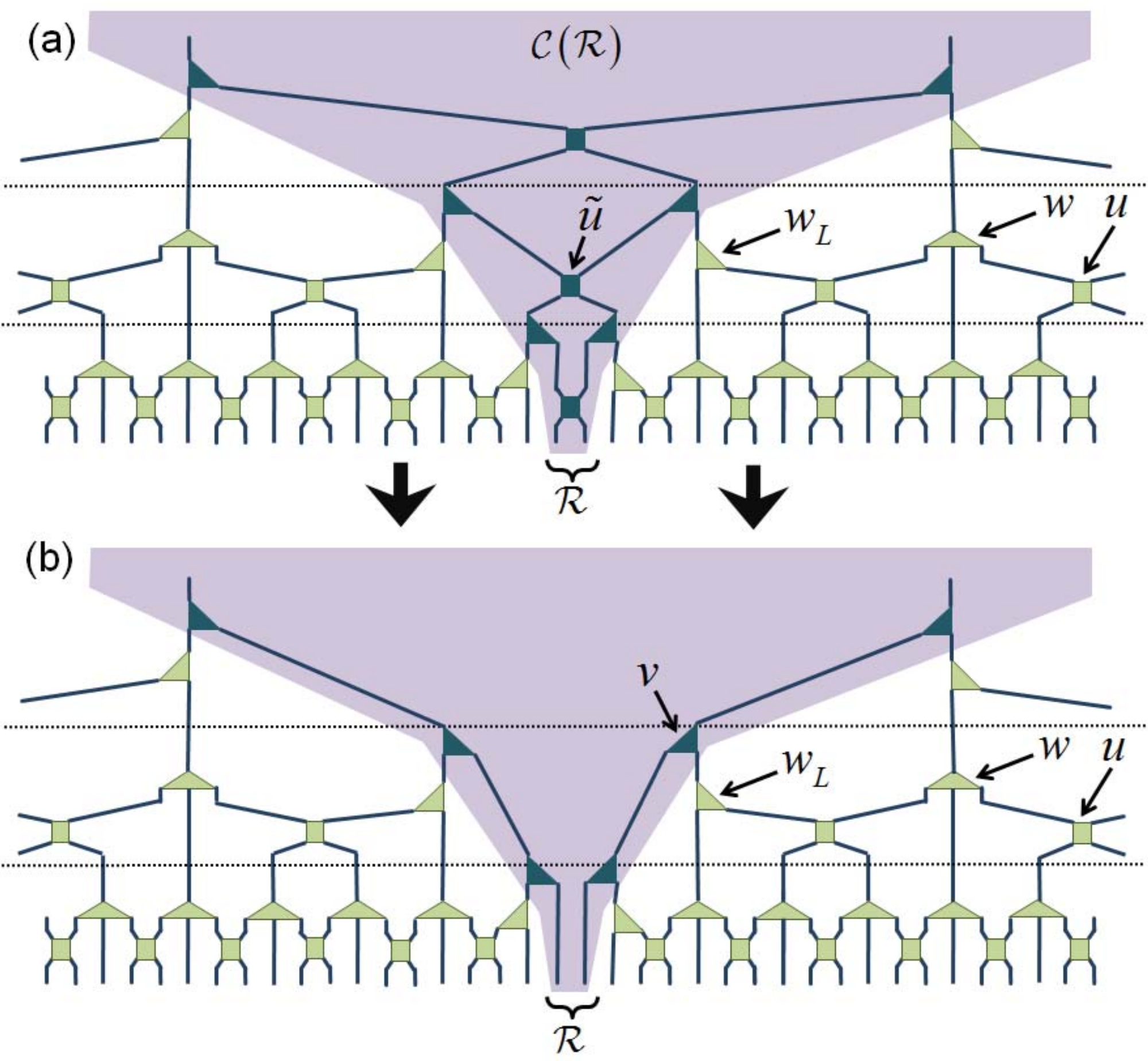}
\caption{Boundary MERA for the ground state $\ket{\psi^{\bound}}$ of Hamiltonian $H^{\bound}$. (a) An impurity MERA can be used as an ansatz for the ground state $\ket{\psi^{\impur}}$ of a homogeneous Hamiltonian $H$ that has an impurity $J_\R^{\impur}$ added on region $\R$, see also Fig. \ref{fig:DefectMERA}. (b) As a special case of the impurity MERA, if the impurity $J_\R^{\impur}$ is chosen such as to remove all interaction between the left and right halves of the chain, as described in Eq. \ref{s3e3}, then the disentanglers $\tilde u$ from (a) can be set to identity. In this way we obtain (two copies of) the boundary MERA, an ansatz for the ground state $\ket{\psi^{\bound}}$ of a semi-infinite system with a single open boundary.}
\label{fig:BoundaryMERA}
\end{center}
\end{figure}

Notice that, since we are dealing with a special case of the impurity Hamiltonian of Eq. \ref{s3e2}, the impurity MERA of Fig. \ref{fig:BoundaryMERA}(a) could be used as an ansatz for its ground state. However, since there is no interaction (and therefore no entanglement) between the left and right semi-infinite halves of the system, we can simplify the impurity MERA by setting the disentanglers $\tilde u$ within the causal cone to identity, resulting in the (doubled) boundary MERA depicted in Fig. \ref{fig:BoundaryMERA}(b). In other words, the theory of minimal updates \cite{Evenbly13} asserts that a modular MERA, consisting of `half' a homogeneous MERA and a single column of boundary tensors $v$, can be used to represent the ground state $\ket{\psi^{\bound}}$ of a homogeneous Hamiltonian with an open boundary,
\begin{equation}
H^{\bound} =  J^{\bound}(0) + \sum\limits_{ r =  0}^{\infty} h^\homog(r,r+1), \label{s3e3b}
\end{equation}
where the additional (and completely unconstrained) one-site term $J^{\bound}$ is included to set the boundary condition. This form of modular MERA for boundary problems, boundary MERA, was first proposed and tested in Ref. \onlinecite{Evenbly10e}. There, however, no theoretical justification of its remarkable success was provided. In Sect. \ref{sect:BenchBound} we expand upon these previous results for boundary MERA, by benchmarking the ansatz both for semi-infinite chains and for finite systems with two open boundaries. Note that a related form of boundary MERA was also proposed in Ref. \onlinecite{Silvi10}.

%%%%%%%%%%%%%%%%%%%%%%%%%%%%%%%%%%%%%
%%%%%%%%%%%%%%%%%%%%%%%%%%%%%%%%%%%%%
%%%%%%%%%%%%%%%%%%%%%%%%%%%%%%%%%%%%%%%
%%%%%%%%%%%%%%%%%%%%%%%%%%%%%%%%%%%%%%%

\subsection{Interfaces} \label{sect:InterfaceMERA}

Next we describe a modular MERA for an interface between two semi-infinite, homogeneous systems $A$ and $B$.

Consider an infinite chain with Hamiltonian
\begin{equation}
H^{\inter}  = H_A^{\homog} + H_B^{\homog} + \alpha ~ J_\R^{\inter}, \label{s3e4}
\end{equation}
where $H_A^{\homog}$ ($H_B^{\homog}$) is the restriction to the left (right) semi-infinite half of the chain of a Hamiltonian for a scale and translation invariant system $A$ ($B$), and where $J_\R^{\inter}$ describes a coupling between $A$ and $B$ across the interface $\R$.

\begin{figure}[!tbh]
\begin{center}
\includegraphics[width=8.5cm]{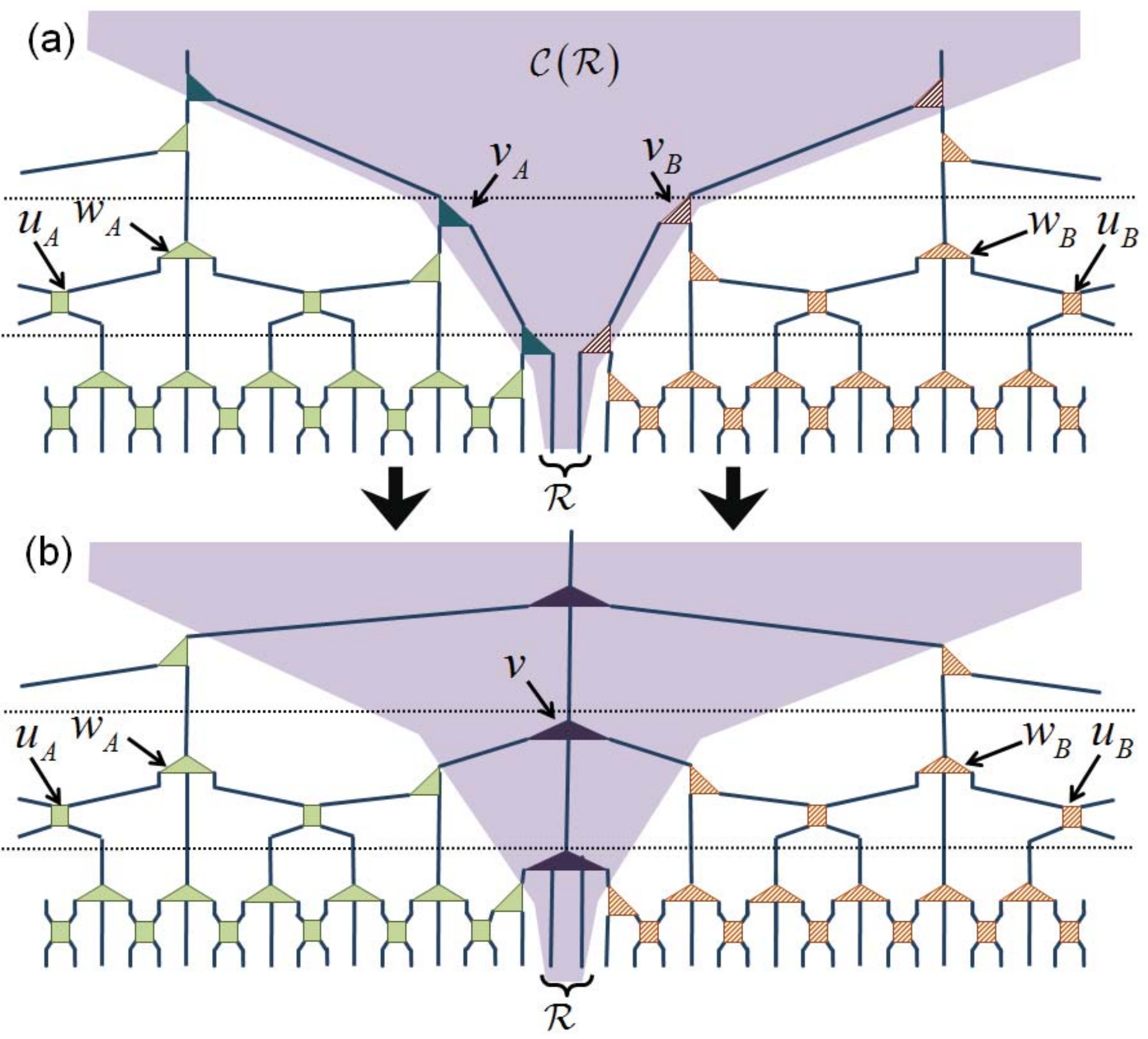}
\caption{Interface MERA. (a) A pair of (different) boundary MERA are used to represent the interface of two systems in the case where there is no coupling across the interface sites, i.e. if $\alpha=0$ in Eq. \ref{s3e4}. (b) If a non-zero interface coupling $\alpha>0$ is introduced, then the MERA from (a) is modified within the causal cone $\C(\R)$ of region $\R$ with the introduction of a new set of isometric tensors $v$. The resulting ansatz is an interface MERA.}
\label{fig:InterfaceMERA}
\end{center}
\end{figure}

If the strength $\alpha$ of the interface coupling is set at $\alpha=0$, then Hamiltonian $H^{\inter}$ reduces to a pair of non-interacting open boundary Hamiltonians of the form described in Eq. \ref{s3e3b}. In this case, the ground state could be represented with two (different) boundary MERAs, as depicted in Fig. \ref{fig:InterfaceMERA}(a). If we now consider switching on the interface coupling, i.e. $|\alpha| > 0$, then the theory of minimal updates asserts that only the inside of the causal cone of $\R$ in Fig. \ref{fig:InterfaceMERA}(a) needs be modified. Similar to the approach with the impurity MERA in Fig. \ref{fig:DefectMERA}(c), we replace the structure within the causal cone by a new set of isometric tensors $v$, which leads to the interface MERA as shown in Fig. \ref{fig:InterfaceMERA}(b). The performance of the interface MERA is benchmarked in Sect. \ref{sect:BenchTwo}.

%%%%%%%%%%%%%%%%%%%%%%%%%%%%%%%%%%%%%
%%%%%%%%%%%%%%%%%%%%%%%%%%%%%%%%%%%%%
%%%%%%%%%%%%%%%%%%%%%%%%%%%%%%%%%%%%%%%
%%%%%%%%%%%%%%%%%%%%%%%%%%%%%%%%%%%%%%%

\subsection{Other defects} \label{sect:FurGen}

\begin{figure}[!tbh]
\begin{center}
\includegraphics[width=8.5cm]{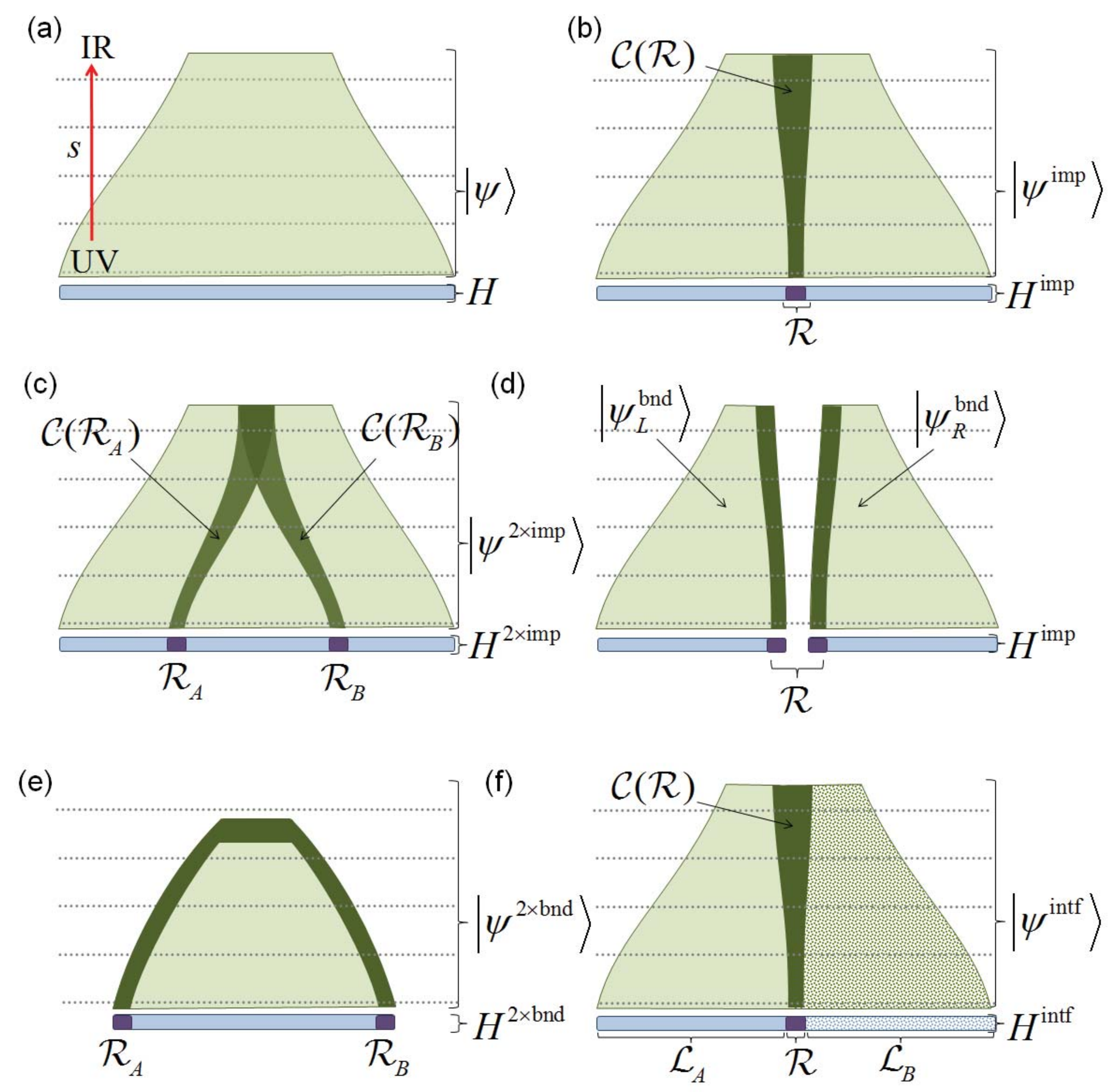}
\caption{Several types of modular MERA, using a schematic representation in which the tensors of the MERA are no longer drawn explicitly: light shading indicates regions of the tensor network occupied with tensors $\{u,w\}$ from a homogeneous system, and dark shading indicates regions occupied by tensors associated to a defect.
(a) MERA for the scale and translation invariant ground state $\ket{\psi^{\homog}}$ of a homogeneous Hamiltonian $H^\homog$.
(b) Impurity MERA for the ground state $\ket{\psi^{\impur}}$ of an impurity Hamiltonian $H^{\impur}$, Eq. \ref{s3e2}, see also Fig. \ref{fig:DefectMERA}.
(c) Modular MERA for the ground state $\ket{\psi^{2\times \impur}}$ of a Hamiltonian $H^{2\times \impur}$ with two impurities localized on disjoint regions $\R_A$ and $\R_B$.
(d) Tensor product of two boundary MERAs for the ground state $\ket{\psi_L^{\bound}}\otimes\ket{\psi_R^{\bound}}$ of an impurity Hamiltonian $H^{\impur}$ in which the impurity is used to remove any interaction between the left and right halves of the chain.
(e) Modular MERA for the ground state $\ket{\psi^{2\times \bound}}$ of the Hamiltonian $H^{2\times \bound}$ for a finite chain with two open boundaries at $\mathcal R_A$ and $\mathcal R_B$.
(f) Interface MERA for the ground state $\ket{\psi^{\inter}}$ of an interface Hamiltonian $H^{\inter}$, Eq. \ref{s3e4}, describing the interface between two two homogeneous systems $A$ and $B$.}
\label{fig:MERAtypes}
\end{center}
\end{figure}

The theory of minimal updates produces a modular MERA also for more complex problems, such as systems involving multiple impurities, or for systems with several types of defects, such a system with both a boundary and an impurity. In the benchmark results of Sect. \ref{sect:Bench} we describe a modular MERA for a system with two impurities, for a finite system with two open boundaries, and for a Y-interface of three semi-infinite quantum spin chains. A summary of several types of modular MERA, together with the corresponding Hamiltonians, is depicted in Fig. \ref{fig:MERAtypes}. Notice that in all instances, the modular MERA is characterized by a small number of tensors that does not scale with the system size. Thus it can be used to address thermodynamically large systems directly, as shall be demonstrated in the benchmark results.

%%%%%%%%%%%%%%%%%%%%%%%%%%%%%%%%%%%%%
%%%%%%%%%%%%%%%%%%%%%%%%%%%%%%%%%%%%%
%%%%%%%%%%%%%%%%%%%%%%%%%%%%%%%%%%%%%%%
%%%%%%%%%%%%%%%%%%%%%%%%%%%%%%%%%%%%%%%

%%%%%%%%%%%%%%%%%%%%%%%%%%%%%%%%%%%%%
%%%%%%%%%%%%%%%%%%%%%%%%%%%%%%%%%%%%%
\begin{figure}[!tbhp]
\begin{center}
\includegraphics[width=8.5cm]{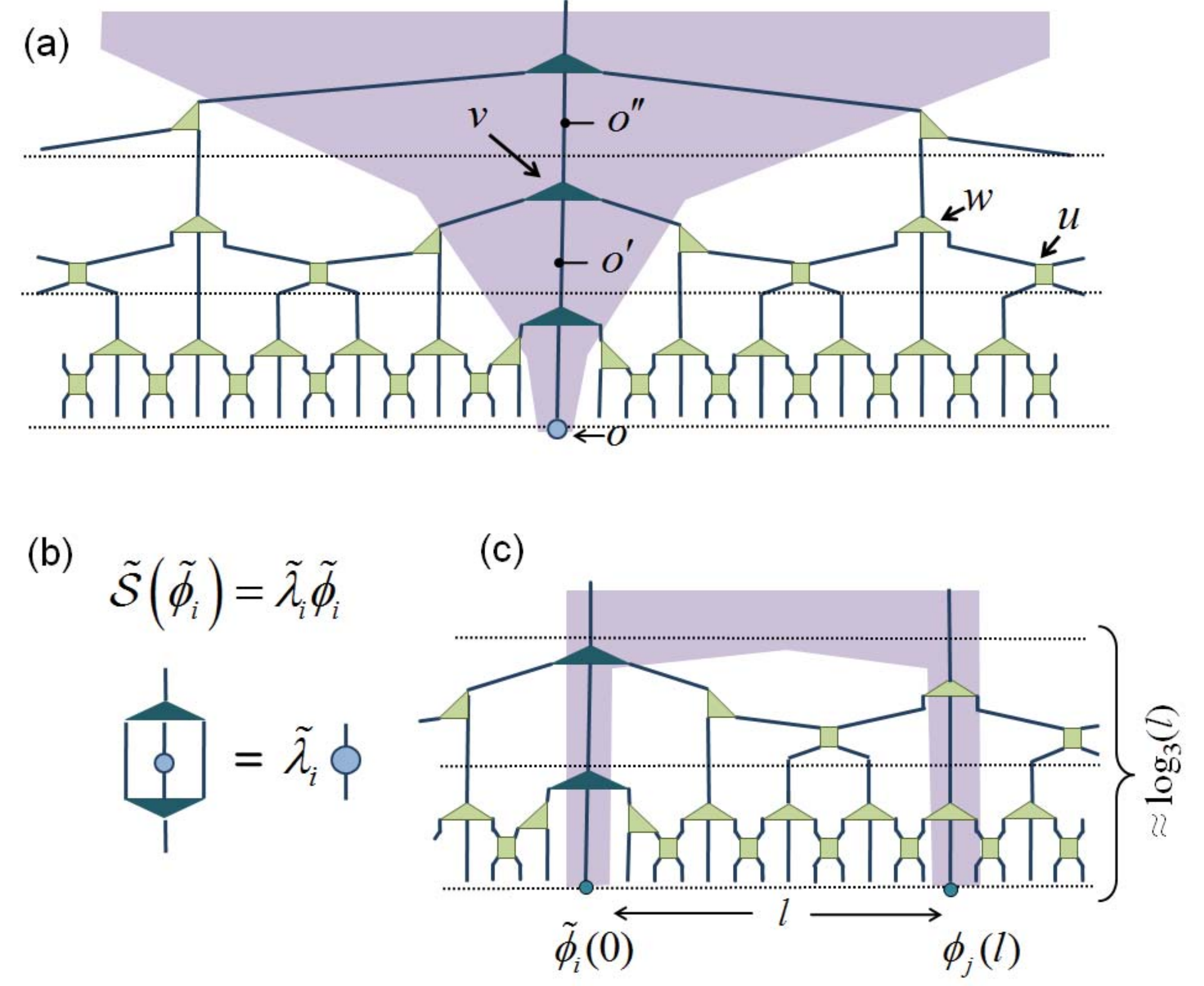}
\caption{(a) A one-site operator $o$ located at the impurity site of an impurity MERA, is coarse-grained into one-site operators $o'$, then $o''$, and so forth. (b) The scaling superoperator $\tilde{\mathcal S}$ associated to the impurity. (c) An operator at the site of the impurity $\tilde \phi_i (0)$ and an operator $\phi_j (l)$ some distance $l$ from the impurity become nearest neighbors after $O (\log_3 (l))$ coarse-graining steps.}
\label{fig:TwoCorrLoc}
\end{center}
\end{figure}
%%%%%%%%%%%%%%%%%%%%%%%%%%%%%%%%%%%%%%%
%%%%%%%%%%%%%%%%%%%%%%%%%%%%%%%%%%%%%%%

\subsection{Extraction of universal properties} \label{sect:CritMod}

Next we explain how to extract the large length scale, universal properties of a defect from the modular MERA. 

We will see that the structure of the ansatz automatically implies (i) the existence of a new set of scaling operators and scaling dimensions associated to the defect [that is, in addition to the (so-called \textit{bulk}) scaling operators and scaling dimensions associated to the host system, see Appx. \ref{sect:ScaleMERA}]; (ii) that the expectation values of local observables differ from those in the absence of the defect by an amount that decays as a power-law with the distance to the defect. These properties, which match those obtained in the context of boundary conformal field theory (BCFT) \cite{Cardy86, Francesco97, Cardy08}, indicate that the modular MERA is a very natural ansatz to describe ground states of quantum critical systems in the presence of a defect, and further justifies the validity of the theory of minimal updates of Ref. \onlinecite{Evenbly13}.

For concreteness, let us consider the impurity MERA in Fig. \ref{fig:DefectMERA}(c), which is fully characterized by the (homogeneous) tensors $\left\{ u, w \right\}$ and the impurity tensor $v$. Let $o$ be a local operator that is measured on the region $\R$ where the impurity is located (which we effectively collapse into a single site). Each layer $U$ of the impurity MERA can be interpreted as a coarse-graining transformation that will map $o$ into a new local operator,
\begin{equation}
o \stackrel{U}{\rightarrow} o' \stackrel{U}{\rightarrow}  o'' \stackrel{U}{\rightarrow} \ldots , \label{s3e5a}
\end{equation}
as also illustrated in Fig. \ref{fig:TwoCorrLoc}(a). The coarse-graining of one-site operators located at the impurity is achieved by means of a scaling superoperator $\tilde{\mathcal{S}}$ associated to the impurity,
\begin{equation}
o' = \tilde{\mathcal{S}} \left( o \right), \label{s3e5}
\end{equation}
where the form of $\tilde{\mathcal{S}}$ is depicted in Fig. \ref{fig:TwoCorrLoc}(b). Notice that $\tilde{\mathcal{S}}$ depends only on the impurity tensor $v$ (i.e. it does not depend on tensors $\left\{ u, w \right\}$). One can diagonalize the impurity superoperator $\tilde{\mathcal{S}}$ (as was done with the scaling superoperator $\mathcal S$ in Appx. \ref{sect:ScaleMERA}) to obtain its scaling operators $\tilde \phi _i$ and scaling dimensions $\tilde \Delta _i$, which are defined as
\begin{equation}
	\tilde{\mathcal{S}} (\tilde \phi_{i}) = \tilde \lambda_{i} \tilde \phi_{i},~~~~~~~\tilde \Delta_{i} \equiv -\log_3 \left( \tilde \lambda_{i} \right). \label{s3e6}
\end{equation}

Let us now evaluate the ground state correlator between an impurity scaling operator $\tilde \phi _i$ located at the site of the impurity ($l=0$), and a bulk scaling operator $\phi _j$ located at site $l$, $\left\langle {\tilde \phi _i   (0)\phi _j  (l)} \right\rangle$, as illustrated in Fig. \ref{fig:TwoCorrLoc}(c). For convenience we choose $l=(3^{s-1}-1)/2$ for a integer $s\geq 0$. After applying one layer of coarse-graining the distance between the scaling operators is reduced to $l'$, $l \rightarrow l' = (l - 1)/3$, which leads to the equality,
\begin{equation}
\left\langle {\tilde \phi _i  (0)\phi _j (l)} \right\rangle  = \tilde \lambda_{i} \lambda_j \left\langle {\tilde \phi_i  (0)\phi _j \left( {(l - 1)/3} \right)} \right\rangle, \label{s3e7}
\end{equation}
where $\tilde \lambda_{i}$ and $\lambda_j$ are eigenvalues of the scaling superoperators $\tilde{\mathcal{S}}$ and $\mathcal{S}$, respectively. After $\log_3 \left((2l+1)/3 \right)$ coarse-graining transformations, the two scaling operators become nearest neighbors in the (effective) lattice. Iterating Eq. \ref{s3e7} that many times, we obtain
\begin{align}
 \left\langle {\tilde \phi _i (0)\phi _j (l)} \right\rangle  &= \tilde C_{i j }  \left( {3^{ - \left( {\tilde \Delta _i    + \Delta _j  } \right)} } \right)^{\log _3 \left( {\frac{{2l + 1}}{3}} \right)}  \nonumber \\
  &= \tilde C_{i j }  \left( {\frac{{2l + 1}}{3}} \right)^{ - \left( {\tilde \Delta _i    + \Delta _j  } \right)}\nonumber \\
  &\approx k_0 \tilde C_{i j }\frac{1}{~l^{\tilde \Delta _i   + \Delta _j }}. \label{s3e8}
\end{align}
In the last step we have ignored a subdominant term that becomes negligible in the large $l$ limit, and have introduced the constant $k_0\equiv (3/2)^{\tilde \Delta _i   + \Delta _j }$. The constant $\tilde C_{i j }$ is defined as the correlator for the scaling operators on adjacent sites,
\begin{equation}
\tilde C_{i j }  \equiv \left\langle {\tilde \phi _i   (0) \phi _j^{} (1)} \right\rangle  = \mbox{Tr} \Big( \big( \tilde \phi_i  \otimes \phi_j  \big)       \rho \Big). \label{s3e9}
\end{equation}
Here $\rho $ is the two-site reduced density matrix on the site of the impurity and the adjacent site.

Eq. \ref{s3e8} reproduces a well-established result from BCFT \cite{Cardy86, Francesco97, Cardy08}: the correlator between a scaling operator at the impurity and a scaling operator outside the impurity decays polynomially with the distance $l$, with an exponent that is the sum of the corresponding impurity scaling dimension $\tilde \Delta_i $ and bulk scaling dimension $\Delta_j$.

Let us now specialize Eq. \ref{s3e8} by setting the impurity scaling operator to the identity, $\tilde \phi _i  = \mathbb I$. This leads to
\begin{equation}
\left\langle {\phi _j (l)} \right\rangle \approx k_0 \tilde C_{\mathbb I j }\frac{1}{~l^{\Delta _j }}, \label{s3e10}
\end{equation}
i.e., the expectation value of a bulk scaling operator $\phi _j$ tends to zero polynomially in distance $d$ from the impurity with an exponent equal to its scaling dimension $\Delta_j$. Recall that in a bulk critical system all bulk scaling operators (with the exception of the identity) have vanishing expectation value, $\left\langle {\phi _j} \right\rangle_\homog =0$. Thus, in the large $l$ limit, the expectation value of arbitrary local operator $o(l)$ located at site $l$ of the impurity MERA differs from its bulk expectation value $\left\langle o \right\rangle _{\homog}$ as,
\begin{equation}
\left\langle {o\left( l \right)} \right\rangle_{\impur}  - \left\langle o \right\rangle _{\homog}  ~~ \approx~~ \frac{1}{l^\Delta}, \label{s3e11}
\end{equation}
where the exponent $\Delta$ of the decay represents the dominant (smallest, non-zero) scaling dimension of the operator $o$ when decomposed in a basis of bulk scaling operators. Eq. \ref{s3e11} shows that in the modular MERA the expectation values of local observables deviate from bulk expectation values everywhere, with a magnitude that decays polynomially with respect to the distance $l$ from the defect.

%%%%%%%%%%%%%%%%%%%%%%%%%%%%%%%%%%%%%%%%%%%%%%%%%%%%%%%%%%%%%%%%%%%%%%%%%%%%%%%%%%%%%%%%
%%%%%%%%%%%%%%%%%%%%%%%%%%%%%%%%%%%%%%%%%%%%%%%%%%%%%%%%%%%%%%%%%%%%%%%%%%%%%%%%%%%%%%%%

\section{Optimization of modular MERA} \label{sect:OptMod}

\begin{figure}[!tbhp]
\begin{center}
\includegraphics[width=8.5cm]{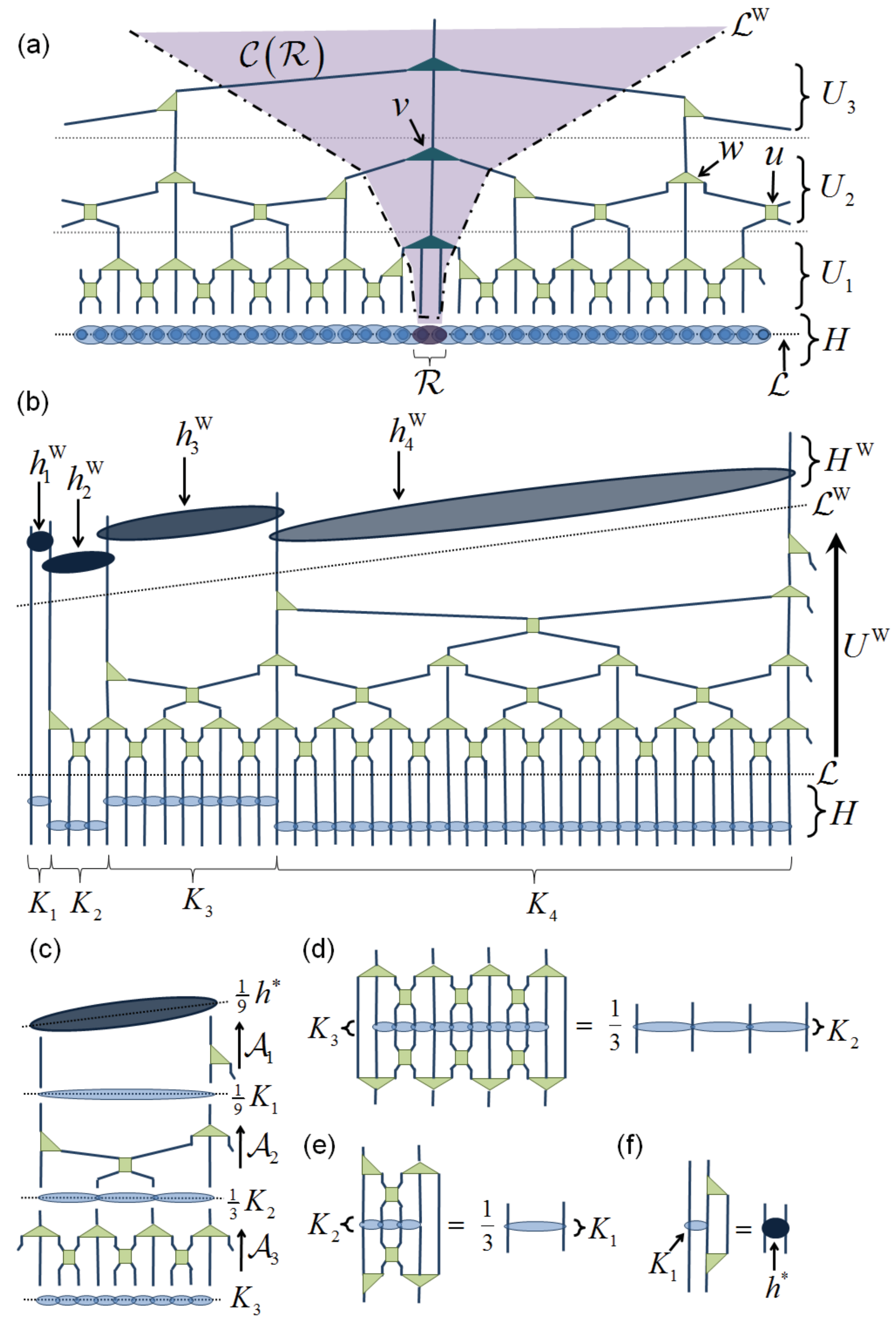}
\caption{(a) An impurity MERA, comprised of bulk tensors $\left\{ u, w \right\}$ and impurity tensors $v$ for a $1D$ lattice $\mathcal L$. The causal cone $\C(\R)$ of the impurity region $\R$ is shaded; the Wilson chain $\L^{\Wilson}$ is the $1D$ lattice formed along the boundary of this causal cone. (b) The inhomogeneous coarse-graining $U^{\Wilson}$ maps the initial Hamiltonian $H$, here partitioned into shells $K_z$ of varying size (see Eq. \ref{eq:shells}), to the effective Hamiltonian $H^{\Wilson}$ defined on the Wilson chain $\mathcal L^{\Wilson}$. (c) A schematic depiction of the coarse-graining of a term from the local Hamiltonian $K_3$, assuming scale invariance of the Hamiltonian $H$, to a local coupling on the Wilson chain, see Eq. \ref{eq:AD6}. (d) Diagrammatic representation of the coarse-graining described in Eq. \ref{eq:AD7} for $s=3$. (e) Diagrammatic representation of the coarse-graining described in Eq. \ref{eq:AD7} for $s=2$. (g) A diagrammatic representation of $\mathcal A_1 (K_1) = h^*$.}
\label{fig:LogScale}
\end{center}
\end{figure}

In this section we describe how the modular MERA can be optimized. For concreteness, we focus on the optimization of the impurity MERA depicted in Fig. \ref{fig:LogScale}(a), noting that other modular MERAs, such as those introduced in Sect. \ref{sect:Modularity}, can be optimized using a similar approach.

In the following, the impurity MERA will be optimized so as to approximate the ground state of an impurity Hamiltonian $H$ of the form,
\begin{equation}
H^{\impur} =  H^\homog + J_\mathcal{R}^\impur, \label{s4e1}
\end{equation}
where $H^\homog  = \sum\nolimits_r {h^\homog (r,r + 1)} $ is the Hamiltonian of a translation invariant, quantum critical host system and the term $J_\mathcal{R}^\impur$ represents a local impurity localized on a region $\R$ of the lattice.
The proposed optimization algorithm is a direct implementation of the theory of minimal updates. First, a scale-invariant MERA for the ground state $\ket{\psi^{\homog}}$ of the host Hamiltonian $H^\homog$ is obtained, which is then modified within the causal cone $\C(\R)$ of region $\R$ in order to account for the impurity $J_\mathcal{R}^\impur$ and obtain the ground state $\ket{\psi^{\impur}}$ of $H^{\impur}$. The three steps for optimizing the impurity MERA are thus as follows:
\begin{enumerate}
	\item The tensors $\{u,w\}$ describing the host system are obtained through optimization of a scale-invariant MERA for the ground state $\ket{\psi^{\homog}}$ of the host Hamiltonian $H^\homog$. \label{step:s1e1}
	\item The original impurity Hamiltonian $H^\impur$, defined on the infinite lattice $\mathcal L$, is mapped to an effective Hamiltonian $H^{\Wilson}$ on a semi-infinite Wilson chain ${\mathcal L^{\Wilson}}$ (to be introduced below),
\begin{equation}
H^{\impur} \stackrel{U^{\Wilson}}{\longrightarrow} H^{\Wilson}, \label{s4e1b}
\end{equation}
through an inhomogeneous coarse-graining $U^{\Wilson}$ defined in terms of tensors $\{u,w\}$. \label{step:s1e2}
	\item The impurity tensors $v$ are obtained through optimization of a tensor network approximation to the ground state $\ket{\psi^{\Wilson}}$ of the effective problem $H^{\Wilson}$ on the Wilson chain. \label{step:s1e3}
\end{enumerate}

The optimization of the MERA for the host Hamiltonian, step \ref{step:s1e1} above, has been covered extensively in e.g. Refs. \cite{Pfeifer09,Evenbly09, Evenbly11b} to which we refer the reader. We now describe in Sect. \ref{sect:LogScale} the details of step \ref{step:s1e2}, and in Sect. \ref{sect:OptLog} the optimization algorithm for step \ref{step:s1e3}.

%%%%%%%%%%%%%%%%%%%%%%%%%%%%%%%%%%%%%
%%%%%%%%%%%%%%%%%%%%%%%%%%%%%%%%%%%%%
%%%%%%%%%%%%%%%%%%%%%%%%%%%%%%%%%%%%%%%
%%%%%%%%%%%%%%%%%%%%%%%%%%%%%%%%%%%%%%%

\subsection{Effective Hamiltonian for the Wilson chain} \label{sect:LogScale}

Consider a MERA on lattice $\mathcal L$, and a region $\R$ with corresponding causal cone $\C(\R)$. We call the \textit{Wilson chain} of region $\mathcal{R}$, denoted $\mathcal{L}^{\Wilson}$, the one-dimensional lattice obtained by following the surface of the causal cone $\mathcal{C}(\mathcal{R})$, see Fig. \ref{fig:LogScale}(a). That is, the Hilbert space for the Wilson chain is built by coarse-graining the Hilbert space of the initial lattice $\mathcal{L}$ with an \textit{inhomogeneous} (logarithmic scale) coarse-graining transformation $U^{\Wilson}$, which is comprised of all the tensors in the MERA that lay outside the causal cone $\mathcal{C}(\mathcal{R})$, see Fig. \ref{fig:LogScale}(b). In the following we describe how the Hamiltonian $H^{\impur}$ defined on lattice $\mathcal L$ is coarse-grained to an effective Hamiltonian $H^{\Wilson}$ on this Wilson chain, which, by construction, can be seen to be only made of nearest neighbor terms,
\begin{equation}
H^{\Wilson} = \sum\limits_{s = 1}^\infty  {h_{s}^{\Wilson} (s,s+1)}.
\end{equation}
Here the nearest neighbor coupling $h_s^{\Wilson}$ depends on $s$. However, below we will see that scale invariance of the host Hamiltonian $H$ implies that for all values of $s$, $h_s^{\Wilson}$ is proportional to a constant coupling $h^*$. Obtaining the effective Hamiltonian $H^{\Wilson}$ for the Wilson chain is a preliminary step to optimizing the impurity tensors $v$.

It is convenient to split the Hamiltonian $H^{\impur}$ intro three pieces,
\begin{equation}\label{eq:split}
    H^{\impur} = H^{L} + H^{R} + {J'}^{\impur}_{\R}
\end{equation}
where ${J'}^{\impur}_{\R}$ collects the impurity Hamiltonian $J^{\impur}$ and the restriction of the host Hamiltonian $H$ on region $\R$, and $H^L$ and $H^R$ contain the rest of Hamiltonian terms to the left and two the right of region $\R$, respectively. For simplicity, we shall only consider explicitly the contribution to the effective Hamiltonian $H^{\Wilson}$ that comes from $H^R$,
\begin{equation}
H^R = \sum\limits_{r = 1}^\infty  h^{\homog}(r,r+1),
\end{equation}
where $r$ measures the distance from the impurity region $\R$. We note that $H^L$ in Eq. \ref{eq:split} yields an identical contribution, whereas ${J'}^{\impur}_{\R}$ is not touched by the coarse-graining transformation $U^{\Wilson}$. Let us rewrite $H^R$ as
\begin{equation}
H^R = \sum\limits_{s = 1}^\infty  K_s, ~~~~~ K_s \equiv \sum\limits_{r = r_s}^{r_{s+1} -1}  h^{\homog}(r,r+1). \label{eq:shells}
\end{equation}
Here $K_s$ denotes the sum of all terms in $H^R$ supported on the sites of lattice $\mathcal L$ that are in the interval $\left[ {r_s}, {r_{s + 1}}\right]$ to the right of $\mathcal R$, where $r_s$ is
\begin{equation}
{r_s} \equiv ( {{3^{s-1 }} + 1} )/2.
\end{equation}
For instance, $K_1$ is the sum of Hamiltonian terms in the interval $\left[ {r_1}, {r_{2}}\right]=[1,2]$, which is actually just a single term,
\begin{equation}
K_1=h^{\homog}(1,2),
\end{equation}
while $K_2$ is the sum of terms in the interval $\left[ {r_2}, {r_{3}}\right]=[2,5]$,
\begin{equation}
K_2=h^{\homog}(2,3) + h^{\homog}(3,4) + h^{\homog}(4,5),
\end{equation}
and so forth. Let $\mathcal A_s$ denote the ascending superoperator that implements one step of coarse-graining of $K_s$ (the explicit forms of $\mathcal A_3$, $\mathcal A_2$ and $\mathcal A_1$ are depicted in Fig. \ref{fig:LogScale}(d-f), respectively). Then the term $h_s^{\Wilson} (s,s+1)$ of the effective Hamiltonian $H^{\Wilson}$ is obtained by coarse-graining $K_{s}$ a total of $s$ times,
\begin{equation}
h_{s}^{\Wilson} (s,s+1) = \left( \mathcal A_1 \circ \mathcal A_2 \circ \cdots \circ \mathcal A_{s} \right) \left( K_{s} \right). \label{eq:AD5}
\end{equation}
As an example, Fig. \ref{fig:LogScale}(c) depicts the coarse-graining of the term $K_3$,
\begin{equation}
h_{3}^{\Wilson} (3,4) = \left( \mathcal A_1 \circ \mathcal A_2 \circ \mathcal A_3 \right) \left( K_3 \right). \label{eq:AD6}
\end{equation}

Through use of Eq. \ref{eq:AD5} one can evaluate all the terms $h_{s}^{\Wilson}$ for $s\geq 1$ that define the effective Hamiltonian $H^{\Wilson}$ on the Wilson chain $\L^{\Wilson}$. Let us now specialize the analysis to the case where the original Hamiltonian on $\mathcal{L}$ is scale invariant (see Appx. \ref{sect:ScaleMERA}). In this case, $K_s$ transforms in a precise way under coarse-graining, namely
\begin{equation}
\mathcal A_{s} \left( K_{s} \right) = \tfrac{1}{3} K_{s-1}, \label{eq:AD7}
\end{equation}
for all $s > 1$. Let us define $h^{*} \equiv \mathcal A_1 \left( K_{1}\right)$. Then all the terms $h_{s}^{\Wilson} (s,s+1)$ of the effective Hamiltonian $H^{\Wilson}$ are seen to be proportional to this same term $h^{*}$,
\begin{equation}
h_{s}^{\Wilson}(s,s+1) = \frac{1}{3^{s-1}} h^{*}(s,s+1), \label{eq:AD8}
\end{equation}
and the effective Hamiltonian $H^{\Wilson}$ for the Wilson chain is,
\begin{eqnarray}
H^{\Wilson} &=& \sum\limits_{s = 1}^\infty \frac{1}{3^{s-1}} {h^*(s,s+1)} \\
&+& \mbox{contributions from $H^L$}\\
&+& {J'}^{\impur}_{\R} \label{eq:AD9}
\end{eqnarray}
That is, in the scale invariant case, we have obtained a nearest neighbor Hamiltonian where each nearest neighbor term is proportional to $h^{*}$, with a proportionality constant that decays exponentially with $s$. [If the scale invariant MERA contained $M$ transitional layers before reaching scale invariance (see Appx. \ref{sect:ScaleMERA}) then the form of the terms in $H^{\Wilson}$ would be position dependent for $s \le M$, and only become proportional to a fixed $h^*(s,s+1)$ for $s>M$.]

The Hamiltonian $H^{\Wilson}$ is analogous to the effective Hamiltonian Wilson obtained, and subsequently solved, in his celebrated solution to the Kondo impurity problem \cite{Wilson75}. This observation was central to the proposal and justification of minimal updates in MERA in Ref. \onlinecite{Evenbly13}.

%%%%%%%%%%%%%%%%%%%%%%%%%%%%%%%%%%%%%
%%%%%%%%%%%%%%%%%%%%%%%%%%%%%%%%%%%%%
\begin{figure}[!tbhp]
\begin{center}
\includegraphics[width=8.5cm]{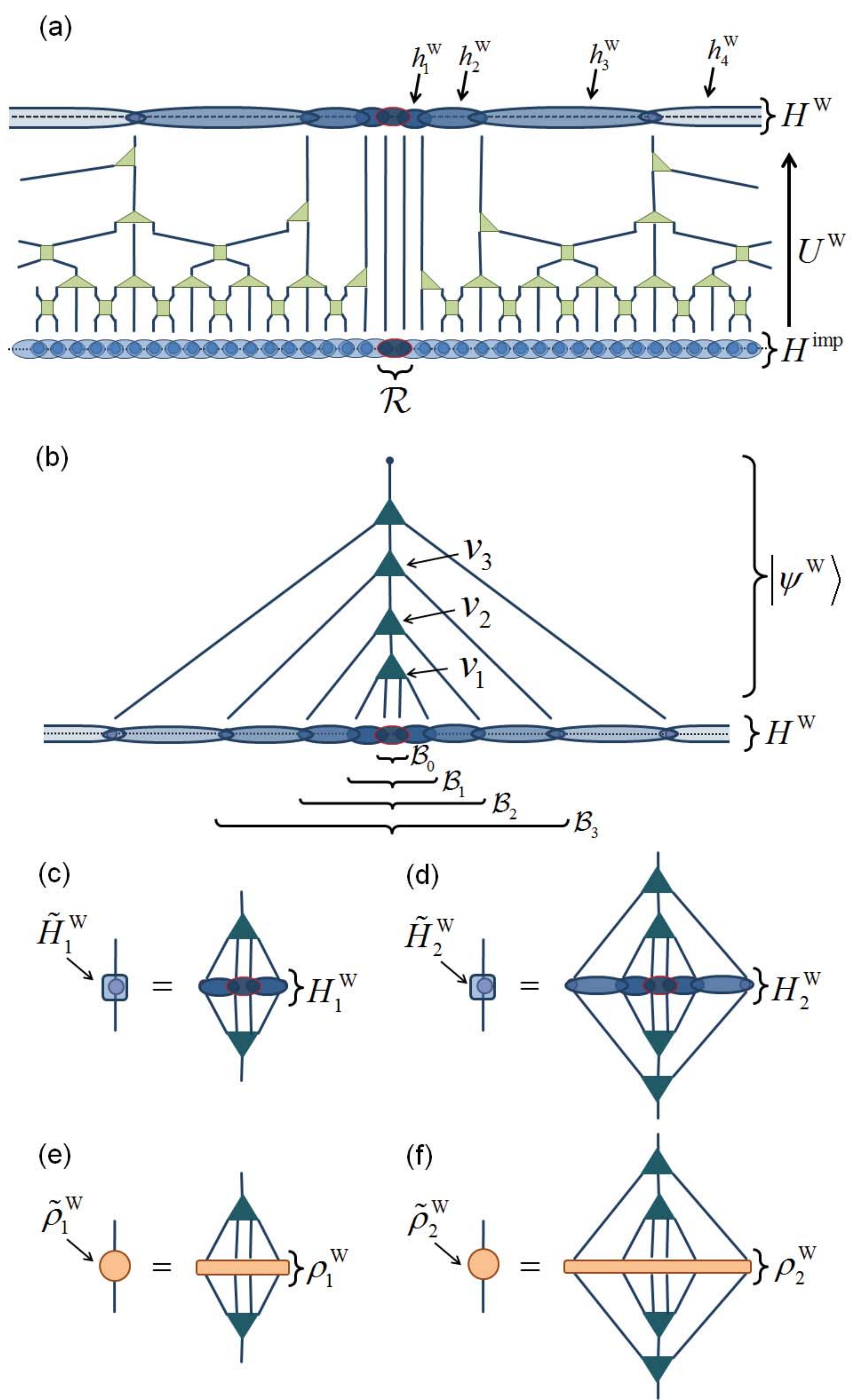}
\caption{(a) The original impurity Hamiltonian, $H^{\impur}=H^{\homog}+J_\R^{\impur}$, defined on lattice $\mathcal L$, is mapped to an effective Hamiltonian $H^{\Wilson}$ defined on the Wilson chain $\L^{\Wilson}$ via the inhomogeneous coarse-graining $U^{\Wilson}$. (b) The set of impurity tensors $v_s$ form a tree tensor network state $\ket {\psi^{\Wilson}}$ on $\L^{\Wilson}$. We denote by $\B_s$ the block of radius $s$ about $\R$. (c-d) The block Hamiltonian $H^{\Wilson}_s$, defined as the part of $H^{\Wilson}$ supported on block $\B_s$, is coarse-grained to the one-site block Hamiltonian $\tilde H^{\Wilson}_s$ using the impurity tensors $\{v_1, v_2, \cdots, v_s\}$. (e-f) The reduced density matrix $\rho^{\Wilson}_s$ on block $\B_s$ is coarse-grained to the one-site reduced density matrix $\tilde \rho^{\Wilson}_s$ using the impurity tensors $\{v_1, v_2, \cdots, v_s\}$.}
\label{fig:DefectAlgA}
\end{center}
\end{figure}
%%%%%%%%%%%%%%%%%%%%%%%%%%%%%%%%%%%%%%%
%%%%%%%%%%%%%%%%%%%%%%%%%%%%%%%%%%%%%%%

\subsection{Optimization of Wilson chain} \label{sect:OptLog}

Once we have constructed the effective Hamiltonian $H^{\Wilson}$ for the (logarithmic scale) Wilson chain $\L^{\Wilson}$, as represented schematically in Fig. \ref{fig:DefectAlgA}(a), we can proceed to optimize for the impurity tensors $v$.

The impurity tensors $v$ form a tensor network known as tree tensor network \cite{Shi06, Tagliacozzo09, Murg10} (TTN) , which we use as a variational ansatz for the ground state $\ket{\psi^{\Wilson}}$ on the Wilson Hamiltonian $H^{\Wilson}$, see Fig. \ref{fig:DefectAlgA}(b). Specifically the impurity tensors $v$ will be obtained through the energy minimization
\begin{equation}\label{eq:mini}
    \min_{v} \bra{\psi^{\Wilson}(v)}H^{\Wilson}\ket{\psi^{\Wilson}(v)}.
\end{equation}
Notice that, if folded through the middle, this TTN is equivalent to a matrix product state (MPS) \cite{Fannes92, Ostlund95, Rommer97}. Therefore, its optimization can be accomplished using standard variational MPS methods \cite{Schollwoeck11}, once they have been properly adapted to a semi-infinite chain.

Here, for concreteness, we describe in detail an optimization algorithm that is similar to the techniques employed in the optimization algorithm for scale invariant MERA \cite{Evenbly09, Pfeifer09, Evenbly11b}. We assume that the state $\ket{\psi^{\Wilson}}$ can be described by the above TTN made of tensors $\{v_s\}_{s=1,2,\cdots}$, where all the tensors for $s > \tilde{M}$ are given by a fixed tensor $v^*$. The number of required transitional tensors $\tilde M$ will in general depend on both the details of the MERA for the state $\ket{\psi^{\homog}}$ of the lattice $\mathcal{L}$ (more specifically, on the number $M$ of transitional layers required before reaching scale invariance, see Appx. \ref{sect:ScaleMERA}), as well as the details of the specific impurity under consideration. In practice the appropriate $\tilde M$ is found heuristically: one starts with a small $\tilde M$, minimizes the energy (using e.g. the algorithm provided below) and then iteratively increases $\tilde M$ until the corresponding optimized energy does no longer depend on $\tilde M$.

In total, $\tilde M+1$ distinct tensors $\left\{ v_1, v_2, \ldots, v_{\tilde M}, v^* \right\}$ need be optimized. This is achieved by iteratively optimizing one tensor at a time, so as to minimize the energy, $E=\bra{\psi^{\Wilson}} H^{\Wilson} \ket{\psi^{\Wilson}}$. If $v_s$ is the tensor to be optimized, then we proceed by computing its linearized environment $\Upsilon_{v_s}$, which is the tensor obtained by removing tensor $v_s$ (but not its conjugate $v_s^{\dagger}$) from the tensor network describing the energy $E=\bra{\psi^{\Wilson}} H^{\Wilson} \ket{\psi^{\Wilson}}$, and that therefore fulfills $E= \mbox{tTr} ( \Upsilon_{v_s} v_s)$, where tTr denotes a tensor trace. An updated $v_s$ that minimizes the energy is then obtained through the singular value decomposition (SVD) of $\Upsilon_{v_s}$. Let us define the nested set of blocks $\B_s \subset \L^{\Wilson}$ as block of radius $s$ around $\R$ with $\B_0 \equiv \R$, see Fig. \ref{fig:DefectAlgA}(b). Then the process of computing linearized environments $\Upsilon_{v_s}$ is simplified by first computing the coarse-grained block Hamiltonians $\tilde H^{\Wilson}_s$ and reduced density matrices $\tilde \rho^{\Wilson}_s$ supported on $\B_s$, as described in Sects.\ref{sect:OptHam} and \ref{sect:OptDen} respectively. Sect. \ref{sect:OptLin} discusses details of the construction of linearized environments and the SVD update, while Sect. \ref{sect:OptAlg} describes how these steps can be composed into the full optimization algorithm.

\subsubsection{Computation of the effective Hamiltonian $\tilde{H}^{\Wilson}_s$} \label{sect:OptHam}

Let us denote by $H^{\Wilson}_s$ the part of the Hamiltonian $H^{\Wilson}$ that is supported on block $\B_s$, and by $\tilde H^{\Wilson}_s$ its effective, one-site version that results from coarse-graining $H^{\Wilson}_s$ by the first $s$ impurity tensors $\{v_1, v_2, \cdots, v_s\}$, see Fig. \ref{fig:DefectAlgA}(c-d) for examples. The block Hamiltonian $\tilde H^{\Wilson}_{s+1}$ for a larger block $\B_{s+1}$ can be computed from the smaller block Hamiltonian $\tilde H^{\Wilson}_{s}$ by
\begin{equation}
\tilde H^{\Wilson}_{s+1} = \tilde \A_{s+1} \left( \tilde H^{\Wilson}_{s} \right) + \tilde \A^L_{s+1} \left( h_s^{\Wilson} \right) + \tilde \A^R_{s+1} \left( h_s^{\Wilson} \right), \label{eq:BlockHam}
\end{equation}
where $\tilde \A_s$ is the one-site impurity ascending superoperator associated to $v_s$, and $\tilde \A^L_{s}$ and $\tilde \A^R_{s}$ are left and right ascending superoperators that add the contributions from the local couplings $h^{\Wilson}_s$ to the block Hamiltonian. The forms of these ascending superoperators are depicted as tensor network diagrams in Fig. \ref{fig:DefectAlgB}(a).

\subsubsection{Computation of the density matrix $\tilde{\rho}^{\Wilson}_s$} \label{sect:OptDen}

\begin{figure}[!tbh]
\begin{center}
\includegraphics[width=8.5cm]{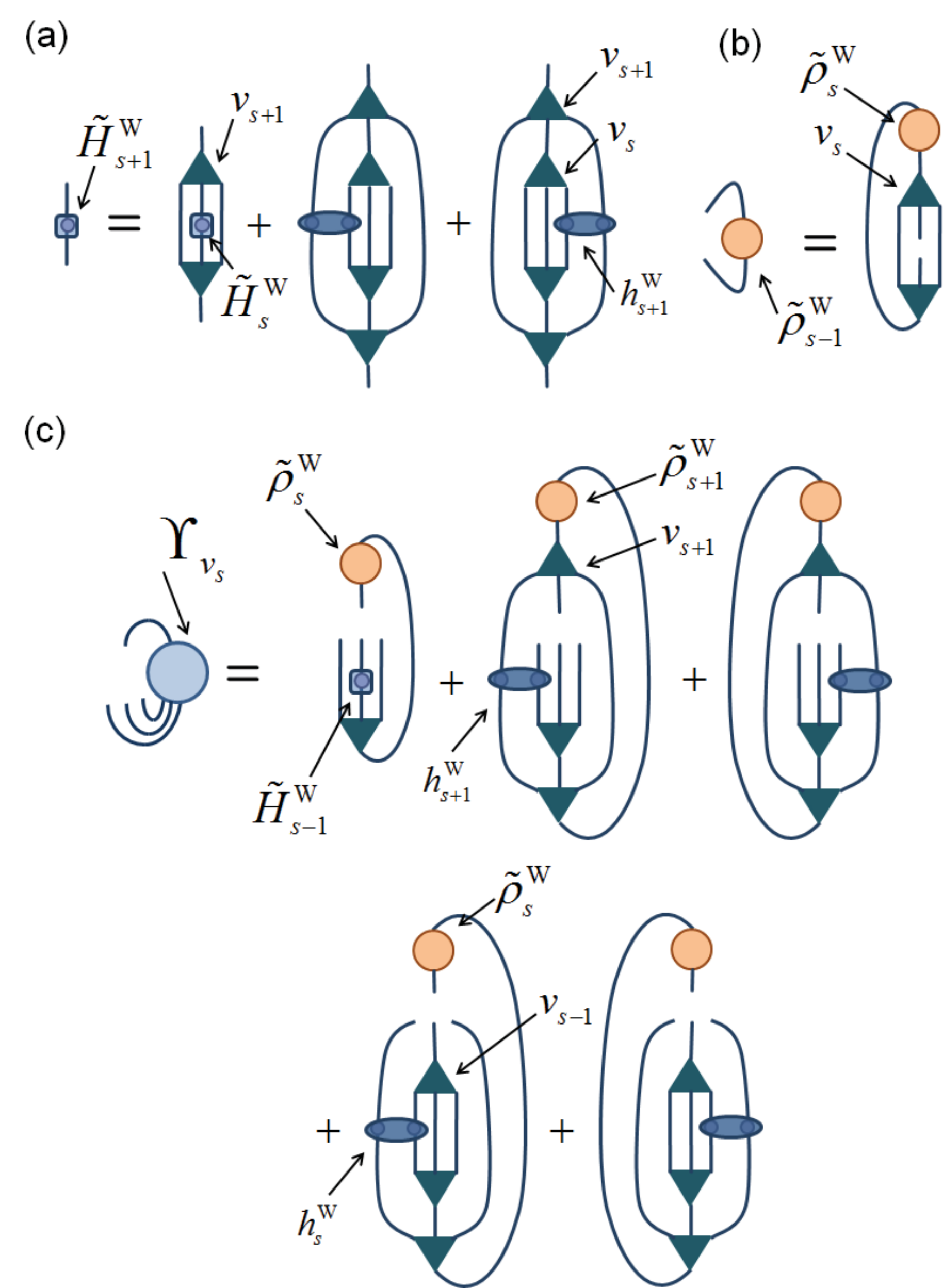}
\caption{The tensor network diagrams for the optimization of the impurity tensor network shown in Fig. \ref{fig:DefectAlgA}(b). (a) The tensor contractions required for evaluating the block Hamiltonian $\tilde H_{s+1}^{\Wilson}$, see also Eq. \ref{eq:BlockHam}. (b) The tensor contraction required for evaluating the reduced density matrix $\tilde \rho_{s-1}^{\Wilson}$ from $\tilde \rho_{s}^{\Wilson}$, see also Eq. \ref{eq:DensityA}. (c) The five contributions to the linearized environment $\Upsilon_{v_s}$ of the impurity tensor $v_s$.}
\label{fig:DefectAlgB}
\end{center}
\end{figure}

We us denote by $\rho^{\Wilson}_s$ the reduced density matrix that is obtained from $\ket{\psi^{\Wilson}}$ by tracing out the sites outside the block $\B_s$, and by $\tilde \rho^{\Wilson}_s$ as its effective, one-site version that results from coarse-graining $\rho^{\Wilson}_s$ with the first $s$ impurity tensors $\{v_1, v_2, \cdots, v_s\}$, see Fig. \ref{fig:DefectAlgA}(e-f) for examples. The one-site density matrix $\tilde \rho _{s - 1}^{\Wilson}$ for a smaller block $\mathcal B_{s-1}$ can be obtained from the density matrix $\tilde \rho _{s}^{\Wilson}$ for the larger region $\mathcal B_{s}$ by fine-graining it with isometry $v_s$, then tracing out the boundary sites. This can be achieved by applying the one-site descending superoperator $\tilde \D_s \equiv \tilde \A_s^{\dagger}$ associated to the impurity tensor $v_s$,
\begin{equation}
\tilde \rho _{s - 1}^{\Wilson} = \tilde \D_s \left( \tilde \rho _s^{\Wilson} \right),\label{eq:DensityA}
\end{equation}
see Fig. \ref{fig:DefectAlgB}(b). Notice that scale invariance, such that $v_s = v_*$ for scales $s > \tilde M$, implies that $\tilde \rho _{s}^{\Wilson}=\tilde \rho_*^{\Wilson}$ for all $s > M$, where the fixed-point density matrix $\tilde \rho_*^{\Wilson}$ satisfies
\begin{equation}
\tilde \rho _*^{\text{W}} = {{\tilde \S}^\dag }\left( {\tilde \rho _*^{\text{W}}} \right).\label{eq:DensityB}
\end{equation}
Here $\tilde \S$ is the one-site scaling superoperator (as introduced in Sect. \ref{sect:CritMod} when studying scale invariant properties of modular MERA), which is just the impurity ascending superoperator $\tilde \A$ constructed from $v^*$. We can thus obtain $\tilde \rho _*^{\text{W}}$ as the dominant eigenvector of ${{\tilde \S}^\dag }$ (e.g. by diagonalizing ${{\tilde \S}^\dag }$). From $\tilde \rho _*^{\text{W}}$, one can then sequentially compute the density matrices $\{\tilde \rho _{\tilde{M}}^{\Wilson}, \tilde \rho _{\tilde{M}-1}^{\Wilson}, \cdots, \tilde \rho _{1}^{\Wilson}\}$ by using Eq. \ref{eq:DensityA} .

%%%%%%%%%%%%%%%%%%%%%%%%%%%%%%%%%%%%%
%%%%%%%%%%%%%%%%%%%%%%%%%%%%%%%%%%%%%
%%%%%%%%%%%%%%%%%%%%%%%%%%%%%%%%%%%%%%%
%%%%%%%%%%%%%%%%%%%%%%%%%%%%%%%%%%%%%%%

\subsubsection{Computation of the linearized environment $\Upsilon_{v_s}$} \label{sect:OptLin}

%The impurity tensors shall be updated using the approach described in Ref. \cite{Evenbly09} where a tensor $v_s$ is updated by first computing its linearized environment $\Upsilon_{v_s}$, and then subsequently taking the singular value decomposition (SVD) of this environment.

Fig. \ref{fig:DefectAlgB}(c) shows the linearized environment $\Upsilon_{v_s}$ for the impurity tensor $v_s$. $\Upsilon_{v_s}$ decomposes into a sum of five terms, each of which corresponds to a small tensor network, and it depends on the effective Hamiltonian $\tilde H^{\Wilson}_{s-1}$, the reduced density matrices $\tilde \rho^{\Wilson}_s$ and $\tilde \rho^{\Wilson}_{s+1}$, the Hamiltonian terms $h^{\Wilson}_s$ and $h^{\Wilson}_{s+1}$, and the impurity tensors $v_{s-1}$, $v_{s}$, and $v_{s+1}$,
\begin{equation}
\Upsilon_{v_s}\left( \tilde H^{\Wilson}_{s-1}, \tilde \rho^{\Wilson}_{s}, \tilde \rho^{\Wilson}_{s+1}, h^{\Wilson}_s, h^{\Wilson}_{s+1}, v_{s-1}, v_{s}, v_{s+1}\right).
\end{equation}

Let us consider first the optimization of $v_s$ for $s \leq \tilde{M}$. In this case, the updated impurity tensor is chosen as $v_s = -V_2 V_1^{\dagger}$, where $V_1$ and $V_2$ are isometric tensors obtained from the SVD of the linearized environment $\Upsilon_{v_s}$, namely $\Upsilon_{v_s} = V_1 S V_2^{\dagger}$, see Ref. \onlinecite{Evenbly09} for further details.

For $s > \tilde{M}$, the impurity tensor $v_s$ is a copy of the impurity tensor $v^*$. In order to update $v^*$ we should construct the environment as the sum of environments for each $s>\tilde{M}$,
\begin{equation}
\Upsilon _{v^* }  = \sum\limits_{s  = \tilde M+1}^\infty  {\Upsilon _{v_s } }.  \label{s4e11}
\end{equation}
Obtaining the environment $\Upsilon _{v^* }$ directly through this infinite summation may only be possible at a very large computational cost. However, since the system is assumed to be scale invariant, the environments ${\Upsilon _{v_s } }$ in Eq. \ref{s4e11} should quickly converge to a fixed environment as we increase $s$. Thus one can obtain an approximate environment $\Upsilon _{v^* } $ of the scaling impurity tensor $v^*$ through a partial summation of Eq. \ref{s4e11},
\begin{equation}
\Upsilon _{v^* }  \approx \sum\limits_{s  = \tilde M+1}^{\tilde M+ 1 + \tau} {\Upsilon _{v_s } }. \label{s4e12}
\end{equation}
The number $\tau+1$ of terms in this partial summation, required in order to obtain a sufficiently accurate environment, will in general depend on the problem under consideration. However, for the numerical results of Sect. \ref{sect:Bench} we find that keeping $\tau \approx 2$ is sufficient in most cases. Once the linearized environment $\Upsilon _{v^* } $ has been computed, the tensor $v^{*}$ is updated by taking the SVD of the environment as in the case $s \leq \tilde{M}$.

\subsubsection{Optimization algorithm} \label{sect:OptAlg}

Let us then review the algorithm to optimize the tensors $\left\{ v_1, v_2, \ldots, v_{\tilde M}, v^* \right\}$ of the TTN of Fig. \ref{fig:DefectAlgA}(b) for the ground state $\ket{\psi^{\Wilson}}$ of the effective Hamiltonian $H^{\Wilson}$. The optimization is organized in sweeps through the TTN, where each sweep consists of a sequence of single tensor updates for each $v_s$, from $s=1$ to $s=\tilde{M}+1$. We iterate these optimization sweeps until the state $\ket{\psi^{\Wilson}}$ has converged sufficiently. 

Recall that the effective Hamiltonian $H^{\Wilson}$ generically takes the form of Eq. \ref{eq:AD9}, with nearest neighbor coupling strengths that decay geometrically with the distance to the origin. Thus, a very good approximation to the ground state of $H^{\Wilson}$ can be obtained using Wilson's numerical renormalization group\cite{Wilson75,Wilson75b} (NRG). Here we use the NRG to initialize the impurity tensors $v_s$, and then apply the variational sweeping to further improve the approximation to the ground state.

Each iteration of the variational sweep is comprised of the following steps:
\begin{enumerate}
	\item Compute the fixed-point density matrix $\tilde \rho _*^{\text{W}}$ through diagonalization of the (adjoint) impurity scaling superoperator ${\tilde \S}^\dag$.
	\item Compute the block density matrices $\tilde \rho _s^{\text{W}}$ for all $s\le \tilde M$ using Eq. \ref{eq:DensityA}.
	\item Sequentially update $v_s$, starting from $s=1$ and proceeding to $s=\tilde M$. For each such values of $s$, first compute the linearized environment $\Upsilon_{v_s}$ and then update the impurity tensor $v_s$ via the SVD of this environment. Then compute the effective Hamiltonian $\tilde H^{\Wilson}_s$ from $\tilde H^{\Wilson}_{s-1}$ using the updated isometry $v_s$, as described in Eq. \ref{eq:BlockHam}.
	\item Update the fixed-point tensor $v^*$: compute an approximate environment $\Upsilon _{v^* }$ as described in Eq. \ref{s4e12}, and then update the fixed-point tensor $v^*$ via the SVD of this environment.	
\end{enumerate}
Notice that this algorithm is analogous to the one introduced to optimize the scale-invariant MERA as described in Ref. \onlinecite{Pfeifer09}.

%%%%%%%%%%%%%%%%%%%%%%%%%%%%%%%%%%%%%%%%%%%%%%%%%%%%%%%%%%%%%%%%%%%%%%%%%%%%%%%%%%%%%%%%
%%%%%%%%%%%%%%%%%%%%%%%%%%%%%%%%%%%%%%%%%%%%%%%%%%%%%%%%%%%%%%%%%%%%%%%%%%%%%%%%%%%%%%%%
%%%%%%%%%%%%%%%%%%%%%%%%%%%%%%%%%%%%%%%%%%%%%%%%%%%%%%%%%%%%%%%%%%%%%%%%%%%%%%%%%%%%%%%%
%%%%%%%%%%%%%%%%%%%%%%%%%%%%%%%%%%%%%%%%%%%%%%%%%%%%%%%%%%%%%%%%%%%%%%%%%%%%%%%%%%%%%%%%
%%%%%%%%%%%%%%%%%%%%%%%%%%%%%%%%%%%%%%%%%%%%%%%%%%%%%%%%%%%%%%%%%%%%%%%%%%%%%%%%%%%%%%%%
%%%%%%%%%%%%%%%%%%%%%%%%%%%%%%%%%%%%%%%%%%%%%%%%%%%%%%%%%%%%%%%%%%%%%%%%%%%%%%%%%%%%%%%%
%%%%%%%%%%%%%%%%%%%%%%%%%%%%%%%%%%%%%%%%%%%%%%%%%%%%%%%%%%%%%%%%%%%%%%%%%%%%%%%%%%%%%%%%
%%%%%%%%%%%%%%%%%%%%%%%%%%%%%%%%%%%%%%%%%%%%%%%%%%%%%%%%%%%%%%%%%%%%%%%%%%%%%%%%%%%%%%%%
%%%%%%%%%%%%%%%%%%%%%%%%%%%%%%%%%%%%%%%%%%%%%%%%%%%%%%%%%%%%%%%%%%%%%%%%%%%%%%%%%%%%%%%%
%%%%%%%%%%%%%%%%%%%%%%%%%%%%%%%%%%%%%%%%%%%%%%%%%%%%%%%%%%%%%%%%%%%%%%%%%%%%%%%%%%%%%%%%

\section{Benchmark results} \label{sect:Bench}

In this section we benchmark the use of the modular MERA for several types of defect in quantum critical systems; specifically we consider impurities, boundaries, and interfaces. In the case of a single impurity, a single boundary, and a simple interface, we use the corresponding modular MERAs introduced in Sects. \ref{sect:ImpurityMERA}, \ref{sect:BoundMERA}, and \ref{sect:InterfaceMERA}. For multiple impurities, two boundaries, and Y-interfaces, we use more complicated modular MERAs that result from a recursive use of the theory of minimal updates, as outlined in Sect. \ref{sect:FurGen}. In several cases, we also specify how to modify the basic optimization algorithm of Sect. \ref{sect:OptMod}.

\subsection{Impurities} \label{sect:BenchImpurity}

We start by benchmarking the use of the modular MERA to describe a quantum critical system in the presence of a single impurity first, and then in the presence of multiple impurities.

\subsubsection{Single impurity} \label{sect:BenchSingle}

Let us first consider a quantum critical system with a Hamiltonian of the form
\begin{equation}
H^{\impur} =  H + J^{\impur}_\R, \label{s5e1}
\end{equation}
where $H$ is a fixed-point Hamiltonian that describes the host system (which is invariant both under translations and changes of scale), and $J^{\impur}_\R$ accounts for an impurity localized on region $\R$ of the lattice. Specifically, we test the impurity MERA in the case where $H$ corresponds to the critical Ising Hamiltonian,
\begin{equation}
H^\textrm{Ising} = \sum\limits_r { - X(r)X(r + 1) + Z(r)}, \label{s5e3}
\end{equation}
where $X$ and $Z$ are Pauli matrices, and the impurity Hamiltonian $J^{\alpha}$ acts on two adjacent lattice sites $r=(0,1)$, where it weakens or strengthens the nearest neighbor term,
\begin{equation}
J^{\alpha}(0,1) =  (1-\alpha) X(0) X(1), \label{s5e4}
\end{equation}
for some real number $\alpha$. The quantum critical Ising model with an impurity of this form, which is in direct correspondence with the $2D$ classical Ising model with a defect line, has been studied extensively in the literature \cite{ID2, ID3, ID4, ID5, ID6, ID7, ID8, ID9, ID10, ID1}. We refer the reader to Ref. \onlinecite{ID1} for a review of the problem.

We optimize the impurity MERA for the ground state $\ket{\psi^{\impur}}$ of this impurity problem using the strategy outlined in Sect. \ref{sect:OptMod}. We fist find tensors $\{u,w\}$ for the ground state of the homogeneous critical Ising model using a scale invariant MERA with bond dimension $\chi=22$. This MERA incorporated both the $\mathbb Z_2$ (spin flip) global on-site symmetry and the reflection symmetry (see Appendix \ref{sect:RefSym}) of $H^\textrm{Ising}$. This optimization required approximately 1 hour of computation time on a 3.2 GHz desktop PC with 12Gb of RAM. The mapping of the initial impurity Hamiltonian $H^{\impur}$ to the effective problem $H^{\Wilson}$ on the Wilson chain $\mathcal L^{\Wilson}$, as described in Sect. \ref{sect:LogScale}, was accomplished in negligible computation time; it is less expensive than a single iteration of the optimization of the scale invariant MERA. Optimization of the impurity tensors $v_s$, as discussed in Sect. \ref{sect:OptLog}, was performed for a range of impurity strengths, namely the two series $\alpha=\left\{0, 0.2, 0.4, 0.6, 0.8, 1 \right\}$ and $\alpha=\left\{ 1/0.8, 1/0.6, 1/0.4, 1/0.2, \infty \right\}$, which required approximately 20 minutes of computation time for each value of $\alpha$.

%%%%%%%%%%%%%%%%%%%%%%%%%%%%%%%%%%%%%
%%%%%%%%%%%%%%%%%%%%%%%%%%%%%%%%%%%%%
\begin{figure}[!tbh]
\begin{center}
\includegraphics[width=8.5cm]{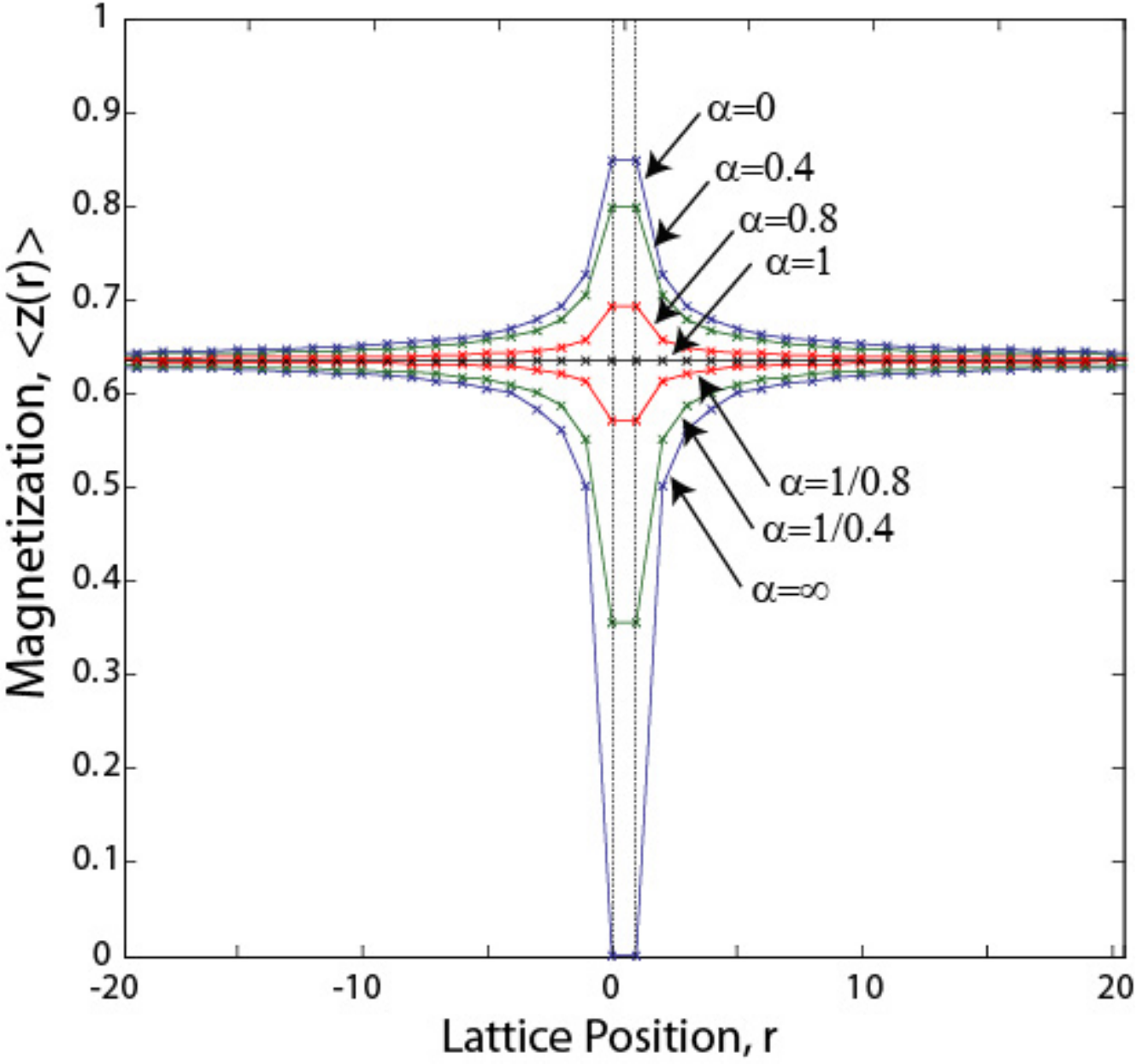}
\caption{The transverse magnetization $\left\langle Z(r) \right\rangle$ evaluated from an impurity MERA ($\times$'s) versus the exact solutions (solid lines) for the critical Ising model with an impurity $J^\alpha$, as described Eq. \ref{s5e4}, located on lattice sites $r=(0,1)$. The magnetization approaches the bulk value $\left\langle Z \right\rangle ^\homog=2/\pi$ polynomially as $|r|^{-1}$, for all values of $\alpha$ considered.}
\label{fig:DefectZMag}
\end{center}
\end{figure}
%%%%%%%%%%%%%%%%%%%%%%%%%%%%%%%%%%%%%%%
%%%%%%%%%%%%%%%%%%%%%%%%%%%%%%%%%%%%%%%

%%%%%%%%%%%%%%%%%%%%%%%%%%%%%%%%%%%%%
%%%%%%%%%%%%%%%%%%%%%%%%%%%%%%%%%%%%%
\begin{figure}[!tbh]
\begin{center}
\includegraphics[width=8.5cm]{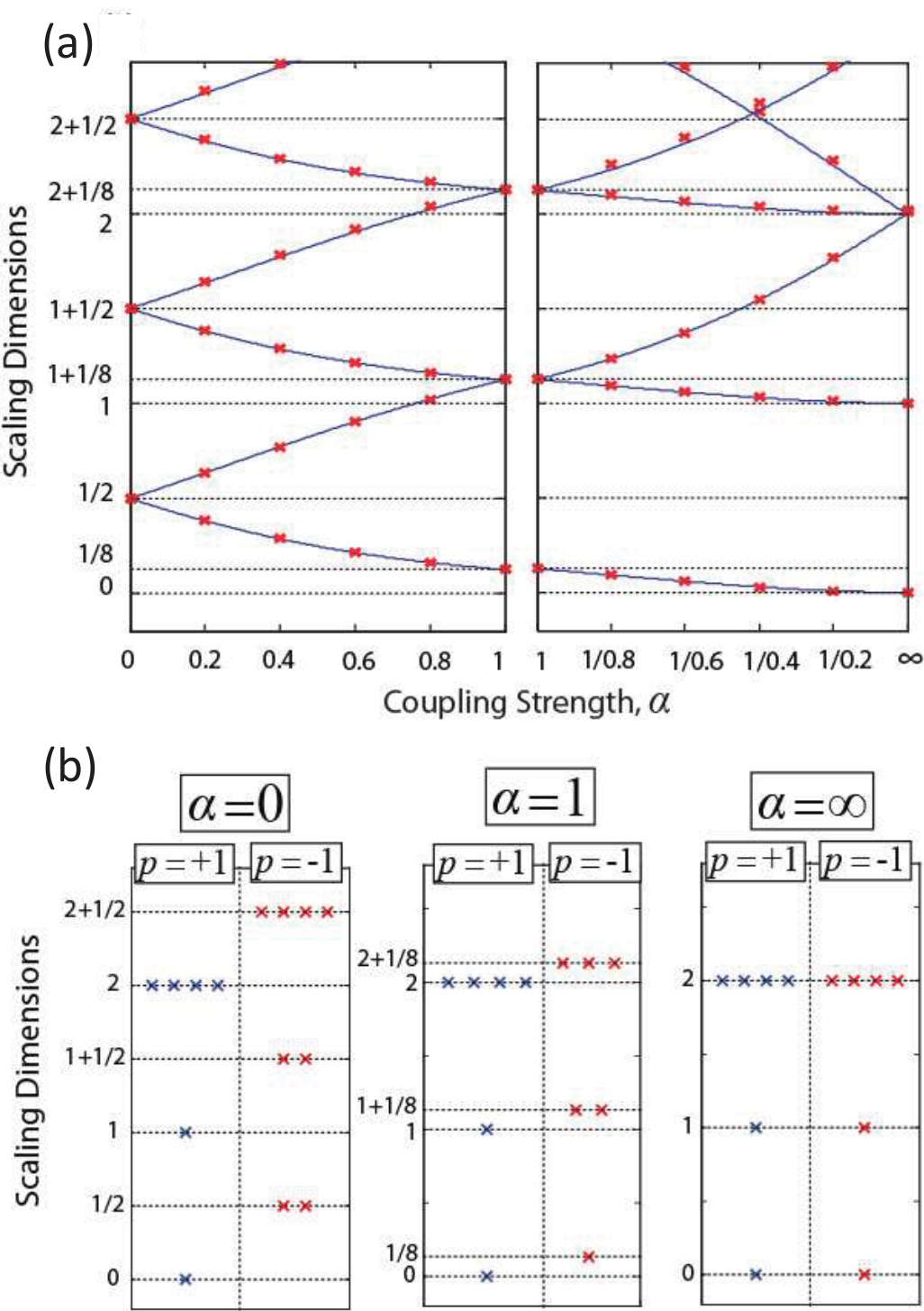}
\caption{(a) Scaling dimensions $\Delta_{\alpha}$ for the critical Ising model with a conformal defect $J^\alpha$, comparing results from the impurity MERA ($\times$'s) with the exact results of Eq. \ref{s5e5} (solid lines). Note that only scaling dimensions in the $p=-1$ parity sector of the $\mathbb Z_2$ global symmetry of the Ising model are plotted, as those in the $p=+1$ parity sector are invariant under addition of the conformal defect. (b) The complete spectrum of scaling dimensions obtained from the MERA, organized according to parity sector $p=\pm 1$, for values of $\alpha=\left\{0,1,\infty\right\}$.}
\label{fig:DefectCritExp}
\end{center}
\end{figure}
%%%%%%%%%%%%%%%%%%%%%%%%%%%%%%%%%%%%%%%
%%%%%%%%%%%%%%%%%%%%%%%%%%%%%%%%%%%%%%%

From the optimized impurity MERA we compute the magnetization profiles $\left\langle Z{(r)} \right\rangle_{\impur}$, as shown Fig. \ref{fig:DefectZMag}, which match the exact profiles (obtained by solving the free fermion problem, see Ref. \onlinecite{ID1}) with high precision. For all defect strengths $\alpha$ considered, the magnetization approaches the constant bulk value $\left\langle Z \right\rangle_\homog =2/\pi$ as $|r|^{-1}$, \emph{i.e.} with scaling dimension $\Delta=1$. This result, consistent with the behavior of modular MERA predicted in Sect. \ref{sect:CritMod}, is in agreement with the scaling of the magnetization $\left\langle Z{(r)} \right\rangle_\impur$ predicted from study of the Ising CFT (where the $Z$ operator is related to the energy density operator $\varepsilon$ of the Ising CFT with scaling dimension $\Delta=1$). For each value of the impurity coupling $\alpha$, we also compute the scaling dimensions $\Delta_\alpha$ associated to the impurity by diagonalizing the impurity scaling superoperator $\tilde{\mathcal S}$, as described Sect. \ref{sect:CritMod}. In Refs. \onlinecite{ID2,ID3,ID1} the spectrum of scaling dimensions for the critical Ising model associated to the impurity $J^\alpha$ have been derived analytically,
\begin{equation}
\Delta _\alpha   = 2\left( {m + \frac{1}{4} + \frac{{\theta _\alpha  }}{\pi }} \right)^2, \label{s5e5}
\end{equation}
where $m$ is a positive integer and $\theta _\alpha$ is a phase associated to the strength of the impurity $\alpha$,
\begin{equation}
\theta _\alpha   = \tan ^{ - 1} \left( {\frac{{1 - \alpha }}{{1 + \alpha }}} \right). \label{s5e6}
\end{equation}
A comparison of the scaling dimensions obtained from MERA and the exact scaling dimensions is presented in Fig. \ref{fig:DefectCritExp}. Remarkably, the impurity MERA accurately reproduces the smallest scaling dimensions (all scaling dimensions $\Delta<2.5$) for the full range of $\alpha$ considered, which include the special cases of (i) an impurity that removes any interaction between the left and right halves of the chain $(\alpha=0)$, (ii) the case with no impurity $(\alpha=1)$, and (iii) an impurity which sets an infinitely strong Ising interaction over two spins $(\alpha=\infty)$.

These results confirm that the impurity MERA accurately approximates the ground state of the impurity system, both in terms of its local expectation values (e.g. magnetization profile $\left\langle Z{(r)} \right\rangle_{\impur}$) and its long distance, universal properties (e.g. scaling dimensions $\Delta_{\alpha}$).

%%%%%%%%%%%%%%%%%%%%%%%%%%%%%%%%%%%%%
%%%%%%%%%%%%%%%%%%%%%%%%%%%%%%%%%%%%%
\begin{figure}[!htbp]
\begin{center}
\includegraphics[width=8.5cm]{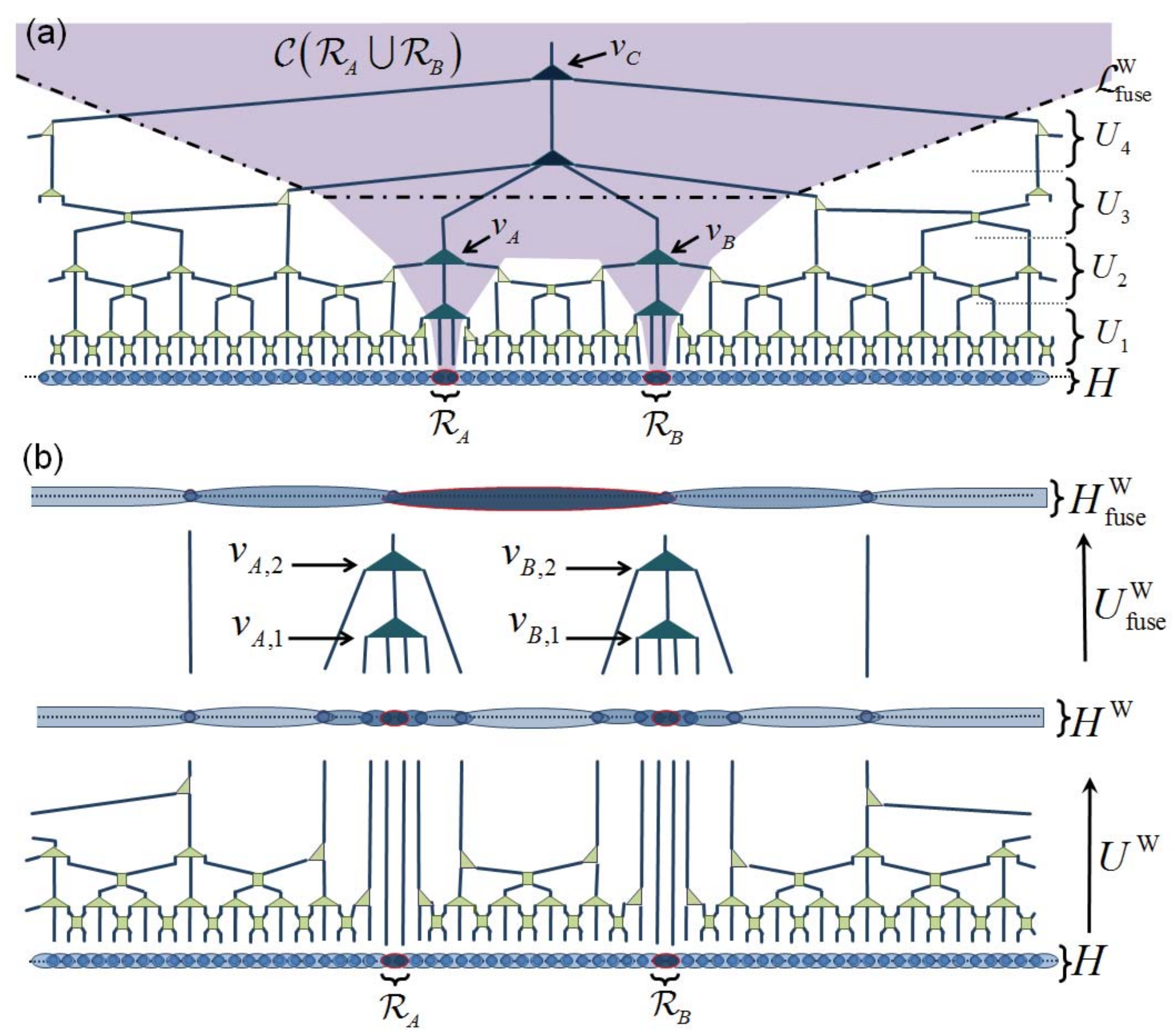}
\caption{(a) An impurity MERA for a system with local impurities on regions $\R_A$ and $\R_B$, here separated by $l=18$ lattice sites. The causal cones of the individual impurities fuse at a depth $s \approx \log_3 (l)$. At small depth, $s<\log_3 (l)$, the MERA has two types of impurity tensor, $v_A$ and $v_B$, one associated to each of the impurities. At greater depth, $s>\log_3 (l)$, the MERA has one type of impurity tensor, $v_C$, associated to a fusion of the two impurities. (b) An inhomogeneous coarse-graining $U^{\Wilson}$, defined from the bulk tensors, maps the original two impurity Hamiltonian $H$ to an effective two impurity Hamiltonian $H^{\Wilson}$. A subsequent coarse-graining $U^{\Wilson}_{\textrm{fuse}}$, defined from the impurity tensors $v_A$ and $v_B$, maps $H^{\Wilson}$ into an effective \emph{single} impurity Hamiltonian $H^{\Wilson}_{\textrm{fuse}}$.}
\label{fig:TwoDefectCG}
\end{center}
\end{figure}
%%%%%%%%%%%%%%%%%%%%%%%%%%%%%%%%%%%%%%%
%%%%%%%%%%%%%%%%%%%%%%%%%%%%%%%%%%%%%%%

\subsubsection{Multiple impurities} \label{sect:BenchMultiple}

Next we consider a system with two impurities, with Hamiltonian
\begin{equation}
H^{\dimpur} = H^\homog + J^{\alpha_A}_{\R_A} + J^{\alpha_B}_{\R_B}, \label{s5e7}
\end{equation}
where $J^{\alpha_A}_{\R_A}$ and $J^{\alpha_B}_{\R_B}$ represent the distinct impurities located on separate local regions $\R_A$ and $\R_B$ of the lattice. 

The two-impurity MERA for the ground state $\ket{\psi^{\dimpur}}$ of Hamiltonian $H^{\dimpur}$ is depicted in Fig. \ref{fig:TwoDefectCG}(a). In this more complex modular MERA the tensors have been modified within the causal cone $\C(\R_A \cup \R_B)$ of the union of regions $\R_A$ and $\R_B$. For length scales $s < \log_3(l)$, where $l$ is the distance separating the two regions ${\R_A}$ and ${\R_B}$, the causal cones $\C(\R_A)$ and $\C(\R_B)$ are distinct, while for length scales $s>\log_3(l)$ the causal cone have fused into a single cone. Thus for short length scales, $s < \log_3(l)$, there are two distinct types of impurity tensor: tensors $v_A$ associated to the impurity $A$ and tensors $v_B$ associated to the impurity $B$. For longer length scales, $s>\log_3(l)$, there is a single type of impurity tensor $v_C$ which is associated to the fusion of the two impurities $A$ and $B$ into a new impurity $C$. The steps for optimizing the two-impurity MERA are as follows:
\begin{enumerate}
	\item Optimize a scale-invariant MERA for the ground state $\ket{\psi}$ of the homogeneous host Hamiltonian $H$ to obtain tensors $\left\{ u, w\right\}$. \label{step:s3e1}
	\item Optimize a (single) impurity MERA for the single impurity Hamiltonian $H^{\impur}=H + J^{\alpha_A}_{\R_A}$ to obtain the impurity tensors $v_A$. \label{step:s3e2}
	\item Optimize a (single) impurity MERA for the single impurity Hamiltonian $H^{\impur}=H + J^{\alpha_B}_{\R_B}$ to obtain the impurity tensors $v_B$. \label{step:s3e3}
	\item Map the original two-impurity Hamiltonian $H^{\dimpur}$ of Eq. \ref{s5e7} to an effective single impurity Hamiltonian $H^{\Wilson}_{\textrm{fuse}}$,
	\begin{equation}
H \stackrel{U^{\Wilson}}{\longrightarrow} H^{\Wilson} \stackrel{U^{\Wilson}_{\textrm{fuse}}}{\longrightarrow} H^{\Wilson}_{\textrm{fuse}}, \label{s5e10} \end{equation}
as depicted in Fig. \ref{fig:TwoDefectCG}(b), where $U^{\Wilson}$ is an inhomogeneous coarse-graining defined in terms of the bulk tensors, and $U^{\Wilson}_{\textrm{fuse}}$ is a coarse-graining defined in terms of the impurity tensors $v_A$ and $v_B$. \label{step:s3e4}
	\item Optimize a TTN for the effective single impurity problem $H^{\Wilson}_{\textrm{fuse}}$ to obtain the impurity tensors $v_C$. \label{step:s3e5}
\end{enumerate}
Thus, by exploiting minimal updates and the modular character of MERA, the two-impurity problem is addressed by solving a sequence of three single impurity problems: two single impurity problems for impurities $A$ and $B$ separately, and a third single impurity problem for the effective impurity $C$ that results from coarse-graining together impurities $A$ and $B$.

To test the validity of this approach, we investigate the case where $H^\homog$ in Eq. \ref{s5e7} is the critical Ising model $H_\textrm{Ising}$ of Eq. \ref{s5e3}, and $J^{\alpha_A}$ and $J^{\alpha_B}$ are each defects of the form described in Eq. \ref{s5e4}. Conformal field theory predicts\cite{ID1} that, when viewed at distances much larger than the separation $l$ between the two impurities, the two-impurity Ising model is equivalent to an Ising model with a single impurity $C$ with effective Hamiltonian $J^{\alpha_C}$. The strength $\alpha_C$ of the fused impurity $C$ relates to the strength $\alpha_A$ and $\alpha_B$ of the original impurities $A$ and $B$ according to \cite{Oshikawa10}
\begin{equation}
\theta _{\alpha_C}  = \theta _{\alpha_A}  + \theta _{\alpha_B}. \label{s5e12}
\end{equation}
where $\theta_\alpha$ is the phase associated to the defect as described by Eq. \ref{s5e6}. We employ the MERA to test a special case of Eq. \ref{s5e12} in which we choose the weight of the second impurity as the inverse of the first, $\alpha_B=1/\alpha_A$, such that $\alpha_C = 1$ is the unique solution to Eq. \ref{s5e12}. In other words, we test the case where the two impurities are predicted to fuse to identity (i.e. no impurity) at large distances.

%%%%%%%%%%%%%%%%%%%%%%%%%%%%%%%%%%%%%
%%%%%%%%%%%%%%%%%%%%%%%%%%%%%%%%%%%%%
\begin{figure}[!htbp]
\begin{center}
\includegraphics[width=8.5cm]{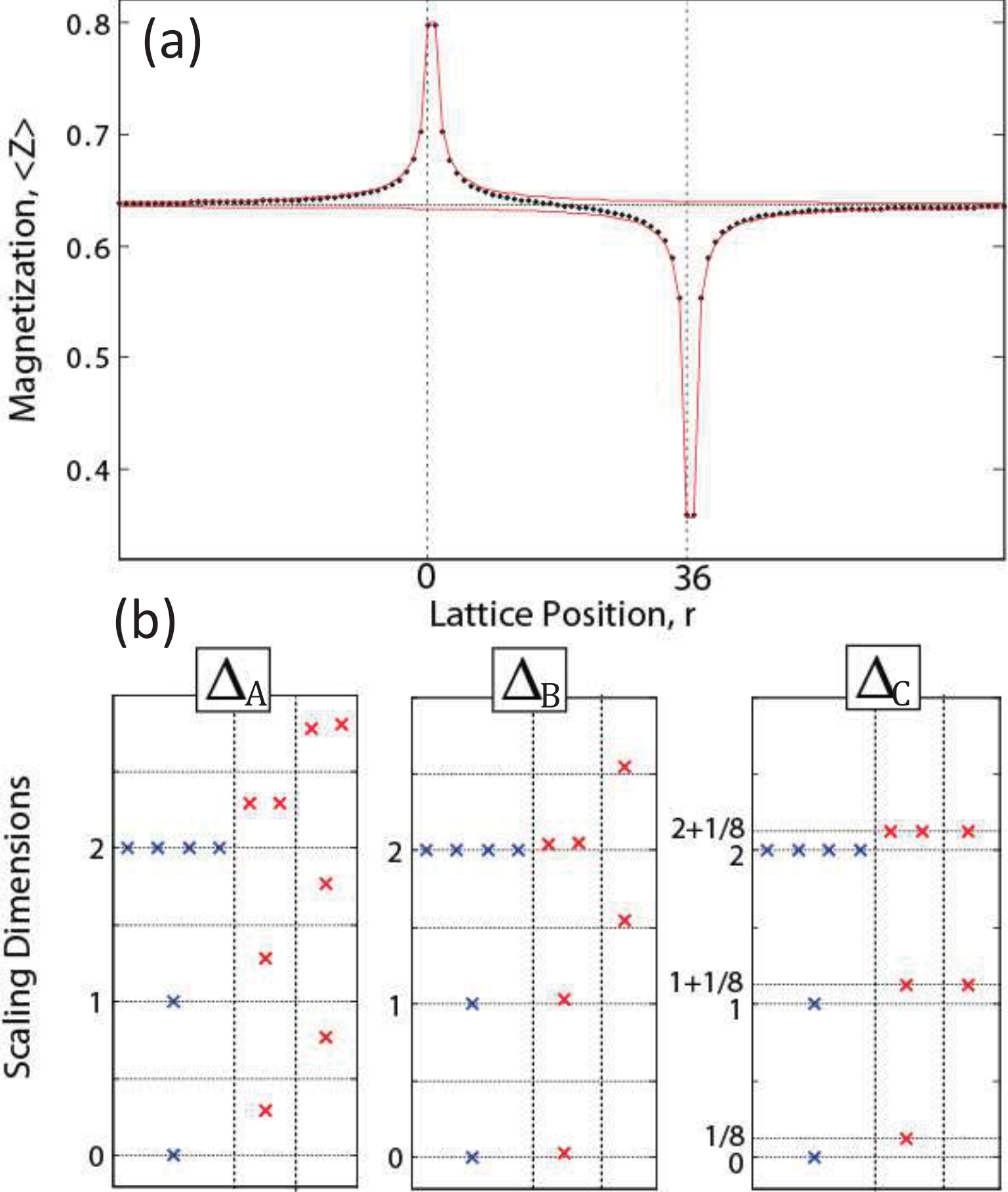}
\caption{(a) Magnetization profile $\left\langle Z(r)\right\rangle$ obtained from the MERA for the ground state of the critical Ising model with two conformal impurities $H^{\alpha_A}$ and $H^{\alpha_B}$ (see Eq. \ref{s5e4}) with strengths $\alpha_A = 0.4$ and $\alpha_B = 1/0.4$ respectively, which are located $l=36$ lattice sites apart. The magnetization profile when both impurities are present is represented with $\bullet$'s, while the two solid lines each represent magnetization profiles when only one of the impurities is present. (b) The spectra of scaling dimensions associated to the impurities: $\Delta_A$ and $\Delta_B$ are the single impurity spectra for impurities of strength $\alpha_A = 0.4$ and $\alpha_B = 1/0.4$ respectively, while $\Delta_C$ is the spectrum arising from the fusion of these conformal impurities. It is seen that $\Delta_C$ matches the scaling dimensions of the bulk (i.e. impurity free) critical Ising model.}
\label{fig:TwoDefectCritExp}
\end{center}
\end{figure}
%%%%%%%%%%%%%%%%%%%%%%%%%%%%%%%%%%%%%%%
%%%%%%%%%%%%%%%%%%%%%%%%%%%%%%%%%%%%%%%

We optimize the two-impurity MERA for the case $\alpha_1=0.4$ and $\alpha_2=1/0.4$, where the impurities are set a distance of $l=36$ sites apart. Tensors $\left\{ w,u \right\}$, and the single impurity tensors $v_A$ and $v_B$ are recycled from the single impurity calculations of Sect. \ref{sect:BenchSingle}. Thus the only additional work to address the two-impurity problem, provided the individual impurities have been previously addressed, is to perform steps \ref{step:s3e4} and \ref{step:s3e5} above, namely producing an effective, single impurity Hamiltonian $H^{\Wilson}_{\textrm{fuse}}$, and then optimizing the impurity tensors $v_C$ for the `fused' impurity $C$. The scaling superoperator $\tilde{\mathcal S}_C$ associated to the fused impurity was diagonalized to obtain the scaling dimensions $\Delta_C$ associated to the fused impurity $C$. These scaling dimensions, together with the magnetization profile $\left\langle Z(r) \right\rangle$ of the two impurity system, are plotted in Fig. \ref{fig:TwoDefectCritExp}. It can be seen that the scaling dimensions $\Delta_C$ reproduce the spectrum of scaling dimensions for the homogeneous Ising model \cite{Francesco97,Henkel99}, as predicted by Eq. \ref{s5e12}, thus indicating that the two-impurity MERA accurately captures the universal properties of the ground state.

The method outlined to address a two-impurity problem can be easily generalized to the case of a system with any finite number of impurities. The many-impurity problem can likewise be reduced to a first sequence of single impurity problems that, under fusion, give rise to a second sequence of single impurity problems, and so on.

\subsection{Boundaries} \label{sect:BenchBound}

Next we benchmark the use of the modular MERA to describe a quantum critical system in the presence of one boundary (semi-infinite chain) and in the presence of two boundaries (finite chain).

\subsubsection{Single boundary (semi-infinite chain)} \label{sect:BenchSemi}

Let us first consider a semi-infinite lattice $\L$ with Hamiltonian $H$,
\begin{equation}
H^{\bound} = J^{\bound} (0)  + \sum_{r = 0}^\infty  h^\homog (r,r + 1) , \label{s6e1}
\end{equation}
where the Hamiltonian term $J^{\bound}$ at site $r=0$ describes the boundary (and can be chosen so as to describe certain types of open boundary conditions, such as `fixed' or `free' open boundary conditions), and $h^{\homog}$ is a nearest neighbor Hamiltonian term such that the Hamiltonian
\begin{equation}
H^{\homog}  = \sum\limits_{r =  - \infty }^\infty  h^{\homog} (r,r + 1) ,\label{s6e1b}
\end{equation}
represents the host system, which is invariant under translations and under changes of scale. The boundary MERA for the ground state $\ket{\psi^{\bound}}$ of Hamiltonian $H^{\bound}$, as described Sect. \ref{sect:BoundMERA}, was initially introduced and tested in Ref. \onlinecite{Evenbly10e}. Here we shall both reproduce and expand upon the results in that paper. A similar construction was proposed also in Ref. \onlinecite{Silvi10}.

\begin{figure}[!tbh]
\begin{center}
\includegraphics[width=8.5cm]{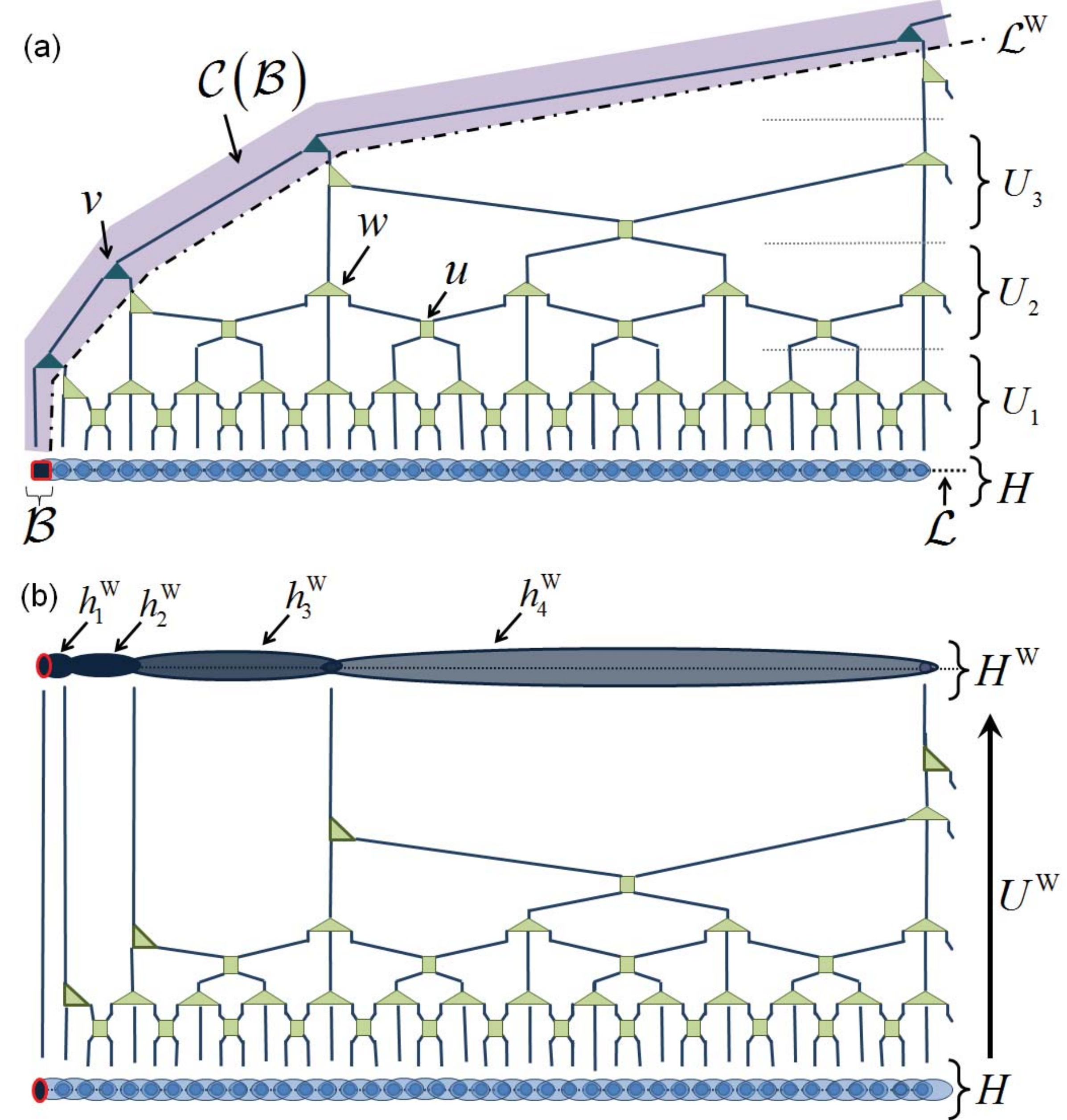}
\caption{(a) A boundary MERA for the semi-infinite chain, each layer $U_s$ of which is described by a pair of bulk tensors $\{ u_s, w_s \}$ and a boundary tensor $v_s$. The causal cone $\C(\B)$ of the boundary $\B$, which only contains boundary tensors $v_s$, is shaded, and the associated Wilson chain $\L^{\Wilson}$ is indicated. (b) An inhomogeneous coarse-graining $U^{\Wilson}$, defined in terms of the bulk tensors, is used to map the original boundary Hamiltonian $H$ to an effective boundary Hamiltonian $H^{\Wilson}$ on the Wilson chain $\L^{\Wilson}$.}
\label{fig:BoundEffective}
\end{center}
\end{figure}

In order to optimize the boundary MERA depicted in Fig. \ref{fig:BoundEffective}(a), which is fully characterized in terms of the tensors $\left\{u,w \right\}$ for the homogeneous system and the tensors $v$ for the boundary, we follow the following steps:
\begin{enumerate}
	\item Optimize tensors $\{u,w\}$ by energy minimization of a MERA for the homogeneous host system with Hamiltonian $H^{\homog}$. \label{step:s4e1}
	\item Map the original boundary Hamiltonian $H^{\bound}$ to the effective boundary Hamiltonian $H^{\Wilson}$ on the Wilson chain $\L^{\Wilson}$, \begin{equation}
H^{\bound} \stackrel{U^{\Wilson}}{\longrightarrow} H^{\Wilson}, \label{s6e2}
\end{equation}
through the inhomogeneous coarse-graining $U^{\Wilson}$, as depicted in Fig. \ref{fig:BoundEffective}(b). \label{step:s4e2}
	\item Optimize the tensors $v$ by energy minimization on the effective Hamiltonian $H^{\Wilson}$. \label{step:s4e3}
\end{enumerate}
These steps can be accomplished with only minor changes to the method presented in Sect. \ref{sect:OptMod}.

%%%%%%%%%%%%%%%%%%%%%%%%%%%%%%%%%%%%%
%%%%%%%%%%%%%%%%%%%%%%%%%%%%%%%%%%%%%
%%%%%%%%%%%%%%%%%%%%%%%%%%%%%%%%%%%%%%%
%%%%%%%%%%%%%%%%%%%%%%%%%%%%%%%%%%%%%%%

%%%%%%%%%%%%%%%%%%%%%%%%%%%%%%%%%%%%%%%
%%%%%%%%%%%%%%%%%%%%%%%%%%%%%%%%%%%%%%%
\begin{figure}[!tbh]
\begin{center}
\includegraphics[width=8.5cm]{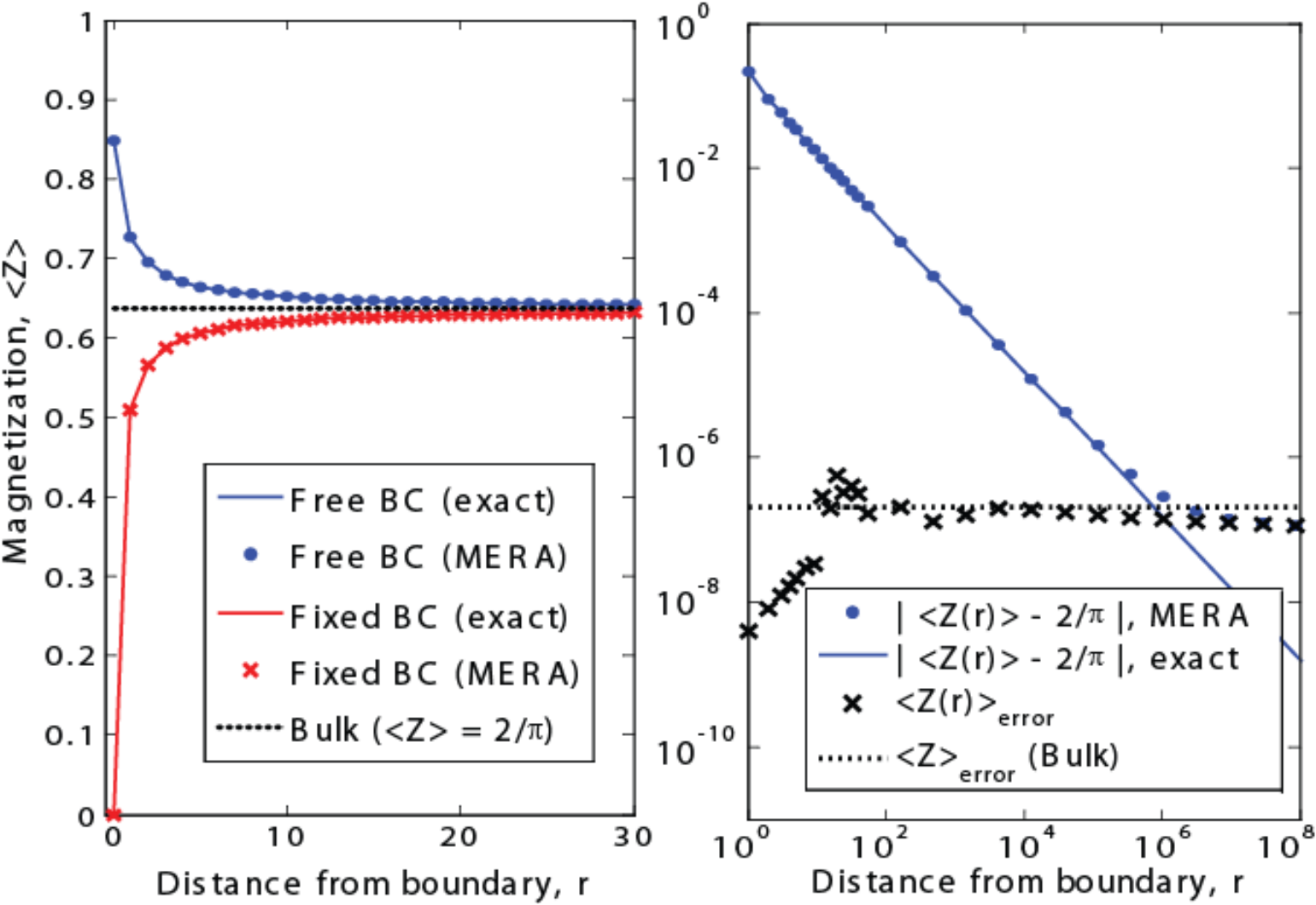}
\caption{Left: expectation value $\langle Z(r) \rangle$ for the critical Ising model with free and fixed BC obtained with a boundary MERA. The exact solution approaches the bulk value $2/\pi$ as $r^{-1}$. Right: error in $\langle Z(r) \rangle$ for free BC (similar to that for fixed BC). The non-vanishing expectation value of bulk scaling operators is accurately reproduced even thousands of sites away from the boundary.}
\label{fig:Zmag}
\end{center}
\end{figure}
%%%%%%%%%%%%%%%%%%%%%%%%%%%%%%%%%%%%%%%
%%%%%%%%%%%%%%%%%%%%%%%%%%%%%%%%%%%%%%%

We consider two quantum critical models for the host Hamiltonian $H^{\homog}$: the critical Ising model $H^{\textrm{Ising}}$ of Eq. \ref{s5e3} and the quantum XX model,
\begin{equation}
	H^\textrm{XX} = \sum\limits_r  X (r) X (r+1) + Y (r) Y(r+1), \label{s6e4}
\end{equation}
where $X$ and $Y$ are Pauli matrices. The boundary condition at site $r=0$ are set either as free boundary, in which case $J^{\bound}=0$ in Eq. \ref{s6e1}, or fixed boundary, $J^{\bound}=\pm X$. Tensors $\{u,w\}$ for the Ising model can be recycled from the calculations of Sect. \ref{sect:BenchImpurity}, while for the quantum XX model they are obtained from a MERA with $\chi=56$ that exploits both reflection symmetry and a global $U(1)$ spin symmetry and required approximately 2 hours of optimization time on a 3.2 GHz desktop PC with 12Gb of RAM. Optimization of the effective boundary problem $H^{\Wilson}$ for the boundary tensors $v$ required less than 10 minutes of computation time for each of the critical models, under each of the boundary conditions tested.

\begin{figure}[!tbh]
\begin{center}
\includegraphics[width=8.5cm]{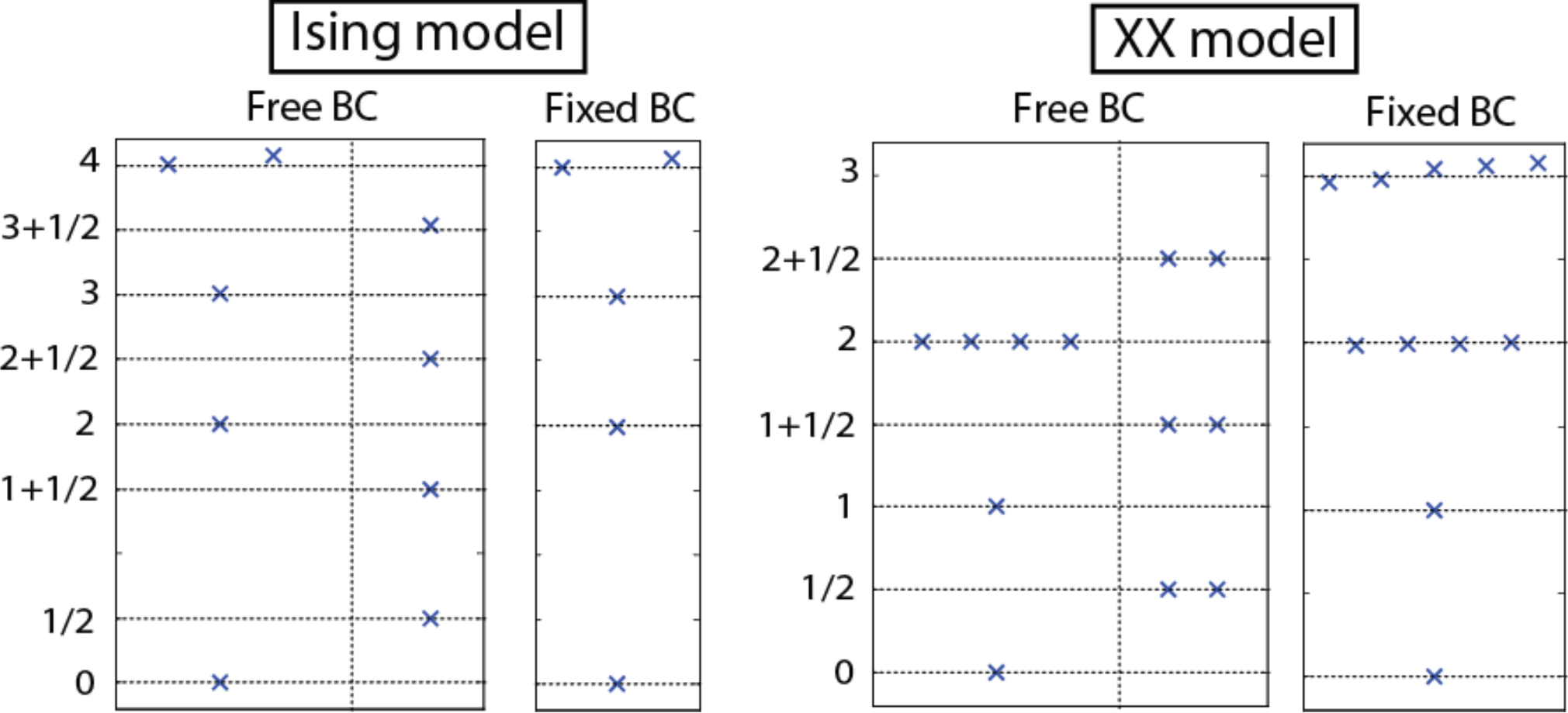}
\caption{A few boundary scaling dimensions, organized in conformal towers, for the quantum Ising and quantum XX models with free and fixed BC. The boundary MERA accurately reproduces the smallest scaling dimensions of each conformal tower.}
\label{fig:ScaleDim}
\end{center}
\end{figure}

Fig. \ref{fig:Zmag} displays the magnetization profile $\langle Z(r) \rangle_{\bound}$ for the Ising model with both free and fixed BC, which are compared against the exact magnetization profiles (obtained using the free fermion formalism),
\begin{align}
&\left\langle {Z(r)} \right\rangle_{{\rm{free}}}  = \frac{2}{\pi }\left( {1 + \frac{1}{{4r + 3}}} \right),\nonumber\\
&\left\langle {Z(r)} \right\rangle_{{\rm{fixed}}}  = \frac{2}{\pi }\left( {1 - \frac{1}{{4r + 1}}} \right). \label{s6e5}
\end{align}
The optimized boundary MERA accurately reproduces the effect of the boundary on the local magnetization even up to very large distances. Specifically, the exact magnetization profile is reproduced within $1\%$ accuracy up to distances of $r\approx 5000$ sites from the boundary. Fig. \ref{fig:ScaleDim} shows the boundary scaling dimensions $\Delta$ for critical Ising and quantum XX models, obtained by diagonalizing the scaling superoperator $\tilde{\mathcal{S}}$ associated to the boundary. The boundary scaling dimensions obtained from the boundary MERA also reproduce the known results from CFT \cite{Francesco97} with remarkable accuracy. For the Ising model the smallest scaling dimensions $( \Delta \le 3) $ are reproduced with less than $0.2\%$ error while for the quantum XX model $(\Delta \le 2.5)$ the error is less than $0.4\%$.

Finally, we analyze the boundary contribution $dE$ to the ground state energy,
\begin{equation}
dE   \equiv \left\langle H^{\bound} \right\rangle  - \frac{1}{2} \left\langle H^\homog \right\rangle, \label{s6e6}
\end{equation}
defined as the difference between the energy $\left\langle H^{\bound} \right\rangle$ of the semi-infinite chain with the boundary term $J^{\bound}$, Eq. \ref{s6e1}, and one half of the ground state energy for the host Hamiltonian on the infinite chain $\left\langle H^\homog \right\rangle$, Eq. \ref{s6e1b}. Since both $\left\langle H^{\bound} \right\rangle$ and $\left\langle H^\homog \right\rangle$ are infinite quantities, we cannot compute $dE$ through the evaluation of the individual terms in Eq. \ref{s6e6}. Instead, we estimate $dE$ by comparing the energy of the first $l$ sites of the semi-infinite chain to the energy of $l$ sites of the infinite homogeneous system, and increase the value of $l$ until the energy difference is converged within some accuracy. For the quantum Ising model on a semi-infinite lattice we obtain the following results: for free BC, a value $dE = 0.18169023$, which is remarkably close to the exact solution\cite{Henkel99}, ${dE}_{\mbox{\tiny exact}}= (1/2 -1/\pi) = 0.18169011...$, and for fixed boundary conditions, a value ${dE} = -0.45492968$ which, based upon the exact solution for finite chains of over a thousand sites, we estimate to carry an error of less than $10^{-6}$.

%%%%%%%%%%%%%%%%%%%%%%%%%%%%%%%%%%%%%%%
%%%%%%%%%%%%%%%%%%%%%%%%%%%%%%%%%%%%%%%
%%%%%%%%%%%%%%%%%%%%%%%%%%%%%%%%%%%%%%%
%%%%%%%%%%%%%%%%%%%%%%%%%%%%%%%%%%%%%%%

\subsubsection{Two boundaries (finite chain)} \label{sect:BenchFinite}

Let us now consider a finite lattice $\mathcal{L}$ made of $N$ sites and with two boundaries, with Hamiltonian
\begin{equation}
H^{\dbound} = J^{\bound}_L (0) + \sum\limits_{r = 0}^{N - 2} {h^\homog (r,r + 1)} + J^{\bound}_R (N-1). \label{s6e8}
\end{equation}
where $J^{\bound}_L$ and $J^{\bound}_R$ at sites $r=0$ and $r=N-1$ describe the left and right boundaries, respectively, and the $h^{\homog}$ is a nearest neighbor Hamiltonian term as in Eq. \ref{s6e1b}. 

\begin{figure}[!tbh]
\begin{center}
\includegraphics[width=8.5cm]{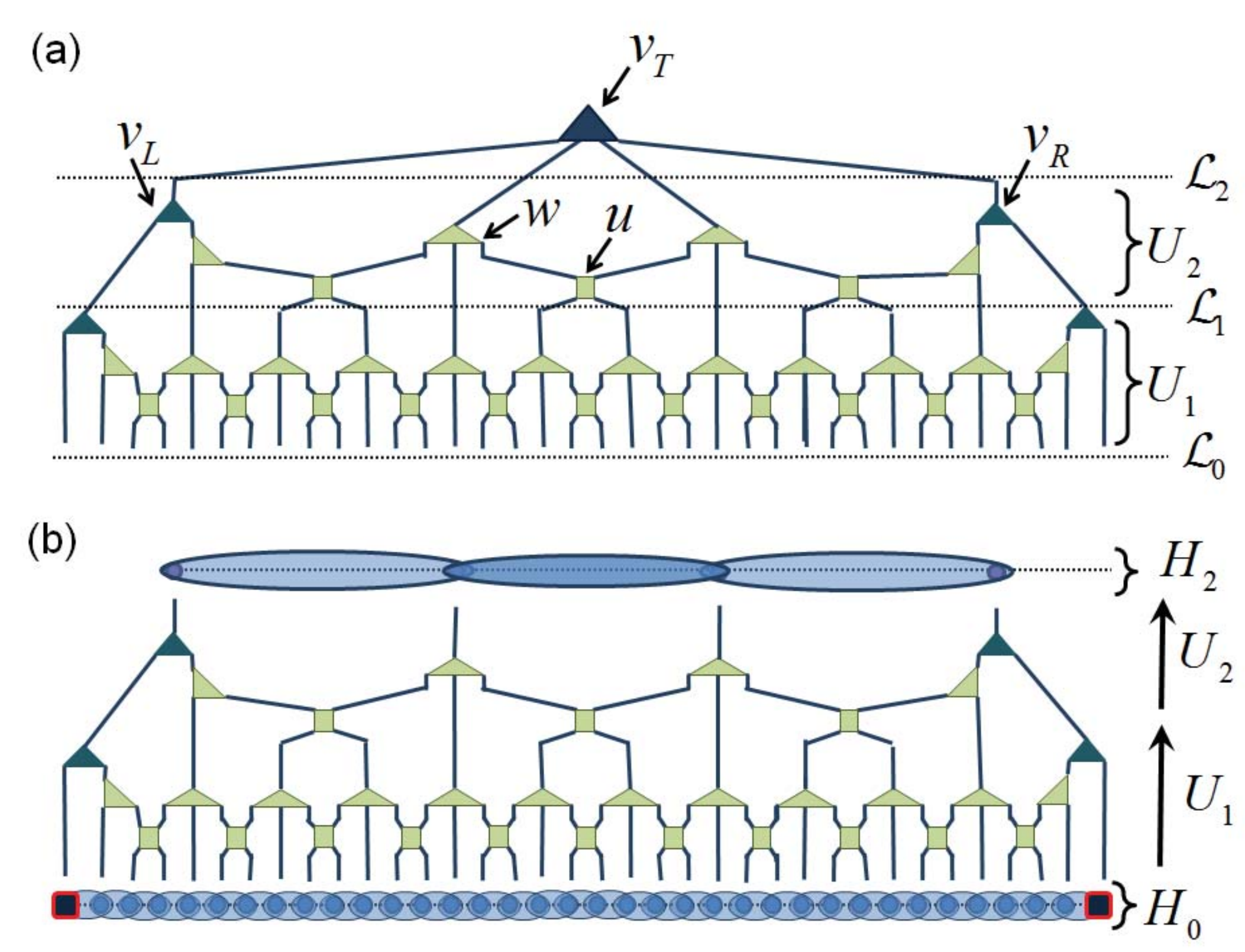}
\caption{(a) An open boundary MERA for finite lattice of $N=36$ sites. The MERA is organized into $T=3$ layers, where each layer $U$ is described by a pair of bulk tensors $\left\{ u,w \right\}$ and (left, right) boundary tensors $v_L$ and $v_R$. The boundary MERA also has a top-tensor $v_T$ at the final level. (b) The original boundary problem $H_0\equiv H^{\dbound}$ defined on an $N$-site lattice $\mathcal L_0$ can be mapped into an effective open boundary problem $H_2$ defined on a $4$-site lattice $\mathcal{L}_2$ through coarse-graining with MERA layers $U_1$ and $U_2$, see also Eq. \ref{s6e9}.}
\label{fig:BoundFinite}
\end{center}
\end{figure}

A two-boundary MERA for the ground state $\ket{\psi^{\dbound}}$ of a finite chain with Hamiltonian $H^{\dbound}$ is depicted in Fig. \ref{fig:BoundFinite}(a). Each layer of tensors consists of tensors $\left\{ u,w \right\}$ in the bulk and tensors $v_L$ and $v_R$ at the left and right boundaries, respectively. The two-boundary MERA is organized into a finite number, $T\approx\log_3 (N)$, of layers, and has an additional tensor $v_T$ at the top. The steps for optimizing this particular form of modular MERA are as follows:
\begin{enumerate}
  \item Optimize tensors $\{u,w\}$ by energy minimization of a MERA for the homogeneous infinite host system with Hamiltonian $H$. \label{step:s5e1}
	\item Optimize the left boundary tensors $v_L$ by energy minimization on an effective semi-infinite, single boundary problem with boundary term $J_{L}^{\bound}$, as described in Sect. \ref{sect:BenchSemi}.
	\item Optimize the right boundary tensors $v_R$ by energy minimization on an effective semi-infinite, single boundary problem with boundary term $J_{R}^{\bound}$, as described in Sect. \ref{sect:BenchSemi}.
	\item Coarse-grain the original boundary problem $H_0 \equiv H^{\dbound}$, defined on the $N$-site lattice $\L_0 \equiv \L$, into an effective boundary problem $H_{T}$ defined on the coarse-grained lattice $\mathcal L_{T}$,
	\begin{equation}
H_{0} \stackrel{U_{1}}{\longrightarrow} H_{1} \stackrel{U_{2}}{\longrightarrow} ~\ldots~ \stackrel{U_{T}}{\longrightarrow} H_{T} \label{s6e9}
\end{equation}
where each $U_s$ is a layer of the two-boundary MERA, as depicted in Fig. \ref{fig:BoundFinite}(b). \label{step:s5e3}
\item Compute the top tensor $v_T$ through diagonalization of the effective Hamiltonian $H_T$ for its ground state or excited states. \label{step:s5e4}
\end{enumerate}
In summary, to treat a finite chain with open boundaries with the MERA, one should first address an infinite system, then two semi-infinite systems, and finally a coarse-grained version of the original Hamiltonian, which is reduced to a small number of sites.

%%%%%%%%%%%%%%%%%%%%%%%%%%%%%%%%%%%%%
%%%%%%%%%%%%%%%%%%%%%%%%%%%%%%%%%%%%%
%%%%%%%%%%%%%%%%%%%%%%%%%%%%%%%%%%%%%%%
%%%%%%%%%%%%%%%%%%%%%%%%%%%%%%%%%%%%%%%

%%%%%%%%%%%%%%%%%%%%%%%%%%%%%%%%%%%%%
%%%%%%%%%%%%%%%%%%%%%%%%%%%%%%%%%%%%%
\begin{figure}[!tbh]
\begin{center}
\includegraphics[width=8.5cm]{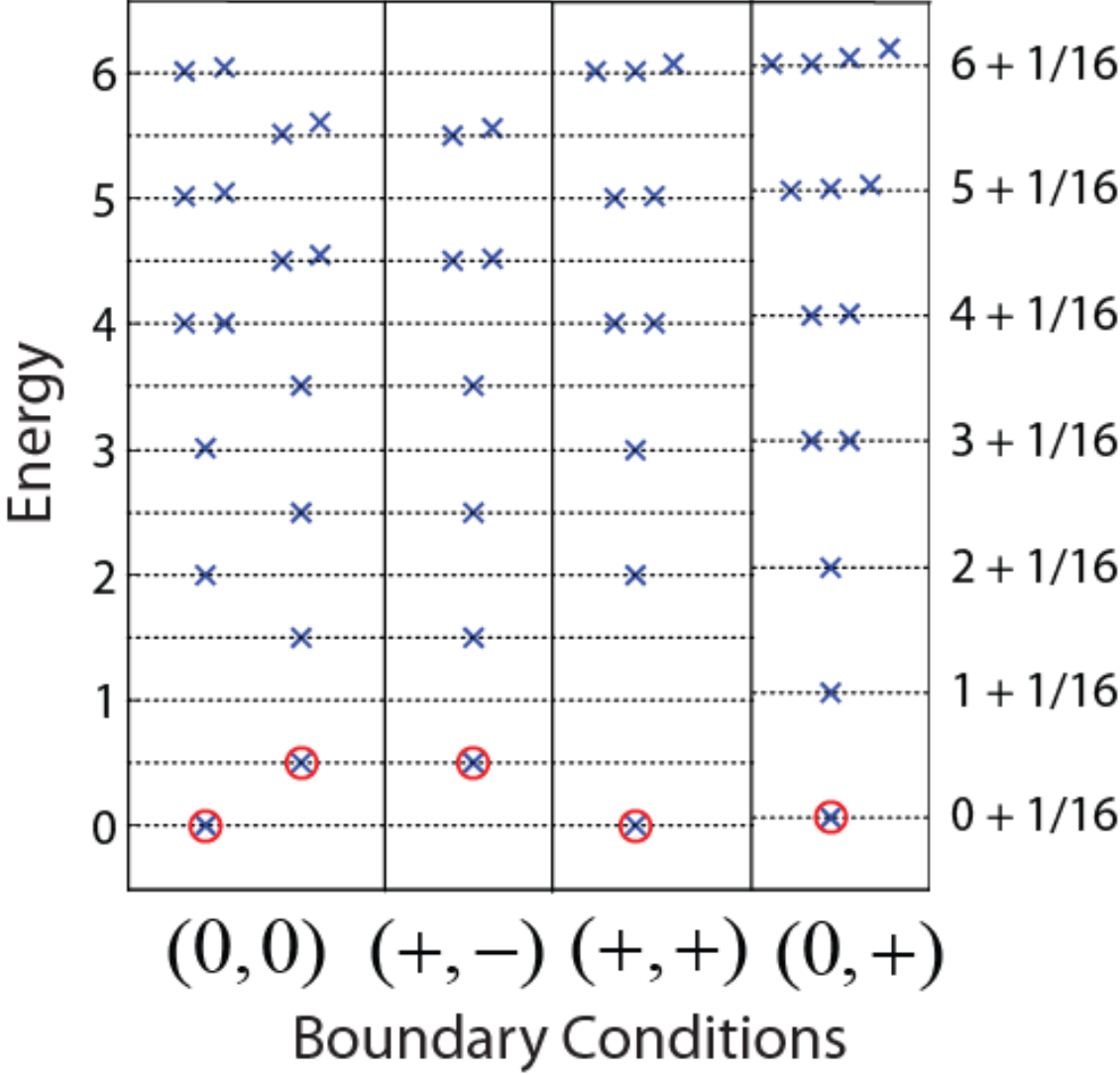}
\caption{Excitation spectra of the quantum Ising model on a finite lattice $\mathcal{L}$ of $N=2916$ sites with different combinations of open boundary conditions. The energy is expressed in units such that the gap between descendants is a multiple of unity. All non-equivalent combinations of open BC are considered. The different open BC are $(0) = \textrm{free}, (+) = \textrm{fixed(up)}, (-) = \textrm{fixed(down)}$.}
\label{fig:FiniteSpect}
\end{center}
\end{figure}
%%%%%%%%%%%%%%%%%%%%%%%%%%%%%%%%%%%%%%%
%%%%%%%%%%%%%%%%%%%%%%%%%%%%%%%%%%%%%%%

To test the validity of the two-boundary MERA to finite systems with open boundary conditions, we investigate the low energy spectrum of the critical Ising model under different fixed and free boundary conditions, as defined Sect. \ref{sect:BenchSemi}. We are able to recycle the tensors $\{u,w\}$ for the homogeneous host system, as well as the boundary tensors $v_L$ and $v_R$ obtained from the previous investigation of semi-infinite Ising chains in Sect. \ref{sect:BenchSemi}. Thus, we only need to perform steps \ref{step:s5e3} and \ref{step:s5e4} above. We proceed by constructing the effective Hamiltonians $H_T$ for a two-boundary MERA with $T=6$ total layers, which equates to a total system size of $N=4 \times 3^T = 2916$ sites, for all non-equivalent combinations of boundary conditions. There are four such non-equivalent combinations: free-free,  fixed(up)-fixed(down), fixed(up)-fixed(up) and free-fixed. The low-energy spectra of the effective Hamiltonians $H_T$ are then computed with exact diagonalization based on the Lanczos method. These low-energy spectra, displayed in Fig. \ref{fig:FiniteSpect}, match the predictions from CFT \cite{Francesco97,Henkel99} to high precision. These results indicate that the two-boundary MERA is not only a good ansatz for the ground states of finite systems with open boundary conditions, but also for their low-energy excited states. Furthermore, only the top tensor $v_T$ of the MERA needs to be altered in order to describe different excited states.

\subsection{Interfaces} \label{sect:BenchInterface}

Next we benchmark the use of the modular MERA to describe the interface between two or more quantum critical systems.

\subsubsection{Interface between two systems} \label{sect:BenchTwo}

Let us first consider the interface between two systems $A$ and $B$, described by an infinite lattice $\L$ with a Hamiltonian of the form
\begin{equation}
H  =  \sum_{r = -\infty}^{-1}  h_A^\homog (r,r + 1) + J^{\inter}(0,1) + \sum_{r = 1}^{\infty}  h_B^\homog (r,r + 1), \label{s7e1}
\end{equation}
where the Hamiltonian term $J^{\inter}$ couples two (left and right) semi-infinite chains $\L_A$ and $\L_B$, $\L = \L_A \cup \L_B$, and the nearest neighbor terms $h_A$ and $h_B$ are such that on an infinite lattice, the Hamiltonians
\begin{eqnarray}
H^{\homog}_{A}  &=& \sum\limits_{r =  - \infty }^\infty  h^{\homog}_{A} (r,r + 1), \\
H^{\homog}_{B}  &=& \sum\limits_{r =  - \infty }^\infty  h^{\homog}_{B} (r,r + 1) ,\label{eq:AB}
\end{eqnarray}
describe homogeneous, quantum critical host systems that are invariant under translations and changes of scale. 

\begin{figure}[!htbp]
\begin{center}
\includegraphics[width=8.5cm]{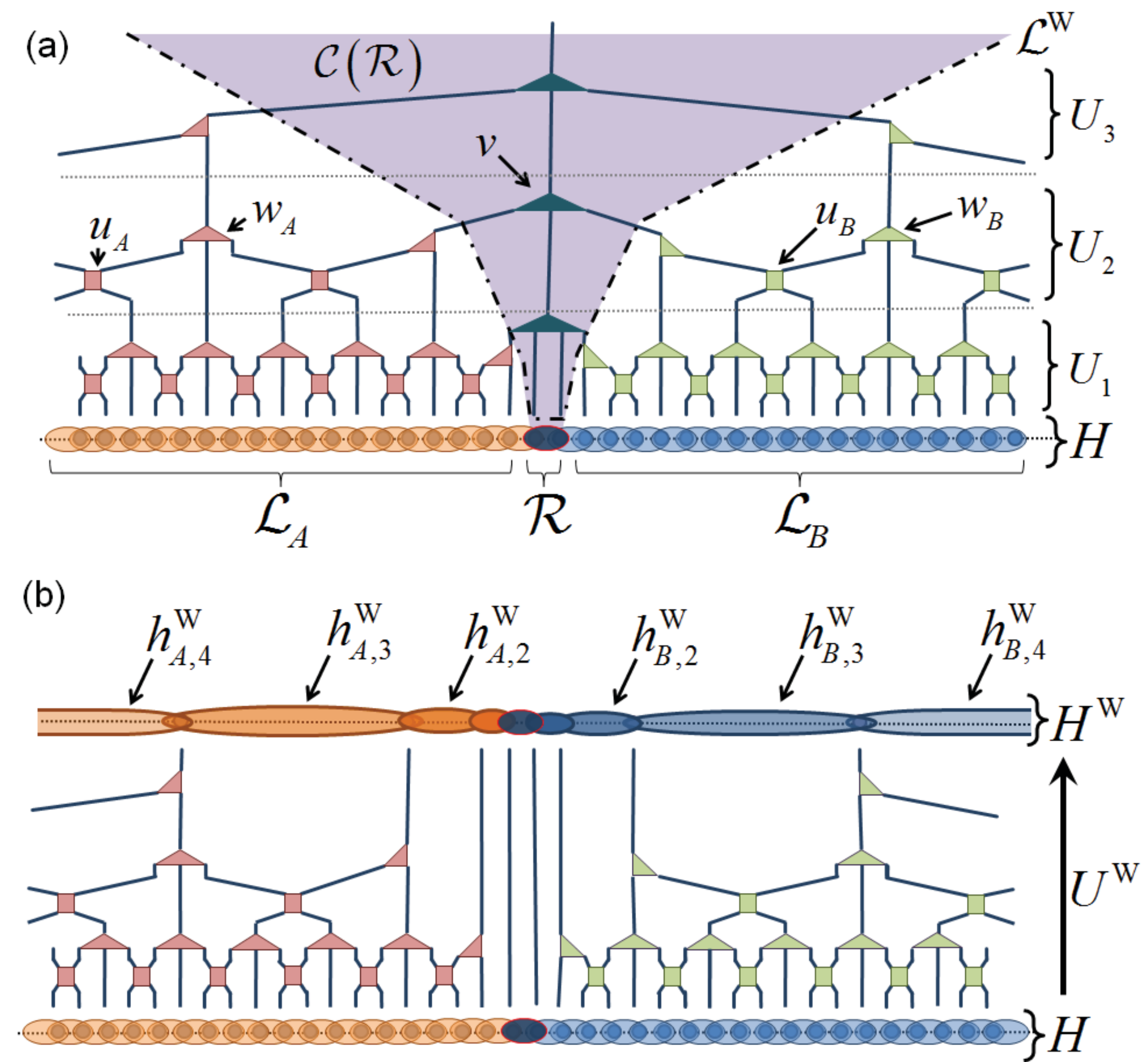}
\caption{(a) An interface MERA is used to describe the interface of a critical system $H_A^{\homog}$, supported on semi-infinite chain $\L_A$, with a different critical system $H_B^{\homog}$, supported on semi-infinite chain $\L_B$. Each layer $U$ of the interface MERA is described by a pair of tensors $\left\{ u_A, w_A \right\}$ associated to host system `$A$', a pair of tensors $\left\{ u_B, w_B\right\}$ associated to host system `$B$', and an interface tensor $v$, which resides in the causal cone $\C(\R)$ of the interface region $\R$. The Wilson chain $\L^{\Wilson}$ associated to the interface $\R$ is indicated. (b) The inhomogeneous coarse-graining $U^{\Wilson}$, defined in terms of the host tensors $\left\{ u_A, w_A\right\}$ and $\left\{u_B, w_B \right\}$, maps original interface Hamiltonian $H$ to an effective interface Hamiltonian $H^{\Wilson}$ defined on the Wilson chain $\L^{\Wilson}$.}
\label{fig:InterfaceEffective}
\end{center}
\end{figure}

The interface MERA for the ground state $\ket{\psi^{\inter}}$ of Hamiltonian $H^{\inter}$, depicted in Fig. \ref{fig:InterfaceEffective}(a), is made of the following tensors: two sets of tensors $\left\{ u_A, w_A \right\}$ and $\left\{ u_B, w_B\right\}$ corresponding to the MERA for the ground state of the host Hamiltonians $H^{\homog}_A$ and $H^{\homog}_B$, respectively, and the interface tensors $v$. Optimization of the interface MERA can be accomplished through a straightforward generalization of the approach described in Sect. \ref{sect:OptMod} for an impurity. The only differences here are that one needs to address first two different homogeneous systems, and that the coarse-graining of $H^{\inter}$ into the effective Hamiltonian $H^{\Wilson}$ on the Wilson chain $\L^{\Wilson}$, see Fig. \ref{fig:InterfaceEffective}(b), uses one set of host tensors $\left\{ u_A, w_A \right\}$ on the left and the other $\left\{ u_B, w_B\right\}$ on the right.

%%%%%%%%%%%%%%%%%%%%%%%%%%%%%%%%%%%%%
%%%%%%%%%%%%%%%%%%%%%%%%%%%%%%%%%%%%%
%%%%%%%%%%%%%%%%%%%%%%%%%%%%%%%%%%%%%%%
%%%%%%%%%%%%%%%%%%%%%%%%%%%%%%%%%%%%%%%

%%%%%%%%%%%%%%%%%%%%%%%%%%%%%%%%%%%%%
%%%%%%%%%%%%%%%%%%%%%%%%%%%%%%%%%%%%%
\begin{figure}[!tbh]
\begin{center}
\includegraphics[width=8.5cm]{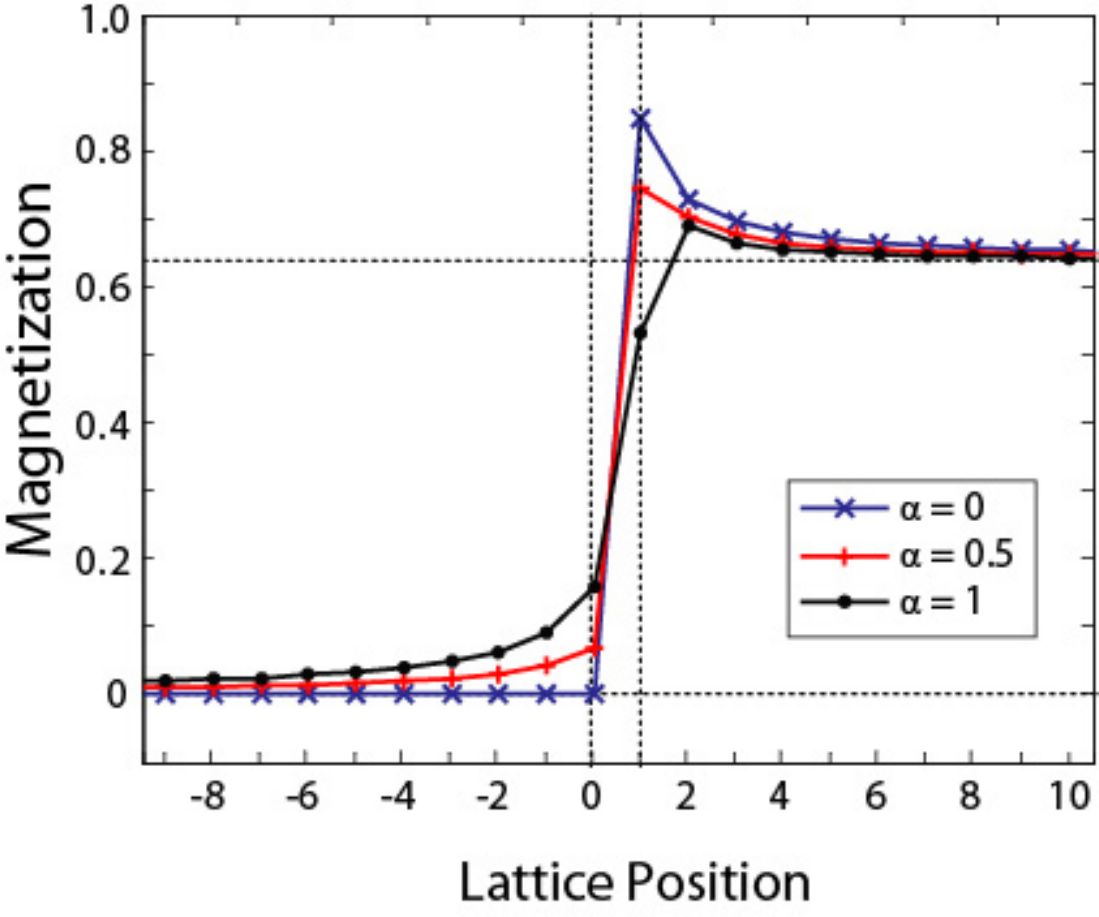}
\caption{The magnetization profile $M(r)$, as defined Eq. \ref{s7e3}, of the interface between a quantum XX model chain (on sites $r\le0$) and the critical Ising chain (on sites $r\ge1$), coupled across the interface $r=(0,1)$. The parameter $\alpha$ relates to the strength of the interface coupling. In all cases the magnetization decays to the bulk value, $M_\homog=0$ for quantum XX and $M_\homog=2/\pi$ for Ising, as $|M(r)-M_\homog| \approx 1/|r|$.}
\label{fig:InterfaceMag}
\end{center}
\end{figure}
%%%%%%%%%%%%%%%%%%%%%%%%%%%%%%%%%%%%%%%
%%%%%%%%%%%%%%%%%%%%%%%%%%%%%%%%%%%%%%%

We test the validity of the interface MERA by choosing as quantum critical systems $A$ and $B$ the quantum XX model in Eq. \ref{s6e4} and the critical Ising model in Eq. \ref{s5e3}, respectively, and as the coupling at the interface the two-site term
\begin{equation}
J^{\inter}(0,1) =  \alpha X(0) X(1), \label{s7e2}
\end{equation}
for several values of $\alpha=\left\{ 0,0.25,0.5,0.75,1 \right\}$. The tensors $\{u_A,w_A\}$ for the quantum XX model and $\{u_B,w_B\}$ for the Ising model are recycled from previous computations in Sect. \ref{sect:BenchBound}. Thus the only additional work required is to produce the effective interface Hamiltonian $H^{\Wilson}$, and then to optimize the interface tensors $v$ by energy minimization over $H^{\Wilson}$. The later, undertaken on a 3.2 GHz desktop PC with 12Gb of RAM, required only approximately 20 minutes of computation time for every value of $\alpha$. Fig. \ref{fig:InterfaceMag} plots the magnetization profile $M(r)$,
\begin{equation}
M(r) \equiv \sqrt {\left\langle {X{(r)} } \right\rangle ^2  + \left\langle {Y{(r)} } \right\rangle ^2  + \left\langle {Z{(r)} } \right\rangle ^2 }, \label{s7e3}
\end{equation}
obtained from the optimized interface MERA.

%%%%%%%%%%%%%%%%%%%%%%%%%%%%%%%%%%%%%
%%%%%%%%%%%%%%%%%%%%%%%%%%%%%%%%%%%%%
\begin{figure}[!tbh]
\begin{center}
\includegraphics[width=8.5cm]{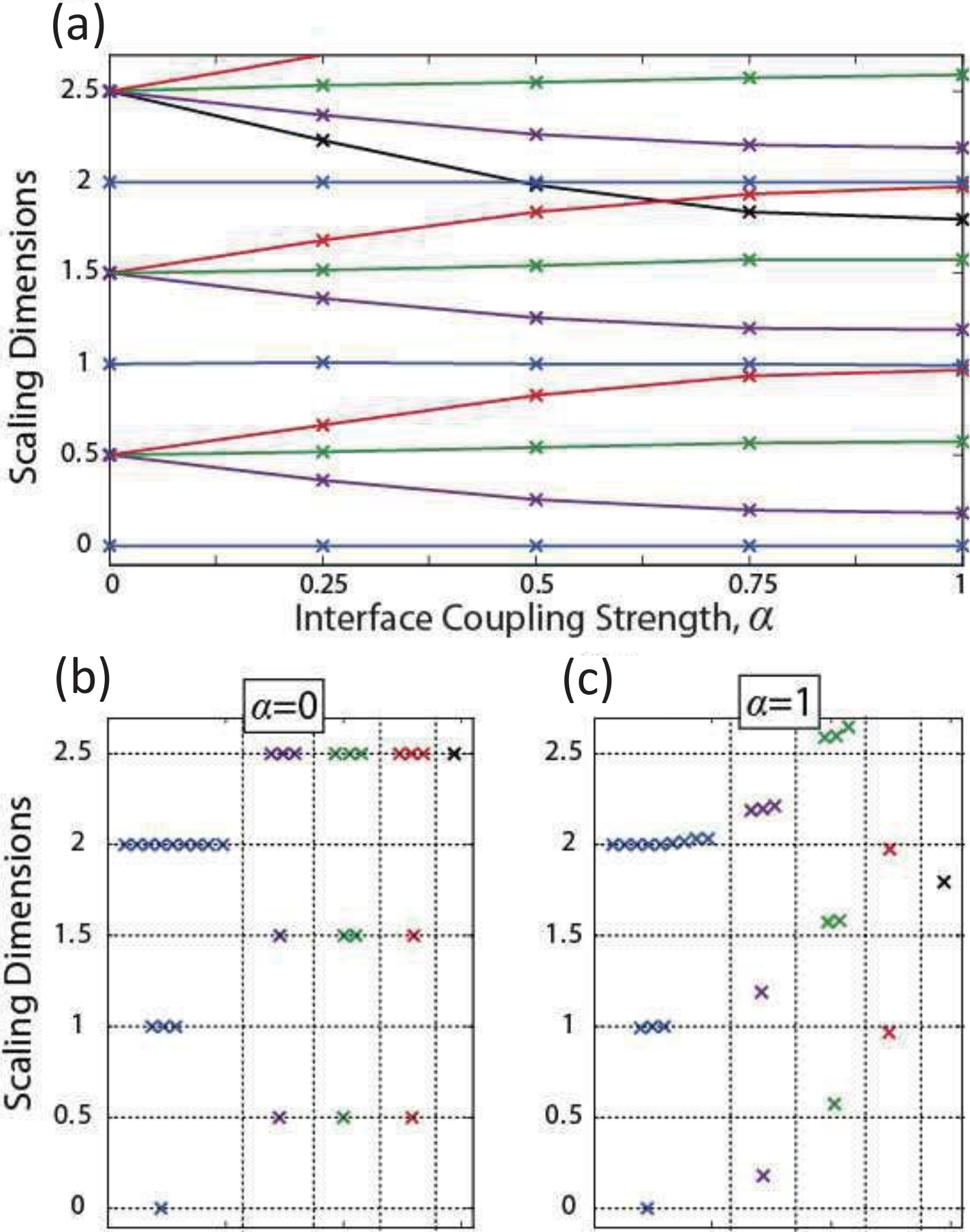}
\caption{(a) Scaling dimensions $\Delta$ associated to the interface of the quantum XX and Ising model, as a function of the coupling strength $\alpha$ across the interface. (b) The scaling dimensions for the interface with no coupling ($\alpha=0$), which take on integer and half-integer values, are seen to be the product of the boundary scaling dimensions for quantum XX and Ising models with free BC. (c) Under interaction strength $\alpha=1$ much of the degeneracy of the $\alpha=0$ case is lifted, yet the scaling dimensions remain organized in conformal towers.}
\label{fig:InterfaceCritExp}
\end{center}
\end{figure}
%%%%%%%%%%%%%%%%%%%%%%%%%%%%%%%%%%%%%%%
%%%%%%%%%%%%%%%%%%%%%%%%%%%%%%%%%%%%%%%

For $\alpha=0$ in Eq. \ref{s7e2} (that is, two decoupled semi-infinite chains), we recover indeed the magnetization profiles for the semi-infinite quantum XX chain and semi-infinite Ising chain with a free boundary, as expected. For $\alpha>0$, the quantum XX chain acquires a non-zero magnetization near the interface, and the magnetization of the Ising chain near the interface is reduced with respect to the case $\alpha=0$. However, away from the interface, the magnetizations still decay polynomially to their values for a homogeneous system: $M_\homog=0$ for the quantum XX model and $M_\homog=2/\pi$ for the critical Ising model.

We also computed the scaling dimensions $\Delta$ associated to the interface, as plotted in Fig. \ref{fig:InterfaceCritExp}, through diagonalization of the scaling superoperator $\tilde{\mathcal S}$ associated to the interface. The exact scaling dimensions are only known to us for the case of interface strength $\alpha=0$ (decoupled case), where one would expect the spectrum of scaling dimensions to be the product of spectra for the open boundary Ising and open boundary quantum XX models on a semi-infinite chains, see Fig. \ref{fig:ScaleDim}. The numerical results of Fig. \ref{fig:InterfaceCritExp} match this prediction. For $\alpha>0$, we no longer have exact scaling dimensions to compare with. However, we see that these are still organized in conformal towers, where the scaling dimensions for descendant fields differ by an integer from the scaling dimensions of the corresponding primary fields \cite{Francesco97}, and where the scaling dimensions of the primary fields depend on $\alpha$. This is a strong indication that the results from the interface MERA are correct. Interestingly, those scaling dimensions that correspond to an integer value for $\alpha=0$, remain unchanged for $\alpha>0$, up to small numerical errors. These are likely to be protected by a symmetry (the interface Hamiltonian has a global $\mathbb Z_2$, spin flip symmetry) similar to the case of the critical Ising impurity model described in Sect. \ref{sect:BenchImpurity}.

%%%%%%%%%%%%%%%%%%%%%%%%%%%%%%%%%%%%%
%%%%%%%%%%%%%%%%%%%%%%%%%%%%%%%%%%%%%
\begin{figure}[!tbh]
\begin{center}
\includegraphics[width=8.5cm]{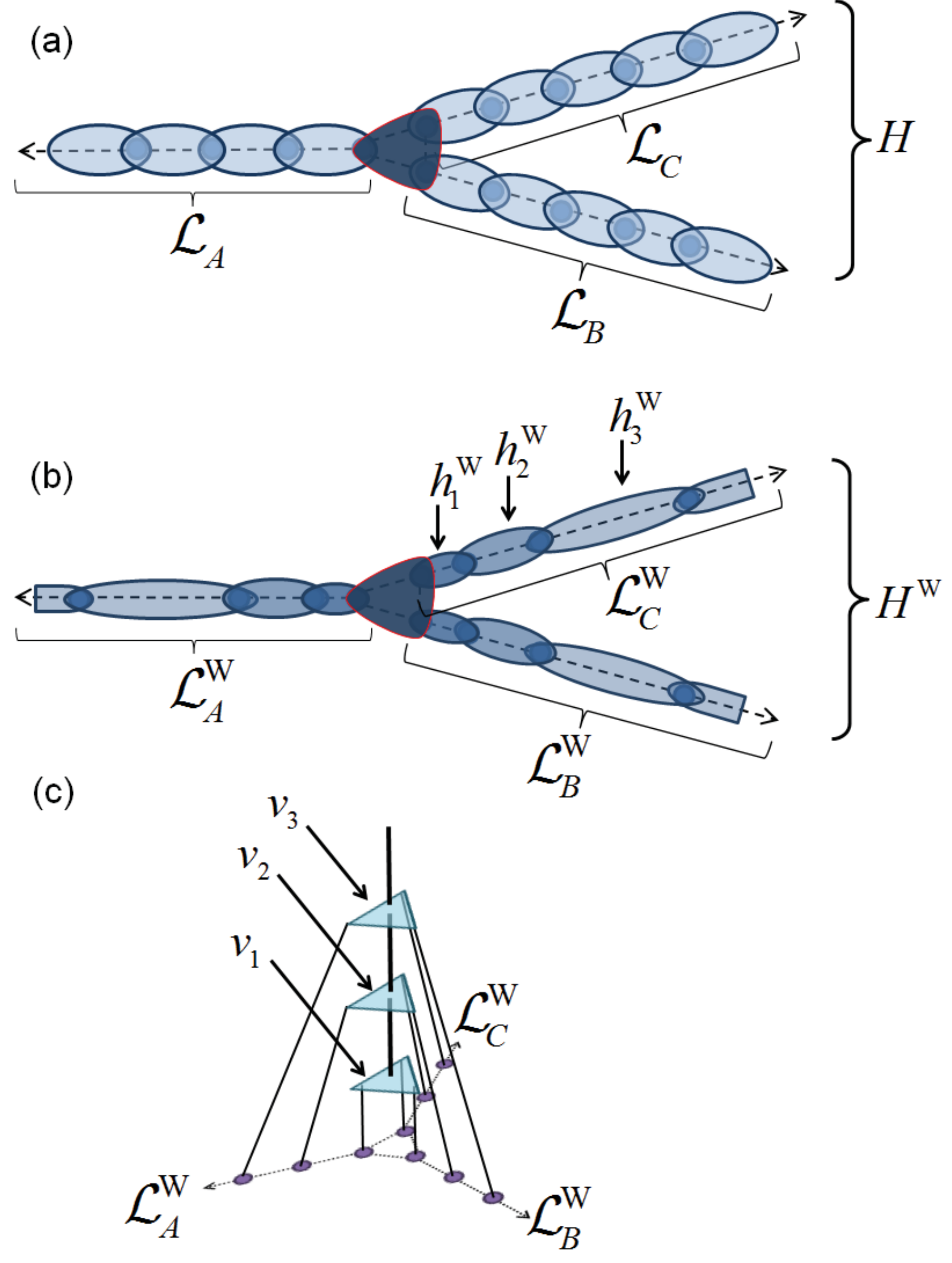}
\caption{(a) A depiction of the Y-interface Hamiltonian $H$, see Eq. \ref{s7e1b}. (b) Under action of the inhomogeneous coarse-graining $U^{\Wilson}$ the Hamiltonian $H$ is mapped to an effective Y-interface Hamiltonian $H^{\Wilson}$ on the Wilson chain. (c) The Y-interface tensors $v_s$, which form a peculiar tree tensor network on the Wilson chain, are obtained through optimization of the effective Hamiltonian $H^{\Wilson}$.}
\label{fig:YMERA}
\end{center}
\end{figure}
%%%%%%%%%%%%%%%%%%%%%%%%%%%%%%%%%%%%%%%
%%%%%%%%%%%%%%%%%%%%%%%%%%%%%%%%%%%%%%%
\subsubsection{Y-interface between three systems} \label{sect:BenchY}

Let us now consider a Y-interface (also called Y-junction) between three systems, as described by a lattice $\L$ made of the union of three semi-infinite lattices $\L_A$, $\L_B$, and $\L_C$, $\L = \L_A \cup \L_B \cup \L_C$, see Fig. \ref{fig:YMERA}(a), with Hamiltonian
\begin{eqnarray}
H^{\Yjun}  &=& J^{\Yjun}(1^A,1^B,1^C) + \sum_{r = 1}^{\infty}  h_A^\homog (r^A,[r + 1]^A)\label{s7e1b}\\
&+& \sum_{r = 1}^{\infty}  h_B^\homog (r^B,[r + 1]^B) + \sum_{r = 1}^{\infty}  h_C^\homog (r^C,[r + 1]^C)\nonumber
\end{eqnarray}
Here we use $r^A$ (and $r^B$, $r^C$) to denote site $r$ of lattice $\L_A$ (respectively, $\L_B$, $\L_C$). The term $J^{\Yjun}$ describes the coupling between the three semi-infinite chains $\L_A$, $\L_B$, and $\L_C$, whereas the nearest neighbor terms $h_A$, $h_B$, and $h_C$ are such that on an infinite lattice, the Hamiltonians
\begin{eqnarray}
H^{\homog}_{A}  &=& \sum\limits_{r =  - \infty }^\infty  h^{\homog}_{A} (r,r + 1), \\
H^{\homog}_{B}  &=& \sum\limits_{r =  - \infty }^\infty  h^{\homog}_{B} (r,r + 1), \\
H^{\homog}_{C}  &=& \sum\limits_{r =  - \infty }^\infty  h^{\homog}_{C} (r,r + 1), \label{eq:ABC}
\end{eqnarray}
describe homogeneous, quantum critical host systems that are invariant under translations and changes of scale. 

The Y-interface MERA for the ground state $\ket{\psi^{\Yjun}}$ of Hamiltonian $H^{\Yjun}$ is a straightforward generalization of the interface MERA considered in Sect. \ref{sect:BenchTwo}. It is characterized by three sets of tensors $\{u_A,w_A\}$, $\{u_B,w_B\}$, and $\{u_C,w_C\}$ that describe the MERA for the ground states of the host Hamiltonians $H^{\homog}_{A}$, $H^{\homog}_{B}$, and $H^{\homog}_{C}$, and a set of tensors $v$ at the Y-interface. Upon optimizing tensors $\{u_A,w_A\}$, $\{u_B,w_B\}$, and $\{u_C,w_C\}$ in three independent optimizations, they are used to map the initial Y-interface Hamiltonian $H^{\Yjun}$ to an effective Hamiltonian $H^{\Wilson}$, see Fig. \ref{fig:YMERA}(b), now by employing three copies of the mapping depicted in Fig. \ref{fig:LogScale}(b). The Y-interface tensors $v$, which are arranged in the TTN structure depicted in Fig. \ref{fig:YMERA}(c), are then optimized to minimize the energy according to the effective Hamiltonian $H^{\Wilson}$ using the approach described in Sect. \ref{sect:OptLog}.

%%%%%%%%%%%%%%%%%%%%%%%%%%%%%%%%%%%%%
%%%%%%%%%%%%%%%%%%%%%%%%%%%%%%%%%%%%%
\begin{figure}[!tbh]
\begin{center}
\includegraphics[width=8.5cm]{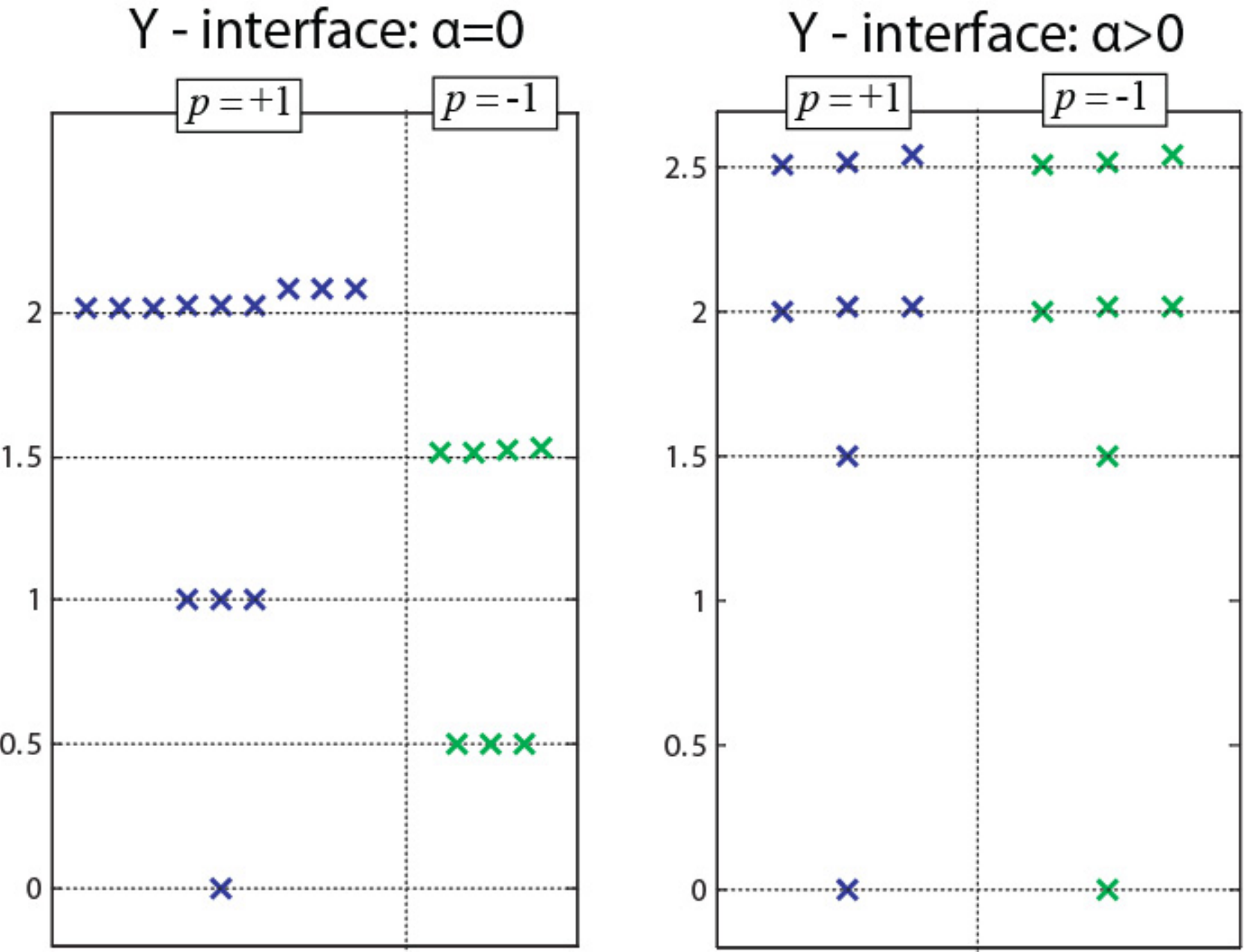}
\caption{The spectrum of scaling dimensions $\Delta$ obtained for the Y-interface of three Ising chains, with $\alpha$ the strength of the coupling at the Y-interface. The scaling dimensions are organized according to parity sectors $p=\pm 1$ of the global $\mathbb Z_2$ symmetry of the Ising model. (left) For the case of $\alpha=0$, i.e. no coupling between different chains, the spectrum is seen to be a product of three times the spectrum of the free boundary Ising chain, see Fig. \ref{fig:ScaleDim}(a), where some numeric error is evident for the larger $\Delta=2$ scaling dimensions. (right) The cases of coupling strength $\alpha=\left\{ 0.25,0.5,0.75,1,1000 \right\} $ all converge to the same spectrum, which symmetric between the $p=\pm 1$ parity sectors.}
\label{fig:YChainCritExp}
\end{center}
\end{figure}
%%%%%%%%%%%%%%%%%%%%%%%%%%%%%%%%%%%%%%%
%%%%%%%%%%%%%%%%%%%%%%%%%%%%%%%%%%%%%%%

We benchmark the Y-interface MERA for an interface of three identical semi-infinite chains, where the each of the chains is a critical Ising model as defined in Eq. \ref{s5e3} and the interface coupling is given by
\begin{eqnarray}
J^{\Yjun} &=&  - \alpha \left[ X(1^A) X(1^B) \right. \\
 &+& \left. X(1^B) X(1^C) + X(1^C) X(1^A) \right], \label{s7e10}
\end{eqnarray}
where the Pauli operators $X(1^A)$, $X(1^B)$, and $X(1^C)$ act on the first site of the semi-infinite lattices $\L_A$, $\L_B$, and $\L_C$ respectively. Once again, tensors $\{u_A,w_A\}$, $\{u_B,w_B\}$, and $\{u_C,w_C\}$ for the critical Ising model are recycled from previous calculations. We optimize the Y-interface tensors $v$ by minimizing the energy of the effective Hamiltonian $H^{\Wilson}$ for interface coupling strengths $\alpha=\left\{ 0, 0.25, 0.5, 0.75, 1, 1000 \right\}$. For each value of $\alpha$ we compute the spectrum of scaling dimensions $\Delta$ associated to the interface by the usual diagonalization of the corresponding scaling superoperator. 

The results for are plotted in Fig. \ref{fig:YChainCritExp}. For $\alpha=0$, which corresponds to three uncoupled semi-infinite Ising chains with free boundary conditions, the spectrum of scaling dimensions obtained from the Y-interface MERA is seen to be indeed the product of three copies of the spectrum of scaling dimensions for free BC Ising model, see Fig. \ref{fig:ScaleDim}, as expected. For all non-zero interface couplings $\alpha>0$, the scaling dimensions converged to an identical spectrum (independent of $\alpha$), with smaller values of $\alpha$ however requiring more transitional layers $\tilde M$ to reach the fixed point, indicating an RG flow to the strong coupling (or large $\alpha$) limit. Indeed, choosing a very large coupling strength, $\alpha=1000$, reproduces the same spectrum of scaling dimensions with only $\tilde M=2$ transitional layers required. Notice that the spectrum obtained for $\alpha>0$, which is identical between $p=\pm 1$ parity sectors of the $\mathbb Z_2$ symmetry of the Ising model, is somewhat similar to that in Fig. \ref{fig:DefectCritExp}(b) for the Ising chain with an infinitely strong bond impurity, $\alpha\rightarrow \infty$, between two sites.

\section{Conclusions} \label{sect:Conclusion}

%%% Modularity

In this manuscript we have built on the theory of minimal updates in holography proposed in Ref. \onlinecite{Evenbly13}, and have argued that a recursive use of the conjectured minimal updates leads to the modular MERA, a surprisingly simple ansatz to describe the ground state of a quantum critical system with defects such as impurities, boundaries, and interfaces. We then have provided compelling numerical evidence that the modular MERA is capable of accurately describing these ground states, by considering a large list of examples.

\subsection{Double conjecture on holographic structure of many-body wave functions}

Notice that the modular MERA is, at its core, a concatenation of two conjectures regarding the structure of the ground state wave-function of quantum critical systems. 

The first conjecture, embodied in the specific of tensors of the MERA, is that the ground state of a quantum critical system contains entanglement that can be removed by means of unitary transformations (disentanglers) acting locally on each length scale \cite{Vidal07}. The second conjecture, the theory of minimal updates \cite{Evenbly13}, is that in order to account for a change of the Hamiltonian in region $\R$, only the tensors inside the causal cone $\C(\R)$ of region $\R$ need to be modified. The results in this paper provide evidence that these two conjectures are correct, and thus teach us about the structure of the ground state wave-function.

\subsection{Computational highlights}

The modular MERA is characterized by a small number of unique tensors that is \emph{independent} of the system size $N$. Similarly, the computational cost of the optimization algorithms is also independent of the system size. As a result, the effects of local defects in an otherwise homogeneous system can be studied directly in the thermodynamic limit, avoiding finite size effects when extracting the universal properties of defects. Furthermore, modularity has the useful implication that tensors can be recycled from one problem to another. For instance, the same tensors $\{u,w\}$ for the homogeneous critical Ising model were used in Sect. \ref{sect:BenchImpurity} for impurity problems, in Sect. \ref{sect:BenchBound} for boundary problems, and in Sect. \ref{sect:BenchInterface} for interface problems. Similarly, the impurity tensors $v$ obtained from a single impurity problem in Sect. \ref{sect:BenchSingle} were later reused in a multiple impurity problem in Sect. \ref{sect:BenchMultiple}.

\subsection{Role of scale and translation invariance}

In this manuscript we have assumed for simplicity that the quantum critical host system was described by a homogeneous Hamiltonian $H$ that was a fixed point of the RG flow, and exploited translation and scale invariance to obtain a MERA for its ground state $\ket{\psi}$ that was fully characterized in terms of just one single pair of tensors $\{u,w\}$. This had the advantage that a finite number of variational parameters (encoded in the pair $\{u,w\}$) was sufficient to completely describe an infinite system. However, the theory of minimal updates does not require scale or translation invariance.

Let us first remove the assumption that the host system is a fixed point of the RG flow. In this case, each layer of tensors of the MERA, corresponding to a different length scale $s$, will be described by a different pair $\{u_s,w_s\}$. Assuming that after some finite scale $M$ the system can effectively be considered to have reached an RG fixed point, characterized by fixed-point tensors $\{u,w\}$, we still obtain a finite description of the ground state of an infinite system in terms of the tensors $\{u_0,w_0,u_1,w_1,\cdots, u_M, w_M\}$ and $\{u,w\}$. The effect of a defect on a region $\R$ can still be accounted for by a modular MERA where the tensors in the causal cone $\C(\R)$ are modified, again by energy minimization over the Wilson Hamiltonian $H^{\Wilson}$ described in Sect. \ref{sect:LogScale}. However, in this case $H^{\Wilson}$ will not have the simple form of Eq. \ref{eq:AD9}, but instead will consist of $s$-dependent terms $h^{\Wilson}_s(s,s+1)$ for $s\leq M$, after which all its terms will be proportional to some coupling $h^*$. This case was briefly mentioned in Sect. \ref{sect:OptMod}.

Let us now also remove the assumption of translation invariance in the host system. Then the MERA for the ground state $\ket{\psi}$ of the host Hamiltonian $H$ requires tensors $\{u(s,r),w(s,r)\}$ that depend both on the scale $s$ and position $r$. In this case the MERA for $\ket{\psi}$ depends on a number of tensors that grows linearly in the system size. In the presence of a defect added to the host Hamiltonian $H$, we can still obtain a modular MERA for the system with the defect by applying a minimal update to the MERA for $\ket{\psi}$. However, in this case we cannot take the thermodynamic limit.

\subsection{Beyond one spatial dimension}

%%% Generalisations to 2D

Although in this manuscript we focused in exploring modularity in $D=1$ spatial dimension, the theory of minimal updates, as proposed in Ref. \onlinecite{Evenbly13}, applies to any spatial dimension $D$, and thus the modular MERA can be also used in systems in $D>1$ dimensions. The algorithms we presented here can be easily generalized to study e.g. a system in $D=2$ dimensions with an impurity (in $D=0$ dimensions). Following the outline described in Sect. \ref{sect:OptMod}, here one would first optimize the MERA for the (impurity free) homogeneous system, and then re-optimize the tensors within the causal cone of the impurity. Notice that, since the causal cone of the impurity is a one-dimensional structure, one would build an effective system (Wilson chain) which is again a semi-infinite chain, as in the $D=1$ case. Instead, the study of a boundary or of an interface in $D=2$ dimensions requires the study of a more complex, $D=2$ effective Hamiltonian, where one dimension corresponds to the extension of the boundary and the other corresponds to scale.

The authors acknowledge Kouichi Okunishi for helpful discussions regarding Wilson's solution to the Kondo problem, and helpful input from Masaki Oshikawa regarding the two-impurity Ising model. Support from the Australian Research Council (APA, FF0668731, DP0878830) is acknowledged. G.E. is supported by the Sherman Fairchild foundation. This research was supported
in part by Perimeter Institute for Theoretical Physics.
Research at Perimeter Institute is supported by the Government of Canada through Industry Canada and by the Province of Ontario through the Ministry of Research and Innovation.

\appendix

\section{Introduction to MERA} \label{sect:MERA}

This appendix contains a brief introduction to entanglement renormalization and the MERA, focusing mostly on a system that is both translation invariant and scale invariant.

\subsection{Coarse graining transformation}\label{sect:CGham}

%%%%% Hilbert space mapping and isometric constraints
We start by reviewing the basic properties of entanglement renormalization and the MERA in a finite, one-dimensional lattice $\mathcal{L}$ made of $N$ sites, where each site is described by a Hilbert space $\mathbb{V}$ of finite dimension $\chi$.

\begin{figure}[!tbh]
\begin{center}
\includegraphics[width=8.5cm]{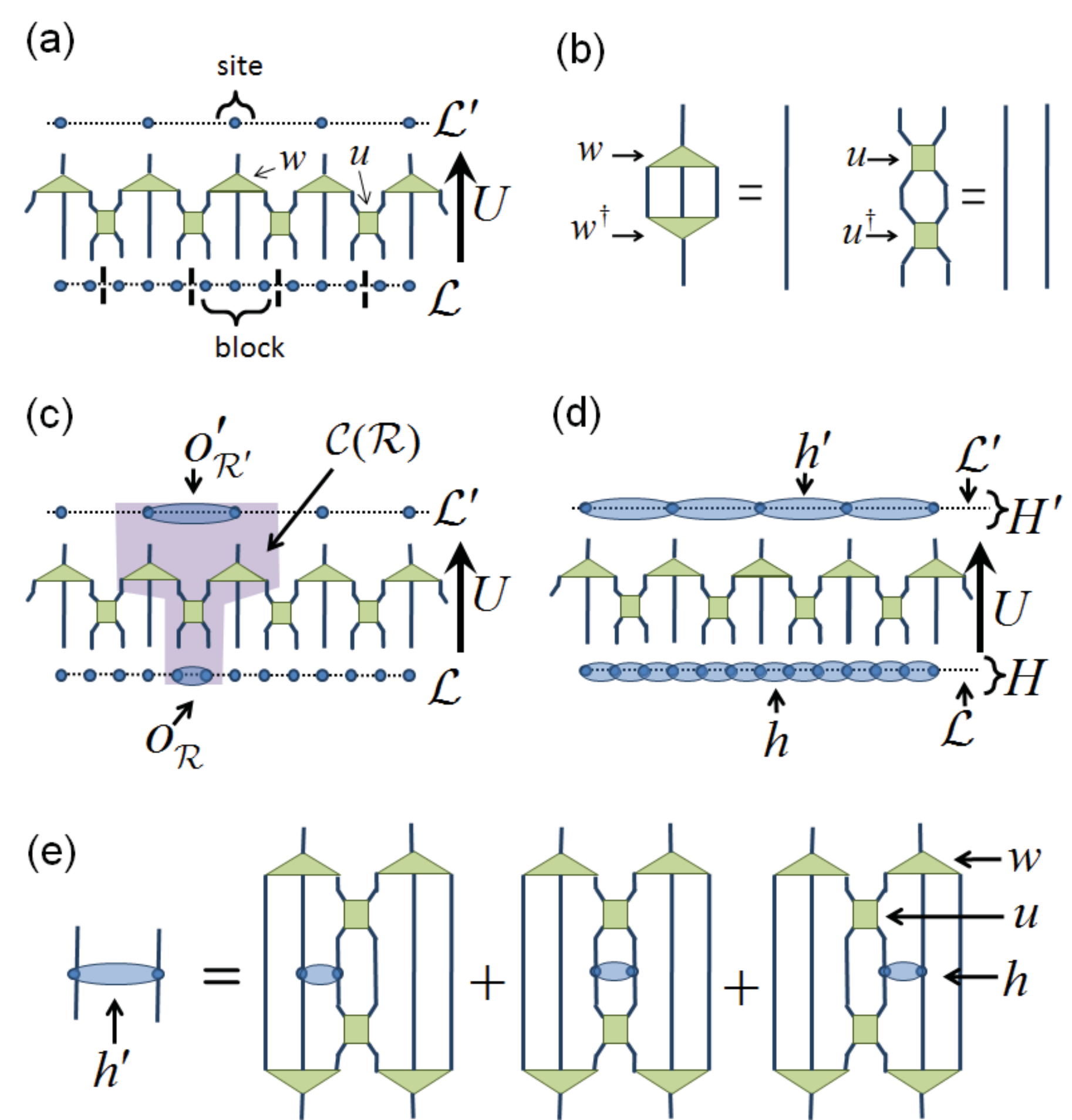}
\caption{(a) The coarse-graining transformation $U$, based on entanglement renormalization, maps a lattice $\L$ made of $N$ sites into a coarse-grained lattice $\L'$ made of $N'=N/3$ sites. (b) The isometries $w$ and disentanglers $u$ that constitute the coarse-graining transformation $U$ are constrained to be isometric, see also Eqs. \ref{eq:B1} and \ref{eq:B2}. (c) An operator $o_{\R}$, supported on a local region $\R \in \L$ made of two contiguous sites, is coarse-grained to a new local operator $o'_{\R'}$, supported on a local region $\R' \in \L'$ made also of two contiguous sites. (d) A nearest neighbor Hamiltonian $H=\sum\nolimits_r h(r,r+1)$ is coarse-grained to a nearest neighbor Hamiltonian $H'=\sum\nolimits_r h'(r,r+1)$. (e) The left, center and right ascending superoperators $\A_L$, $\A_C$ and $\A_R$ can be used to compute the new coupling $h'$ from the initial coupling $h$, see also Eq. \ref{eq:B5}.}
\label{fig:HamIntro}
\end{center}
\end{figure}

Let us consider a coarse-graining transformation $U$ that maps blocks of three sites in $\L$ to single sites in a coarser lattice $\L'$, made of $N'=N/3$ sites, where each site in $\L'$ is described by a vector space $\mathbb{V}'$ of dimension $\chi'$, with $\chi '\le \chi^3$, see Fig. \ref{fig:HamIntro}(a). Specifically, we consider a transformation $U$ that decomposes into the product of local transformations, known as disentanglers $u$ and isometries $w$. Disentangles $u$ are unitary transformations that act across the boundaries between blocks in $\L$,
\begin{equation}
u^{\dagger}:\mathbb V^{ \otimes 2}  \mapsto \mathbb V^{ \otimes 2},\;\;\; u^\dag  u = \mathbb I^{ \otimes 2},\label{eq:B1}
\end{equation}
where $\mathbb I$ is identity on $\mathbb{V}$, while isometries $w$ implement an isometric mapping of a block of three sites in $\L$ to a single site in $\L'$,
\begin{equation}
w^{\dagger}:\mathbb V^{ \otimes 3}  \mapsto \mathbb V',\;\;\; w^\dag  w = \mathbb I',\label{eq:B2}
\end{equation}
where $\mathbb I'$ is the identity operator on $\mathbb{V}'$. The isometric constraints on disentanglers $u$ and isometries $w$ are expressed pictorially in Fig. \ref{fig:HamIntro}(b).

%%%%%%%%%%%%%%%%%%%%%%%%%%%%%%%%%%%%%
%%%%%%%%%%%%%%%%%%%%%%%%%%%%%%%%%%%%%
%%%%%%%%%%%%%%%%%%%%%%%%%%%%%%%%%%%%%%%
%%%%%%%%%%%%%%%%%%%%%%%%%%%%%%%%%%%%%%%

%%%%% Locality: local operators, causal cones, Hamiltonian
An important property of the coarse-graining transformation $U$ is that, by construction, it preserves \emph{locality}. Let $o_{\R}$ be a local operator defined on a region $\R$ of two contiguous sites of lattice $\mathcal{L}$. This operator transforms under coarse-graining as,
\begin{equation}
o_{\R} \stackrel{U}{\longrightarrow} o'_{\R'},\label{eq:B3}
\end{equation}
where the new operator $o'_{\R'}$ is supported on a region $\R'$ of two contiguous sites in lattice $\mathcal{L}'$, see Fig. \ref{fig:HamIntro}(c). The coarse-grained operator $o'_{\R'}$ remains local due to the specific way in which transformation $U$ decomposes into local isometric tensors $u$ and $w$. Indeed, in $U^{\dagger} o_{\R} U$, most tensors in $U$ annihilate to identity with their conjugates in $U^{\dagger}$. The causal cone $\C(\R)$ of a region $\R$ is defined as to include precisely those tensors that do not annihilate to identity when coarse-graining an operator supported on $\R$, and it thus tracks how region $\R$ itself evolves under coarse-graining.

In particular, a local Hamiltonian $H$ on $\L$ will be coarse-grained into a local Hamiltonian $H'$ on $\L'$,
\begin{equation}
H=\sum\limits_r h(r,r+1) \stackrel{U}{\longrightarrow} H'=\sum\limits_r h'(r,r+1), \label{eq:B4}
\end{equation}
see Fig. \ref{fig:HamIntro}(d). The local coupling $h'$ of the coarse-grained Hamiltonian $H'$ can be computed by applying the (left, center, right) ascending superoperators $\mathcal A_L$, $\mathcal A_C$ and $\mathcal A_R$ to the coupling $h$ of the initial Hamiltonian,
\begin{equation}
h' = \mathcal A_L \left(h\right) + \mathcal A_C \left(h\right) + \mathcal A_R \left(h\right), \label{eq:B5}
\end{equation}
see Fig. \ref{fig:HamIntro}(e).

%%%%% MERA
The coarse-graining transformation $U$ can be repeated $T\approx \log_3(N)$ times to obtain a \emph{sequence} of local Hamiltonians,
\begin{equation}
H_0 \stackrel{U_1}{\longmapsto}  H_{1} \stackrel{U_2}{\longmapsto}  \cdots \stackrel{U_{T}}{\longmapsto}  H_{T}, \label{eq:B6}
\end{equation}
where each of the local Hamiltonian $H_s$ is defined on a coarse-grained lattice $\L_s$ of $N_s=N/(3^s)$ sites. Notice the use of subscripts to denote the level of coarse-graining, with the initial lattice $\mathcal L_0 \equiv \mathcal L$ and Hamiltonian $ H_0 \equiv H$. The final coarse-grained Hamiltonian $H_T$ in this sequence, which is defined on a lattice $\L_T$ of $N_T\approx 1$ sites, can be exactly diagonalized so as to determine its ground state $\ket{\psi_T}$. As a linear (isometric) map, each transformation $U_s$ can also be used to fine-grain a quantum state $|\psi_s\rangle$ defined on $\L_{s}$ into a new quantum state $|\psi_{s-1}\rangle$ defined on $\L_{s-1}$,
\begin{equation}
\left| {\psi_{s-1} } \right\rangle  = U_s \left| {\psi_s } \right\rangle.
\end{equation}
Thus a quantum state $|\psi_0\rangle$ defined on the initial lattice $\mathcal L_0$ can be obtained by fine graining state $\ket{\psi_T}$ with the transformations $U_s$ as,
\begin{equation}
\left| {\psi_0 } \right\rangle  = U_1 U_2  \cdots U_T \left| {\psi_T } \right\rangle. \label{eq:B7}
\end{equation}
If each of the transformations $U_s$ has been chosen as to properly preserve the low energy subspace of the Hamiltonian $H_{s-1}$, such that $H_s$ is a low-energy effective Hamiltonian for $H_{s-1}$, then $\left| {\psi_0 } \right\rangle$ is a representation of the ground state of the initial Hamiltonian $H_0$. More generally, the multi-scale entanglement renormalization ansatz (MERA) is the class of states that can be represented as Eq. \ref{eq:B7} for some choice of $\{ U_1, U_2, \ldots, U_T \}$ and $\ket{\psi_T}$.

For a generic choice of local Hilbert space dimensions $\chi_0, \chi_1, \cdots, \chi_{T-1}$ (where $\chi_0 \equiv \chi$), only a subset of all states of lattice $\mathcal{L}$ can be represented in Eq. \ref{eq:B7}, whereas the choice $\chi_s = \chi^{3^s}$ allows for a (computationally inefficient) representation of any state of the lattice.

\subsection{Scale invariant MERA}\label{sect:ScaleMERA}

We now move to discussing the MERA for a quantum critical system that is both scale invariant and translation invariant. We describe how universal information of the quantum critical point can be evaluated, by characterizing the scaling operators and their scaling dimensions. We also review the power-law scaling of two-point correlators. In this appendix, fixed-point objects (e.g. $U$, $H$, $\{u,w\}$, etc) are denoted with a star superscript (as $U^*$, $H^*$, $\{u^*,w^*\}$, etc), whereas in the main text of this manuscript we did not use a star superscript to ease the notation.

\begin{figure}[!tbh]
\begin{center}
\includegraphics[width=8.5cm]{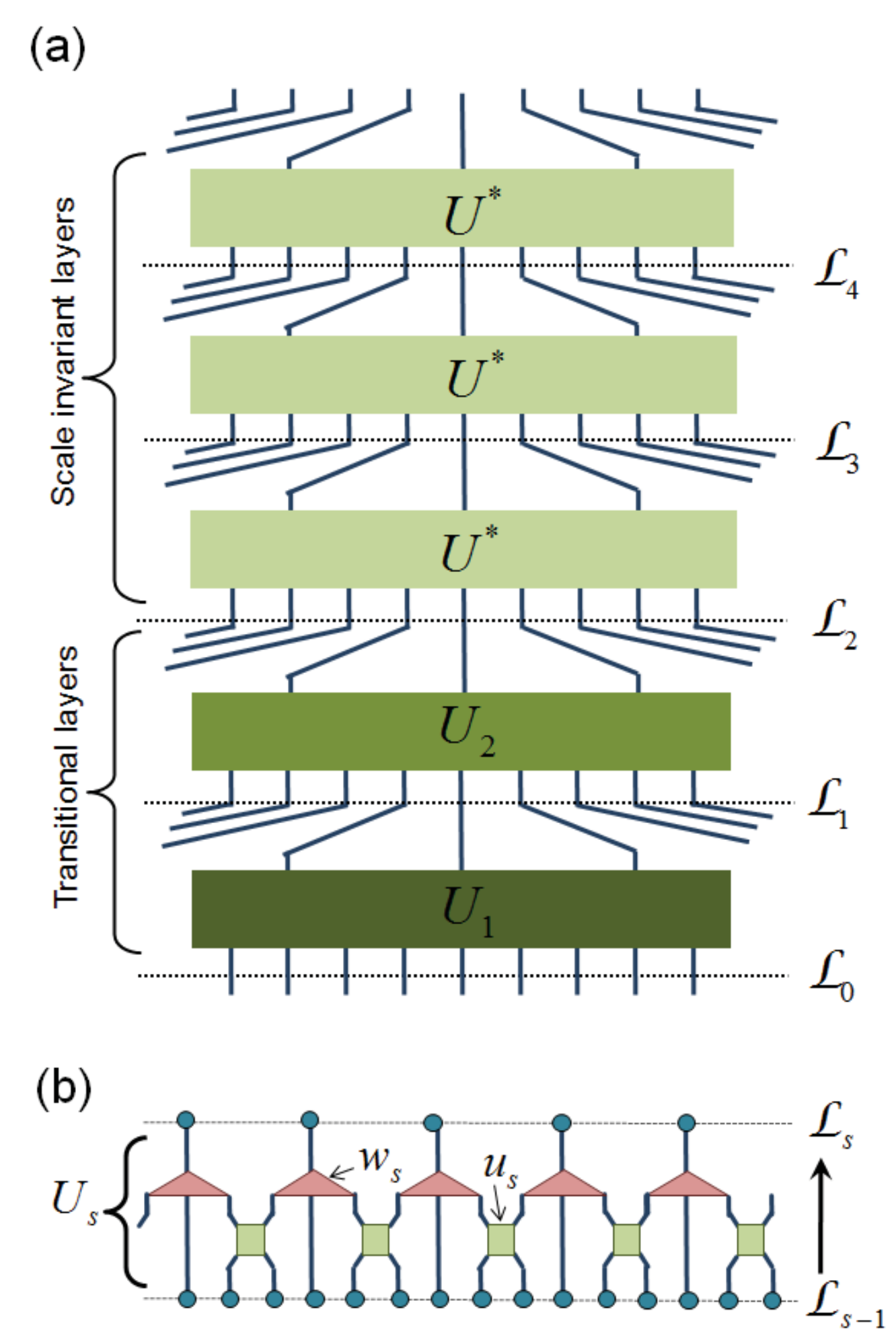}
\caption{(a) A scale invariant MERA consists of some number $M$ of transitional layers with coarse-graining maps $\{U_1, U_2,\ldots, U_M \}$, here $M=2$, followed by an infinite sequence of scaling layers, with a scale invariant map $U^*$. (b) Each $U_s$ of the scale invariant MERA is a coarse-graining transformation composed of local tensors $\{u_s, w_s \}$.}
\label{fig:ScaleIntro}
\end{center}
\end{figure}

Let $\L_0$ be an infinite lattice and let $H_0$ denote a translation invariant, quantum critical Hamiltonian. We assume that this Hamiltonian tends to a fixed point of the RG flow of Eq. \ref{eq:B6}, such that all coarse-grained Hamiltonians $H_s$ are proportionate to a fixed-point Hamiltonian $H^*$ for some sufficiently large $s$. Specifically, the coarse-grained Hamiltonians in the scale invariant regime are related as $H_s=H_{s-1}/\Lambda$, where $\Lambda=3^z$ with $z$ is the dynamic critical exponent of the Hamiltonian (i.e. $z=1$ for a Lorentz invariant quantum critical point). Equivalently, the local couplings that define that Hamiltonians are related as $h_s=h_{s-1}/\Lambda$. For concreteness, let us assume that the initial Hamiltonian $H_0$ reaches the scale invariant (Lorentz invariant) fixed point after $s=2$ coarse-grainings, such that its RG flow can be written,
\begin{equation}
H_0 \stackrel{U_1}{\longmapsto} H_{1} \stackrel{U_2}{\longmapsto} H^* \stackrel{U^*}{\longmapsto} \frac{1}{3} H^* \stackrel{U^*}{\longmapsto} \frac{1}{9} H^* \stackrel{U^*}{\longmapsto}\cdots, \label{eq:B8}
\end{equation}
where $U^*$ represents the scale invariant coarse-graining transformation for $H^*$. In this case, the ground state $\ket{\psi_0}$ of the Hamiltonian $H_0$ can be represented by the infinite sequence of coarse-graining transformations,
\begin{equation}
\left| {\psi_0 } \right\rangle  = U_1 U_2 U^* U^* U^* \cdots  \label{eq:B9}
\end{equation}
see Fig. \ref{fig:ScaleIntro}. The class of states that can be represented as Eq. \ref{eq:B9} are called scale invariant MERA. The scale-dependent transformations before scale invariance, here $U_1$ and $U_2$, correspond to transitional layers of the MERA. These are important to diminish the effect of any RG irrelevant terms potentially present in the initial Hamiltonian, which break scale invariance at short distances. In general, the number $M$ of transitional layers required will depend on the specific critical Hamiltonian under consideration. [Strictly speaking, scale invariance is generically only attained after infinitely many transitional layers, but in practice a finite number $M$ of them often offers already a very good approximation]. We call the fixed-point coarse-graining transformation $U^*$ scale invariant. Notice that the scale invariant MERA, which describes a quantum state on an infinite lattice, is defined in terms of a small number of unique tensors. Each transitional map $U_s$ is described by a pair of tensors $\{u_s, w_s \}$ and the scale invariant map $U^{*}$ is described by the pair $\{u^*, w^* \}$.

%%%%%%%%%%%%%%%%%%%%%%%%%%%%%%%%%%%%%
%%%%%%%%%%%%%%%%%%%%%%%%%%%%%%%%%%%%%
%%%%%%%%%%%%%%%%%%%%%%%%%%%%%%%%%%%%%%%
%%%%%%%%%%%%%%%%%%%%%%%%%%%%%%%%%%%%%%%

\begin{figure}[!htbp]
\begin{center}
\includegraphics[width=8.5cm]{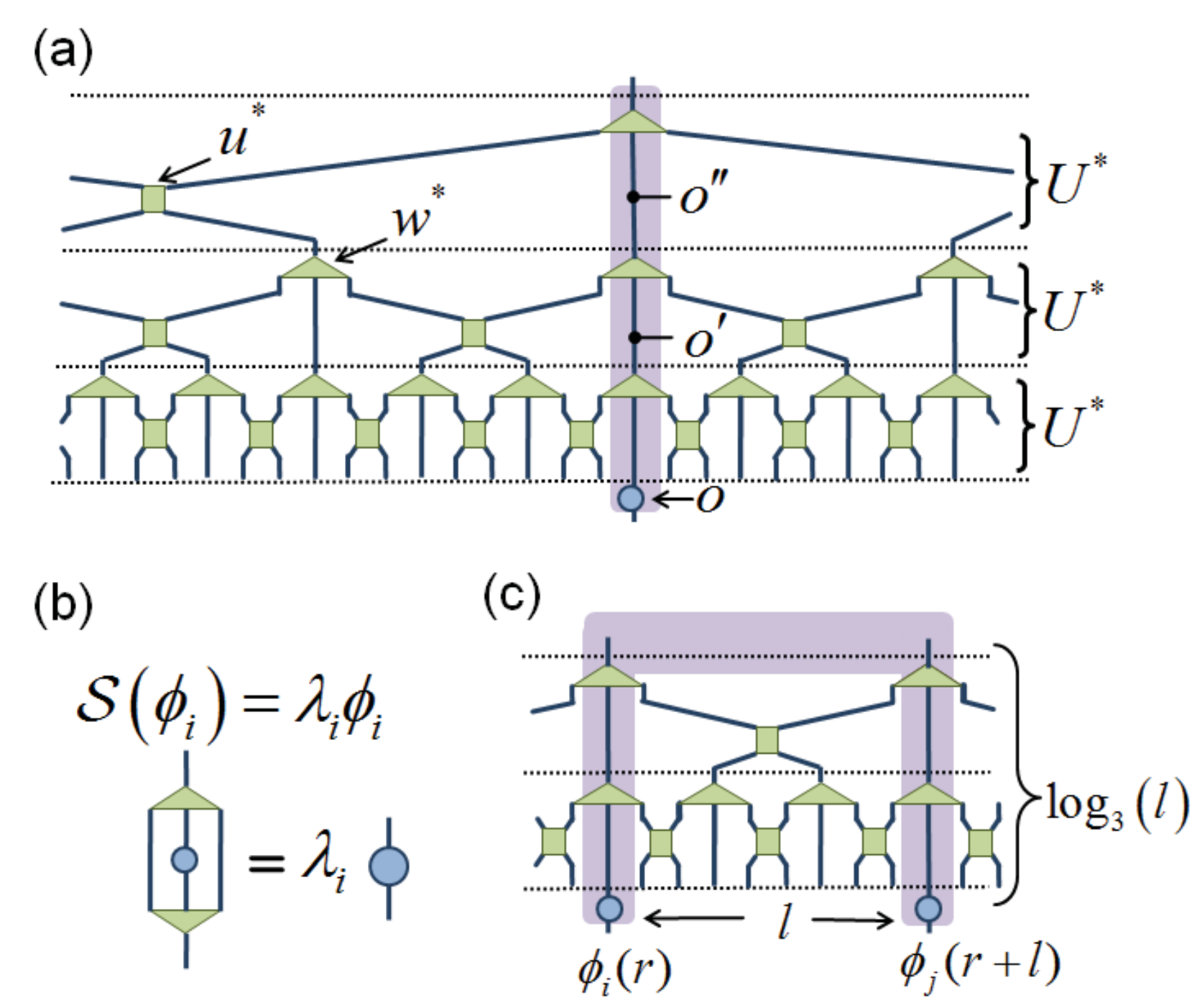}
\caption{(a) Scale invariant MERA composed of an infinite sequence of scale invariant maps $U^*$, which are defined in terms of a single pair of tensors $\{ u^*, w^* \}$. A one-site operator $o$ is coarse-grained into new one-site operators $o'$ and $o''$. (b) The scaling superoperator $\mathcal S$ acts covariantly upon scaling operators $\phi_i$, see also Eq. \ref{eq:B10b}. (c) Two scaling operators $\phi_i$ and $\phi_j$ that are separated by $l$ lattice sites are coarse-grained onto neighboring sites after $\log_3 (l)$ maps $U^{*}$.}
\label{fig:TwoCorr}
\end{center}
\end{figure}

We now discuss how scaling operators and their scaling dimensions can be evaluated from the scale-invariant MERA. This is covered in more detail in e.g. Refs. \onlinecite{Pfeifer09, Montangero09, Evenbly11b}. For simplicity, let us consider a scale invariant MERA with no transitional layers, that is composed of an infinite sequence of a scale invariant map $U^*$, described by a single pair $\{u^*,w^*\}$. As shown in Fig. \ref{fig:TwoCorr}(a), a one-site operator $o$, placed on certain lattice sites, is coarse-grained under the action of layer $U^*$ into new one-site operator $o'$. This coarse-graining is implemented with the one-site scaling superoperator $\mathcal S$,
\begin{equation}
o'	= \mathcal{S}\left( o \right), \label{eq:B10}
\end{equation}
where $\mathcal S$ is defined in terms of the isometry $w^*$ and its conjugate, see also Fig. \ref{fig:TwoCorr}(b). The (one-site) scaling operators $\phi_{i}$ are defined as those operators that transform covariantly under action of $\mathcal S$,
\begin{equation}
	\mathcal{S}(\phi_{i}) = \lambda_{i} \phi_{i},~~~~~~~\Delta_{i} \equiv -\log_3 \lambda_{i}, \label{eq:B10b}
\end{equation}
where $\Delta_{i}$ is the scaling dimension of scaling operator $\phi_{i}$. As is customary in RG analysis, the scaling operators $\phi_{i}$ and their scaling dimensions $\Delta_{i}$ can be obtained through diagonalization of the scaling superoperator $\S$.

One can obtain explicit expressions for two-point correlation functions of the scale invariant MERA based upon their scaling operators, as we now describe. Let us suppose that two scaling operators $\phi_{i}$ and $\phi_{j}$ are placed on special sites $r$ and $r+l$ that are at a distance of $l = 3^q$ sites apart for positive integer $q$, as shown in Fig. \ref{fig:TwoCorr}(c). The correlator $\left\langle \phi_i \left( r \right) \phi_j \left( r+l\right) \right\rangle$ can be evaluated by coarse-graining the scaling operators until they occupy adjacent sites, where the expectation value
\begin{equation}
	C_{ij} \equiv \left\langle \phi_{i}(r) \phi_{j}(r +1)\right\rangle
	= \textrm{Tr} \big( (\phi_{i}\otimes \phi_{j}) {\rho} \big). \label{eq:B12c}
\end{equation}
can then be evaluated with the local two-site density matrix $\rho$ (which is the same at every level of the MERA due to scale invariance).

For each level of coarse-graining applied to the scaling operators $\phi_{i}$ and $\phi_{j}$, we pick up a factor of the eigenvalues of the scaling operators, as described Eq. \ref{eq:B10b}, and the distance $l$ between the scaling operators shrinks by a factor of 3, see Fig. \ref{fig:TwoCorr}(c), which leads to the relation
\begin{equation}
\left\langle \phi_i \left( r \right) \phi_j \left( r+l\right) \right\rangle = \lambda_i \lambda_j ~ \left\langle \phi_i \left( r \right) \phi_j \left( r+l/3\right) \right\rangle. \label{eq:B11}
\end{equation}
Notice that the scaling operators are coarse-grained onto adjacent sites after $T = \log_3 |l|$ levels, thus through iteration of Eq. \ref{eq:B11} we have
\begin{align}
 \left\langle {\phi _i  (r )\phi _j  (r+l )} \right\rangle  &= \left( {\lambda _i  \lambda _j  } \right)^{\log _3 |l|} \left\langle {\phi _i  (r)\phi _j  (r+1)} \right\rangle \nonumber \\
  &= \left( {3^{ - \Delta _i  } 3^{ - \Delta _j  } } \right)^{\log _3 |l|} C_{i j }  \nonumber \\
  &= \frac{{C_{i j } }}{{\left| {l } \right|^{\Delta _i   + \Delta _j  } }} . \label{eq:B12}
\end{align}
where constant $C_{\alpha\beta}$ is the expectation value of the correlators evaluated on adjacent sites,
\begin{equation}
	C_{ij} \equiv \left\langle \phi_{i}(r) \phi_{j}(r +1)\right\rangle
	= \tr \big( (\phi_{i}\otimes \phi_{j}) {\rho} \big). \label{eq:B12b}
\end{equation}
Thus it is seen that the correlator of two scaling operators $\phi_{i}$ and $\phi_{j}$ scales polynomially in the distance between the operators, with an exponent that is the sum of their corresponding scaling dimensions $\Delta_i$ and $\Delta_j$, in agreement with predictions from CFT \cite{Francesco97, Henkel99}.

Notice that Eq. \ref{eq:B12} was derived from structural considerations of the MERA alone and, as such, holds regardless of how the tensors in the scale invariant MERA have been optimized. This argument is only valid for the chosen special locations $r$ and $r+l$. For a generic pair of locations, the polynomial decay of correlations may only be obtained after proper optimization (for instance, via energy minimization) of the MERA so as to approximate the ground state of a translation invariant, quantum critical Hamiltonian $H$.

%%%%%%%%%%%%%%%%%%%%%%%%%%%%%%%%%%%%%
%%%%%%%%%%%%%%%%%%%%%%%%%%%%%%%%%%%%%
%%%%%%%%%%%%%%%%%%%%%%%%%%%%%%%%%%%%%%%
%%%%%%%%%%%%%%%%%%%%%%%%%%%%%%%%%%%%%%%

%%%%%%%%%%%%%%%%%%%%%%%%%%%%%%%%%%%%%
%%%%%%%%%%%%%%%%%%%%%%%%%%%%%%%%%%%%%
\begin{figure}[!tbh]
\begin{center}
\includegraphics[width=5cm]{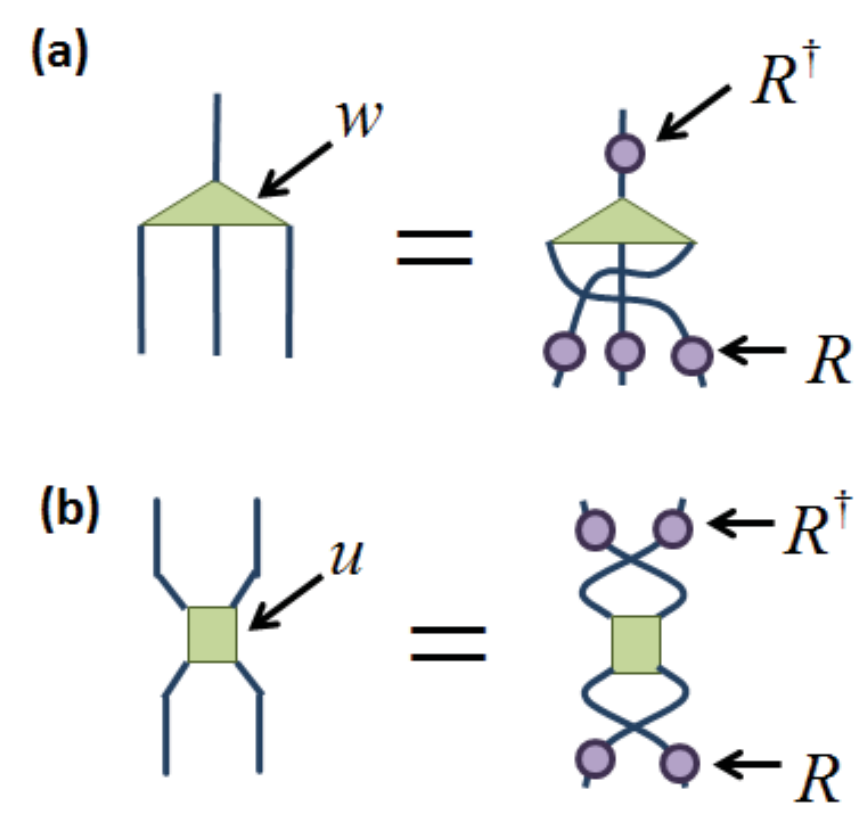}
\caption{(a) The definition of reflection symmetry for a ternary isometry $w$, which involves spatial permutation of indices as well as enacting a unitary matrix $R$ on each index. (b) The definition of reflection symmetry for a disentangler $u$.}
\label{fig:RefSym}
\end{center}
\end{figure}
%%%%%%%%%%%%%%%%%%%%%%%%%%%%%%%%%%%%%%%
%%%%%%%%%%%%%%%%%%%%%%%%%%%%%%%%%%%%%%%

\section{Reflection symmetry} \label{sect:RefSym}

In this appendix we describe how symmetry under spatial reflection can be exactly enforced into the MERA. This is done by directly incorporating reflection symmetry in each of the tensors of the MERA (note that an equivalent approach, dubbed inversion symmetric MERA, was recently proposed in Ref. \onlinecite{Kao13}). Such a step was found to be key in applications of the modular MERA to quantum critical systems with a defect, as considered in Sect. \ref{sect:Bench}. Indeed, we found that in order for the modular MERA to be an accurate representation of the ground state of a quantum critical system with a defect, the homogeneous system (that is, the system in the absence of the defect) had to be addressed with a reflection invariant MERA.

Let us describe how the individual tensors of the MERA, namely the isometries $w$ and disentanglers $u$, can be chosen to be reflection symmetric, i.e.
\begin{equation}
w = {\rm{Rft}}\left( w \right),\; \; u = {\rm{Rft}}\left( u \right), \label{s8e1}
\end{equation}
see Fig. \ref{fig:RefSym}. Here ${\rm{Rft}}\left( \cdot \right)$ is a superoperator that denotes spatial reflection, which squares to the identity. The spatial reflection on a tensor involves permutation of its indices, as well as a `reflection' within each index, as enacted by a unitary matrix $R$ such that $R^2 = I$. The latter is needed because each index of the tensor effectively represents several sites of the original system, which also need to be reflected (permuted). Matrix $R$ has eigenvalues $p=\pm 1$ corresponding to reflection symmetric and reflection antisymmetric states, respectively. It is convenient, though not always necessary, to work within a basis such that each $\chi$-dimensional index $i$ decomposes as $i=(p,\alpha_p)$, where $p$ labels the parity ($p=1$ for even parity and $p=-1$ for odd parity) and $\alpha_p$ labels the distinct values of $i$ with parity $p$. In such a basis, $R$ is diagonal, with the diagonal entries corresponding to the eigenvalues $p=\pm 1$.

Let us turn our attention to the question of how reflection symmetry, as described in Eq. \ref{s8e1}, can be imposed on the MERA tensors. For concreteness, we consider an isometry $w$ (analogous considerations apply to a disentangler). Notice that we cannot just symmetrize $w$ under reflections directly,
\begin{equation}
w' = \frac{1}{2} \left( w+ \textrm{Rft}(w) \right), \label{s8e2}
\end{equation}
because the new, reflection symmetric tensor $w'$ will no longer be isometric. Instead, we can include an additional step in the optimization algorithm that symmetrizes the \emph{environment} of the tensors before each tensor is updated. In the optimization of the MERA \cite{Evenbly09b}, in order to update an isometry $w$ one first computes its linearized environment $\Upsilon _w $. Now, to obtain an updated isometry that is reflection symmetric, we first symmetrize its environment,
\begin{equation}
\Upsilon _w  \mapsto \Upsilon '_w  = \Upsilon _w  + {\rm{Rft}}\left( {\Upsilon _w } \right). \label{s8e3}
\end{equation}
In this way we ensure that the updated isometry $w'$ (which is obtained through a SVD of $\Upsilon '_w$, see Ref. \onlinecite{Evenbly09b}), is reflection symmetric, yet also retains its isometric character. Likewise the environments $\Upsilon _u $ of disentanglers $u$ should also be symmetrized.

%%%%%%%%%%%%%%%%%%%%%%%%%%%%%%%%%%%%%
%%%%%%%%%%%%%%%%%%%%%%%%%%%%%%%%%%%%%
\begin{figure}[!tb]
\begin{center}
\includegraphics[width=8.5cm]{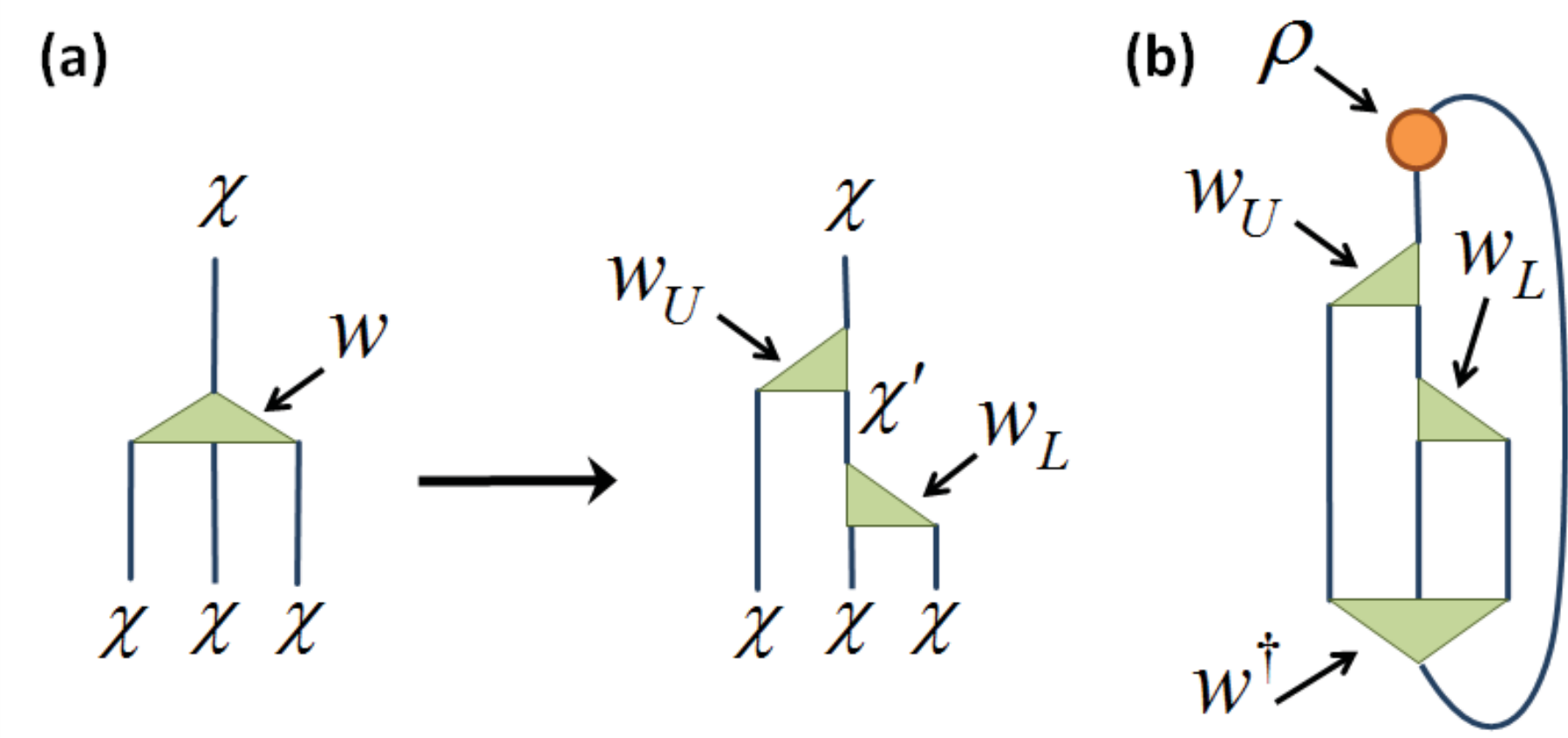}
\caption{(a) A isometry $w$ from the ternary MERA, which coarse-grains three $\chi$-dimensional lattice sites into a single $\chi$ dimensional lattice site, is decomposed into upper and lower binary isometries, $w_{U}$ and $w_{L}$. The index connecting the upper and lower binary isometries is chosen at an independent dimension $\chi '$. (b) The upper and lower binary isometries $w_{U}$ and $w_{L}$ should be chosen to maximize their overlap with the ternary isometry $w$ against the one-site density matrix $\rho$, see Eq. \ref{s9e1}.}
\label{fig:IsoSplit}
\end{center}
\end{figure}
%%%%%%%%%%%%%%%%%%%%%%%%%%%%%%%%%%%%%%%
%%%%%%%%%%%%%%%%%%%%%%%%%%%%%%%%%%%%%%%

\section{Decomposition of isometries} \label{sect:IsoDecomp}

In the formulation of modular MERA described in Sect. \ref{sect:Modularity} it was convenient to decompose some of the isometries $w$ of the MERA used to describe the homogeneous system into pairs of upper and lower isometries $w_{U}$ and $w_{L}$, as depicted in Fig. \ref{fig:IsoSplit}(a). In this section we discuss how this can be accomplished.

Let $\chi$ denote the bond dimension of the indices of the isometry $w$, and let $\chi'$ denote the index connecting the upper and lower isometries $w_{U}$ and $w_{L}$. Since $\chi'$ effectively represents two sites with bond dimension $\chi$, we have that the isometric character of $w_U$ requires $\chi'\le \chi^2$. We should perform this decomposition such that it does not change the quantum state described by the MERA (perhaps to within some very small error). Therefore the best choice of upper $w_{U}$ and lower $w_{L}$ isometries follows from maximizing their overlap with the isometry $w$ against the one-site density matrix $\rho$. That is, we choose them such that they maximize
\begin{equation}
\rm{Tr} \left( \rho  w_U w_L w ^\dag  \right), \label{s9e1}
\end{equation}
see Fig. \ref{fig:IsoSplit}(b). Given the density matrix $\rho$ and isometry $w$, one can obtain $w_U$ and $w_L$ by iteratively maximizing the above trace over each of the two tensors, one at a time. Ideally, we would like the decomposition of $w$ into the product of $w_U$ and $w_L$ to be exact, that is, such that such that $\rm{tr} \left( \rho  w_U w_L w ^\dag  \right)=1$. This is typically only possible for $\chi' = \chi^2$. However, in practice we find that for choice of bond dimension $\chi '$ between one or two times the dimension $\chi$, i.e. $\chi < \chi '< 2\chi$, the above trace is already $1-\epsilon$ with $\epsilon$ negligibly small. The use of a $\chi'$ smaller than $\chi^2$ results in a reduction of computational costs.

\begin{thebibliography}{99}
\bibitem{Proper}
Without the action of disentanglers $u$, short-range entanglement is preserved under coarse-graining, leading to an effective description that still contains some of the original, small scale degrees of freedom. As a result, the coarse-graining transformation is not a proper realization of the RG. Indeed, two many-body systems that differ in irrelevant short-range details but behave identically at low energies [that is, two many-body systems that flow to the same fixed-point of the RG] will flow to different fixed-points of the coarse-graining transformation, because after being coarse-grained they still retain small scale details that reveal their origin.
\end{thebibliography}
\end{document}